\documentclass[10pt]{article}%

\usepackage[a4paper,top=2.5cm,left=2cm,bottom=2.5cm,right=2cm]{geometry}
\usepackage{amsmath}
\usepackage{amssymb}
\usepackage{amsthm} 
\usepackage{mathabx} 
\usepackage{graphicx}
\usepackage{hyperref} 
\usepackage[authoryear,sort&compress]{natbib} 
\usepackage{subfigure}  
\usepackage{color}
\usepackage{ulem} 

\RequirePackage[utf8]{inputenc}
\RequirePackage[T2A]{fontenc}
\RequirePackage[english, russian]{babel}

\newtheorem{proposition}{Proposition} 
\newtheorem{lemma}{Lemma} 
\newtheorem{property}{Property} 
\newtheorem{axiom}{Axiom} 
\newtheorem{definition}{Definition} 

\theoremstyle{remark}

\usepackage{lineno}

\begin{document}
\title{Proper Interpretation of Heaps' and Zipf's Laws}

\author{Kim Chol-jun  \\%
\small \textit{Faculty of Physics, \textbf{Kim Il Sung} University, Taesong District, Pyongyang City, DPR Korea }
\\
\small postal code:+850
\\
}
\maketitle
\begin{abstract}
We derive Heaps’ and Zipf’s laws, two fundamental laws in language processes such as text or speech, on the basis of the property of the process: the uniform distribution of words. We exclude the traditional research trend that Zipf’s law is considered as an internal property in the process without proof and other laws are derived from the law. We show that the uniform distribution of words in the process can explain Heaps’ law and furthermore Zipf’s law. This can give an axiomatic interpretation of the laws.
\end{abstract}

\paragraph{keywords:}Zipf's law; Heaps' law; arithmetic growth; power-law distribution; Growth processes; Random/ordered microstructure; Stochastic processes

\paragraph{}
\textbf{Zipf’s law of word frequency in text is a most typical example of power-law distribution. Zipf’s law has been reported in various phenomena, including economic, astrophysical, and so on, and its mechanisms have been developed diversely. Most authors believed and tried to explain those phenomena by a unique mechanism such as the ``rich-get-richer’’ process. However, I would claim that different phenomena occur from different mechanisms so that Zipf’s law in the language process is based on an arithmetic growth of words while other phenomena may be caused by a geometric growth. It has been pursued to develop a mechanism proper to the language process, though not omnipotent.}

\section{Introduction}\label{sec:intro}
The distribution of words in language processes such as text or speech is a problem difficult to study only in a blind statistical way. Despite the fact that tabletop experiments are possible, or even that we can trace one by one word in text, its mechanism is being discussed. The distribution of word frequency seems to have no centrality but a skewness. This distribution is attributed to a power-law distribution called Zipf’s law. 

The power-law distribution has been studied for over a century. \cite{Estoup1916} and \cite{Zipf1932} found that the distribution of word frequency in novels is a power-law one, i.e., the number of word types is reciprocal to word frequency. Many phenomena in economics and astrophysics and so on have been found to follow the power-law distribution \citep[e.g. see][]{Newman2005}. Most of them had also been attributed to Zipf’s law, despite Zipf’s law is originally related to the word distribution.  

Several generative models have been proposed to explain power-law distributions, including Zipf’s law for word frequency. One of the most famous models is the ``preferential attachment'' process \citep{Yule1924, Simon1955, Barabasi1999}. Originally, this model had been developed for the distribution of size of biological genera or species. However, such systems give a geometric growth while the number of words in text increases arithmetically. Furthermore, the original “preferential attachment" model assumed the rate of new words as constant throughout text \citep[e.g. see][]{Li2016}, which is inconsistent with the observation as to text, in particular Heaps’ law \citep{Heaps1978} that says that the rate of new words decreases as going to the end of text. Though the model can be modified so that this rate varies in text, the rule of manipulation to choose words in the process is not verified in practice.

A competitive model might be the ``principle of least effort'' or an optimization model. This model was originally by \cite{Estoup1916} and \cite{Zipf1936, Zipf1949} and developed by \cite{Mandelbrot1953} and later works \citep[e.g. see][]{Ferrer i Cancho2003}. However, the model seems less intuitive and casts a suspicion on people. In fact, it is similar to the principle of least action in physics, which, however, has a unique solution, while text, though of the same content, can vary from person to person, from language to language or even from time to time by a single speaker. Additionally, Zipf’s law is detected in all the languages, including even those that have no articles or prepositions, as in English  \citep{Yu2018}. Zipf’s law has been reported to appear in children’s utterance or schizophrenic speech \citep{Zipf1942, Alexander1998}, for which it is afraid if they try to optimize their speaking. Additionally, \cite{Linders2023} proposed a theory that language might have been optimized along human evolution and/or growth. However, according to them, the distribution of words uttered from humans with low intelligence should be greatly different from Zipf’s law or should have not formed Zipf’s law yet. \cite{Williams2015} showed Zipf’s law for clauses, phrases, words and even for graphemes and letters.

\cite{Yu2018} analyzed texts in 50 languages and obtained commonly appeared 3-segment property. They explained this property by the so-called ``cognitive model,'' which, however, adopts a similar modeling to or seems to be even less practical than ``preferential attachment,'' including jumps of parameters between frequent and rare words. A full version of cognitive or psychological interpretation seems to have not been completed yet. According to \cite{Piantadosi2014}, we could be skeptical of the opinion that Zipf’s law is an internal linguistic property. Zipf’s law appears in other human’s products, such as music. And the words merely meaning planets or months follow the power-law distribution. N-gram also is not a meaningful language process, but giving Zipf’s law. Do such phenomena or human’s activity have a relation to the optimization principle or cognitive process in brain?

Another important model is a ``Monkey’s typing randomly'' or ``intermittent silence'' which was proposed by \cite{Miller1957} and studied on  \citep[e.g. see][]{Li1992}. In this model, a sequence of randomly selected characters turns out to be a word if it is separated by blank spaces inserted randomly as well. It has been shown that the word of randomly generated characters has a power-law distribution similar to Zipf’s law. The model seems to be based only on a blind statistical consideration without sophisticated manipulation. However, the power-law distribution originated from a morphological trend that ``short words’’ are more frequent but less abundant than ``long words''. That trend, in turn, originated only from the way of making words from characters in the model but cannot be reproduced in real text.  

A graphical analysis also has been studied, which, however, seems more complex but less effective \citep[e.g. see][]{Natale2018}. Other manipulative or artificial processes, such as SSR process \citep{Corominas-Murtra2015} or P\'{o}lya's urn model \citep{Polya1930} were studied \citep[also see][]{Natale2018, Tria2014}, but it is not certain if those processes could represent the language process properly. The distribution might be broken if even only an item of those manipulations were to be changed. According to the previous works, Zipf’s law seems so pertinacious or sticky that it appeared in arbitrary text or speech.  

As aforementioned, the power-law distribution had also been found in economic phenomena such as income taxes \citep{Pareto1896}, the population in city \citep{Auerbach1913, Zipf1949} or firm size distribution \citep{Axtell2001}. Based on Gibrat's model \citep{Gibrat1931}, \cite{Gabaix1999} gave an explanation of Zipf's law for population in city. \cite{Chol-jun2022} developed a model called ``geometrically growing system'' that can explain the power-law distribution appearing in broad fields such as demography, epidemiology or econometrics. However, those models seem to have no relevance to glottometrics. In fact, the increase of word occurrences in text seems arithmetic but not geometric. \cite{Marzo2021} analyzed Zipf’s law appearing in various phenomena where the law is evolving variously. Such an evolution of power-law distribution is interpreted in the frame of the geometrically growing system by \cite{Chol-jun2024}. 

To summarize, all the aforementioned models seem to be improper for the language process. As most of previous works pursued, we keep an attitude to interpret Zipf’s law from a more basic assumption or property. Even though Zipf’s law might be an internal property of the language process, the law is statistical and should originate from more basic statistical property of the language process. To find out this basic property, we should delve into text and filter and sort words. In this context, we, first of all, analyze texts, find more basic statistical properties in text and build a practicable mathematical model based on the observation.

\section{Uniform Distribution of Word in Language Process}\label{sec:phen}
First, let's define symbols. Let $x$ stand for the position of a word within text, $X$ for the whole length or the tokens of the text, $y$ for the type number (simply a type) of a newly appeared word and $Y$ for the whole word types within the text. In text analysis, we encounter terms ``token'' and ``type'': the term ``token'' means the total number of words in text, while the term ``type'' means the number of distinct words in text. In other words, if a word appears repeatedly, the token increases while the type is not changed. If a word appears $k$ times in the text, where $k$ is the number of the word or word frequency, its positions form a series: $x_1(y), x_2(y), \dots, x_k(y)$. However, for Heaps’ law, we only concern with the first position so that we denote $x_1(y)$ simply by $x(y)$. The mean span or the periodicity of the former series is denoted by $\bar{x}(y)$. On the other hand, $n(k)$ implies the number of types (or, simply, types) of all words appeared $k$-times throughout the text. Then the type number of such words form a series: $y_1\vert k, y_2\vert k, \dots, y_{n(k)}\vert k$. 

In order to make a proper interpretation of Heaps’ or Zipf’s laws, we need to analyze real language processes. We have analyzed texts and could pick some properties. 

	\begin{property} \label{th:prop1}
	The words are uniformly distributed in their position within text.
	\end{property}
	
\begin{figure*}
\centering
\subfigure[]{\includegraphics[width=0.42\textwidth]{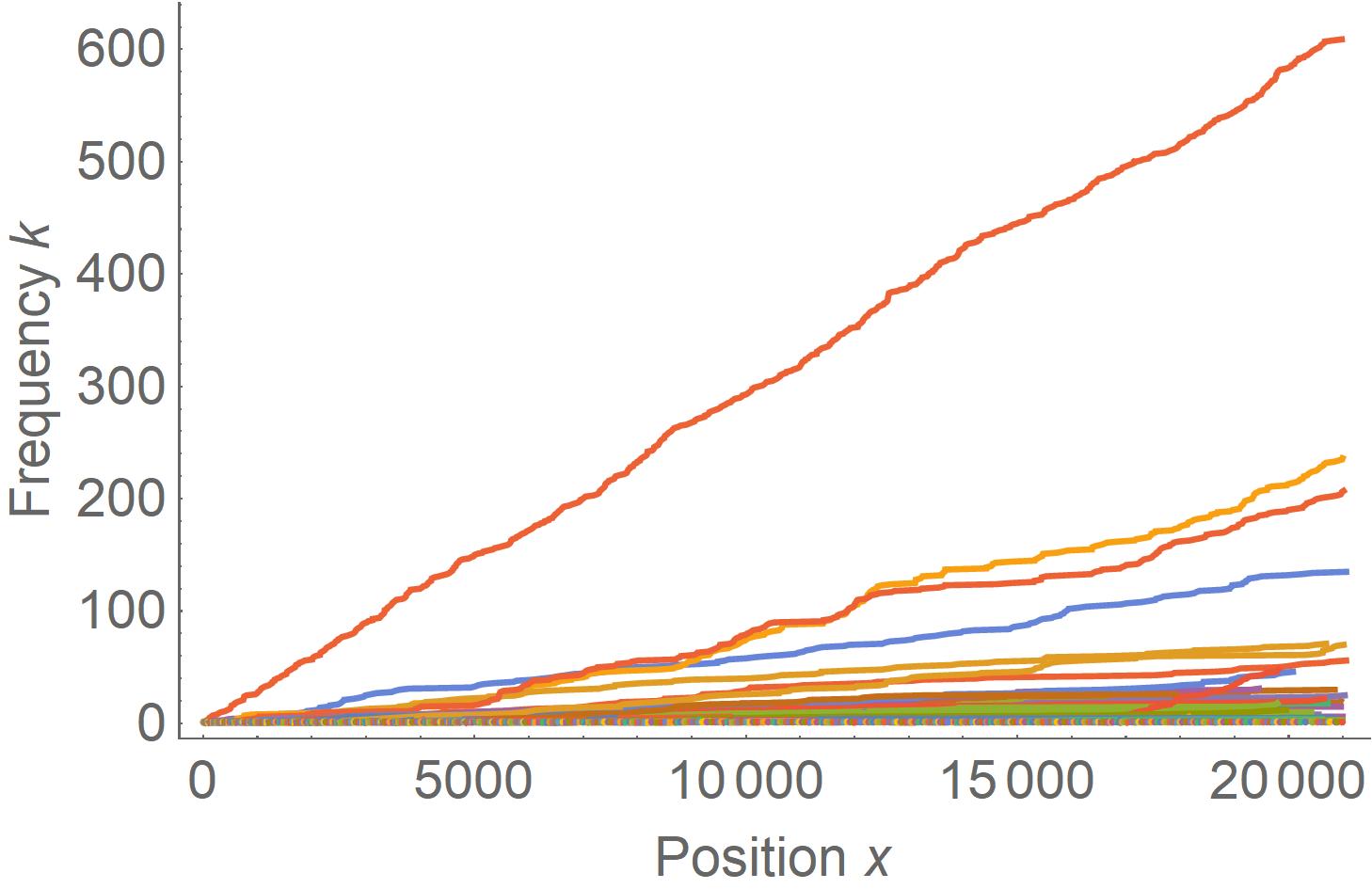}}
\subfigure[]{\includegraphics[width=0.42\textwidth]{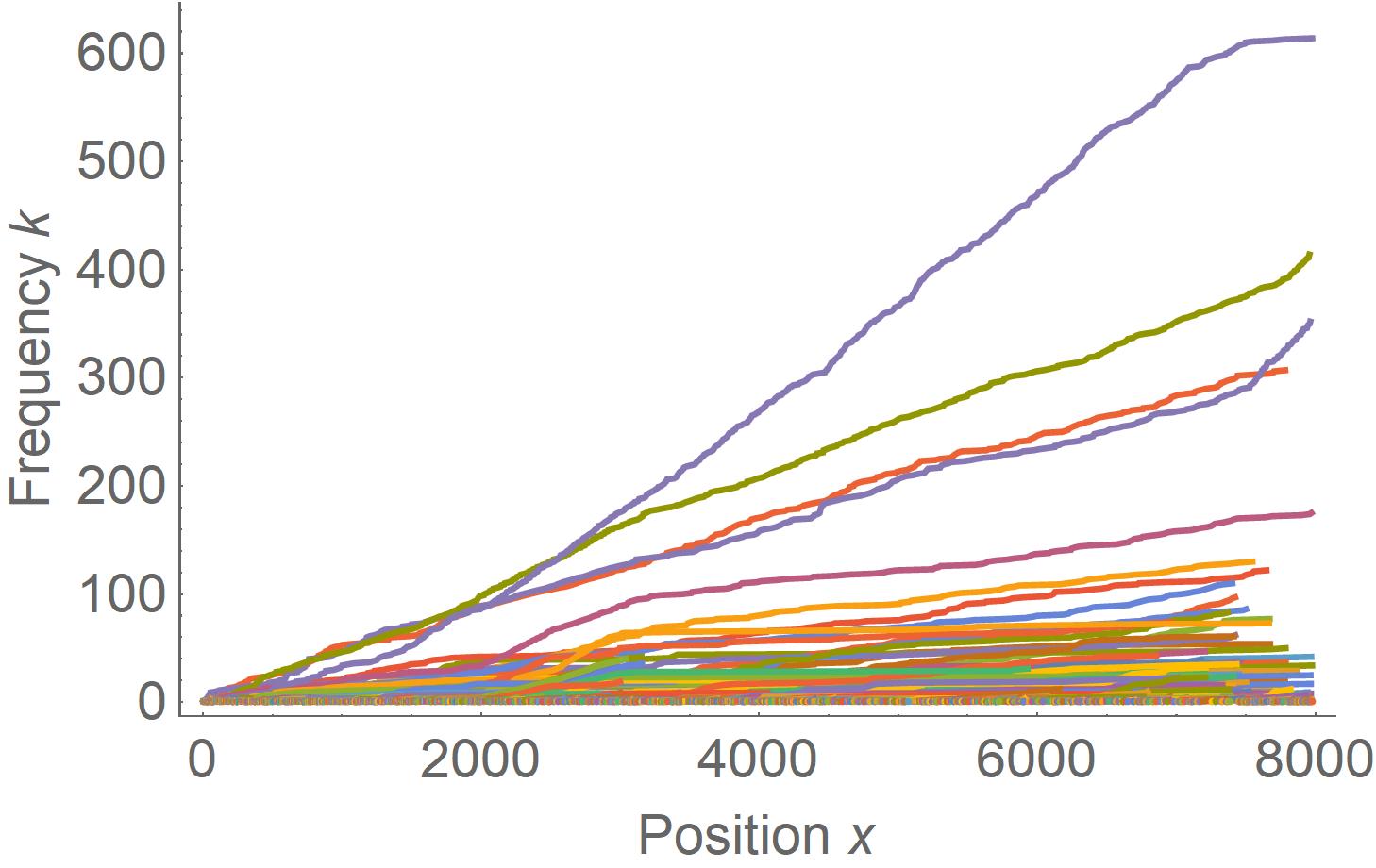}} \\
\subfigure[]{\includegraphics[width=0.23\textwidth]{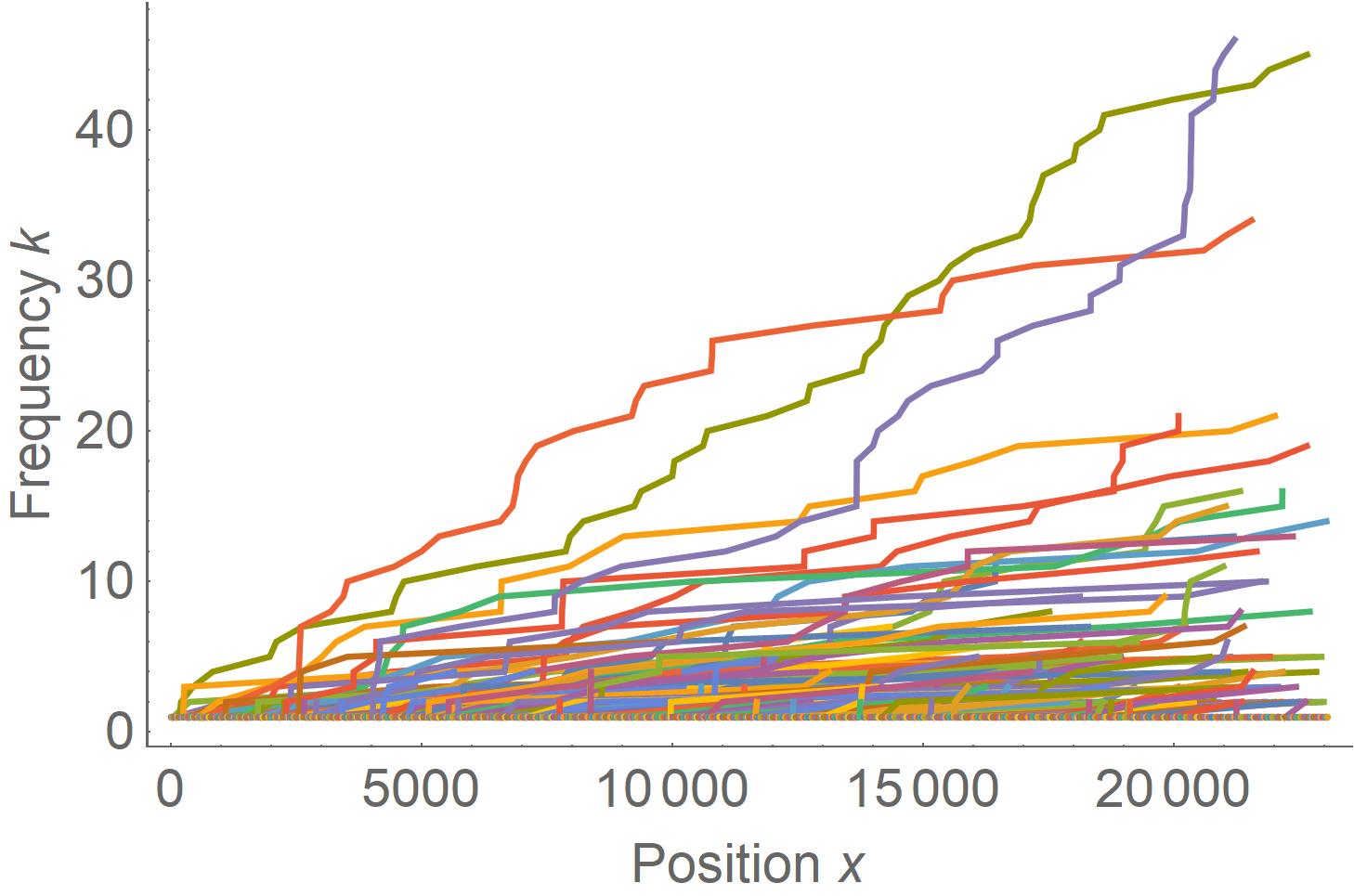}}
\subfigure[]{\includegraphics[width=0.23\textwidth]{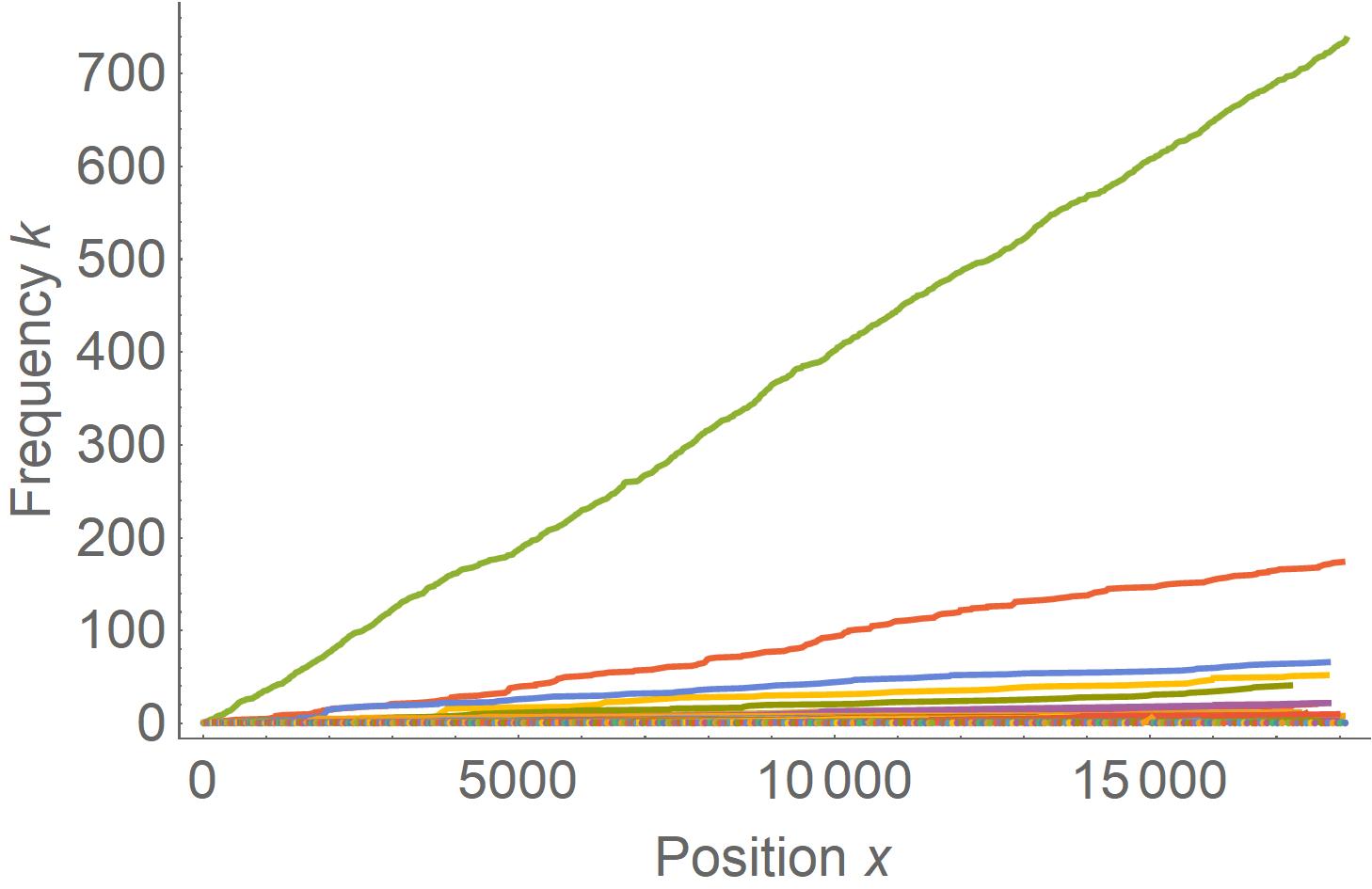}}
\subfigure[]{\includegraphics[width=0.23\textwidth]{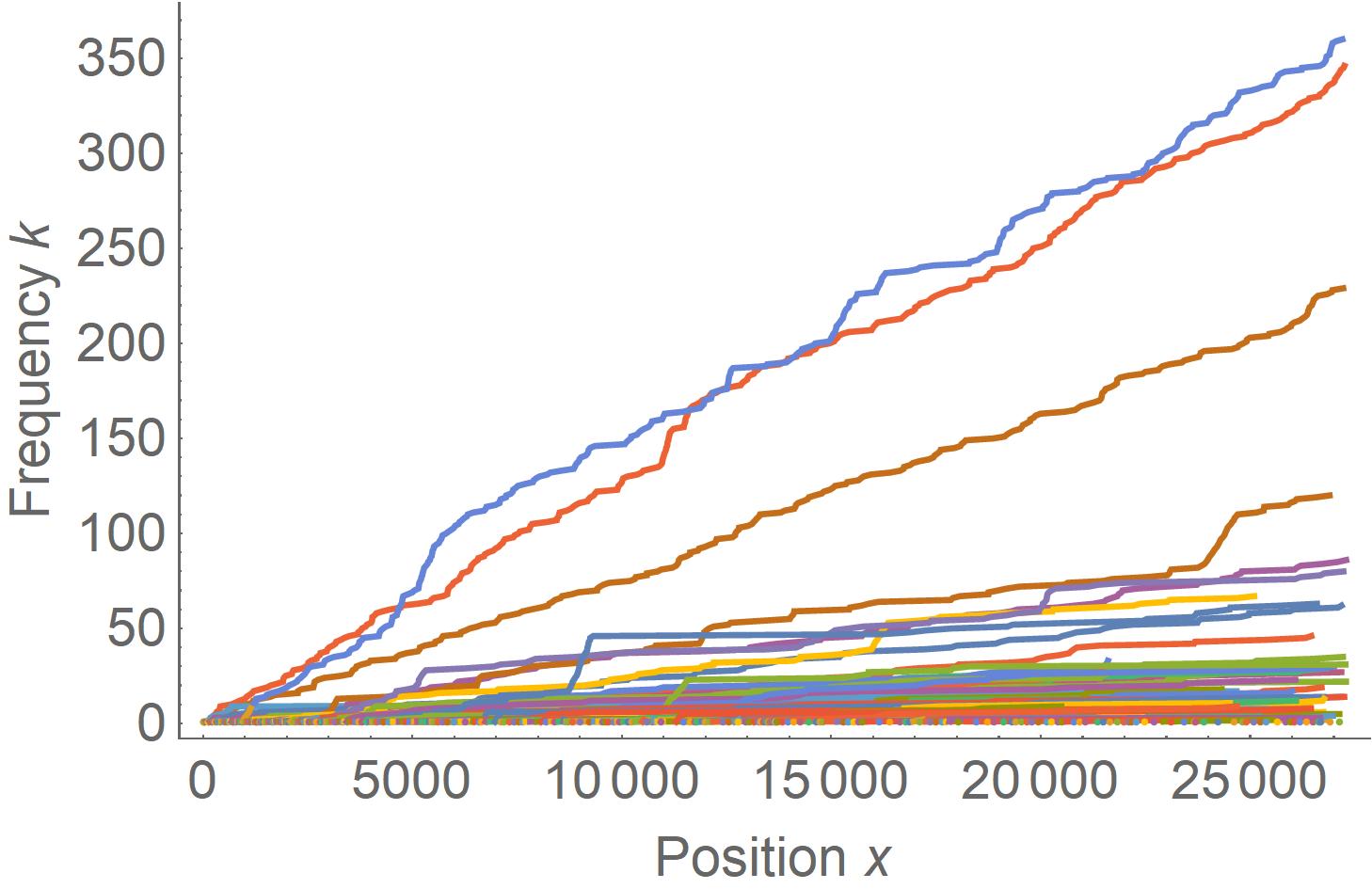}}
\subfigure[]{\includegraphics[width=0.23\textwidth]{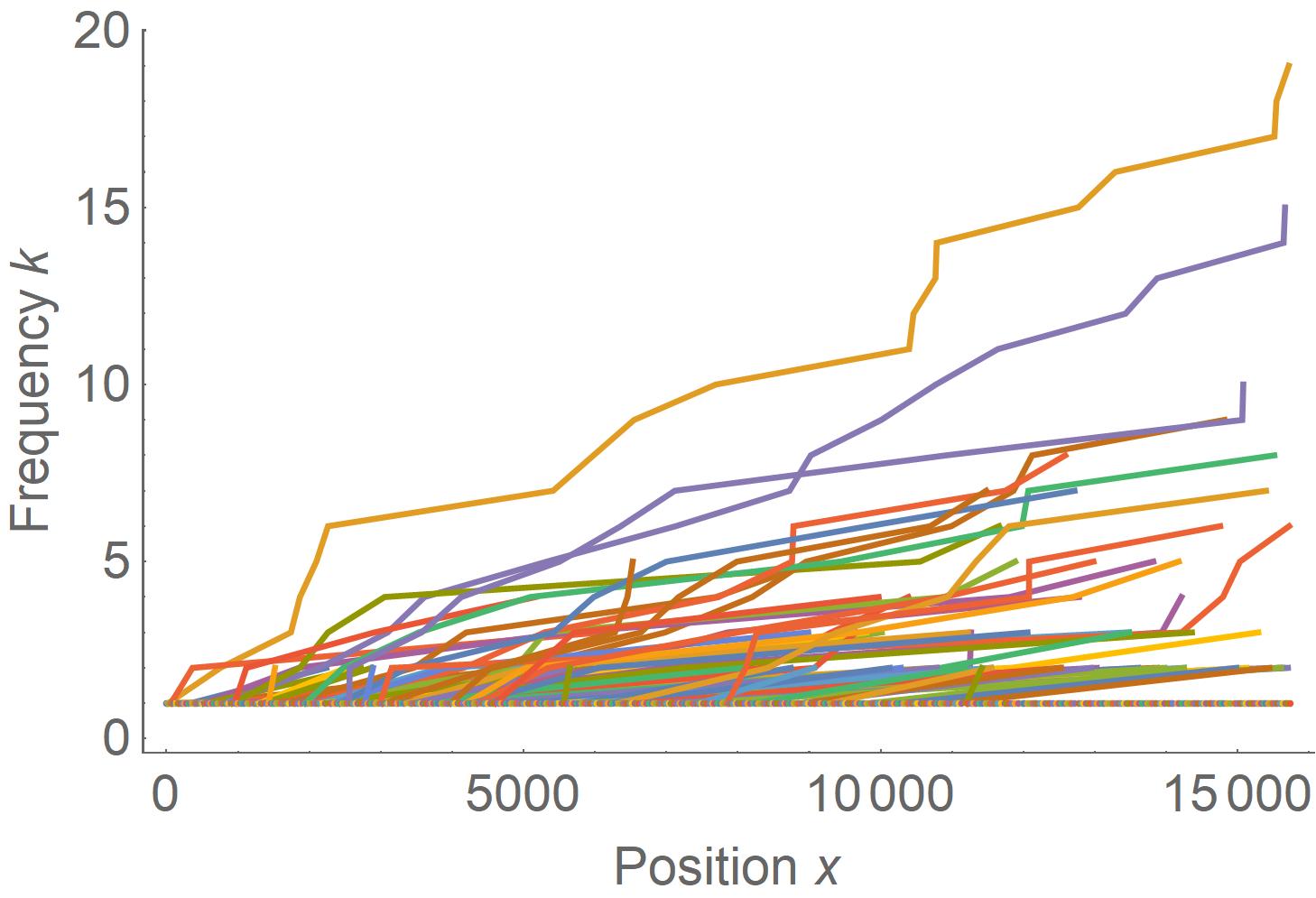}}
\caption{\label{fig:LinearInc}The growth curve of words. (a) the excerpt of $Moby Dick$ and (b) $Cj2019$ and (c) Arabic, (d) Hindi, (e) Lao and (f) Polish texts in Leipzig Corpora. Each curve corresponds to each word. It seems sure that the number of occurrences of almost frequent words grows linearly with the tokens of text. This also implies that frequent words might be uniformly distributed throughout the text.} 
\end{figure*}

The first point that we can claim might be a uniform distribution of words in language process. As we can easily observe, the number of frequent words, including grammatical words, such as ``the'' and keywords of content, increases linearly along with the length of text. Fig.~\ref{fig:LinearInc} shows the analysis of an excerpt in $Moby Dick$ written by Herman Melvillea and the author’s debut paper \cite{Chol-jun2019} (hereafter, simply $Cj2019$), which might be in poor English so that could be a kind of challenge to Zipf’s law found in normal or native English literatures. Additionally, we analyze texts with diverse alphabets from diverse sources selected in Leipzig Corpora Collection (\url{https://wortschatz.uni-leipzig.de/en/download}): Arabic, Hindi, Lao and Polish texts collected from web and news. The observation gives us a possibility to extend the assumption of the uniform distribution from frequent words to the rarest words. 

\begin{figure*}
\centering
\subfigure[]{\includegraphics[width=0.42\textwidth]{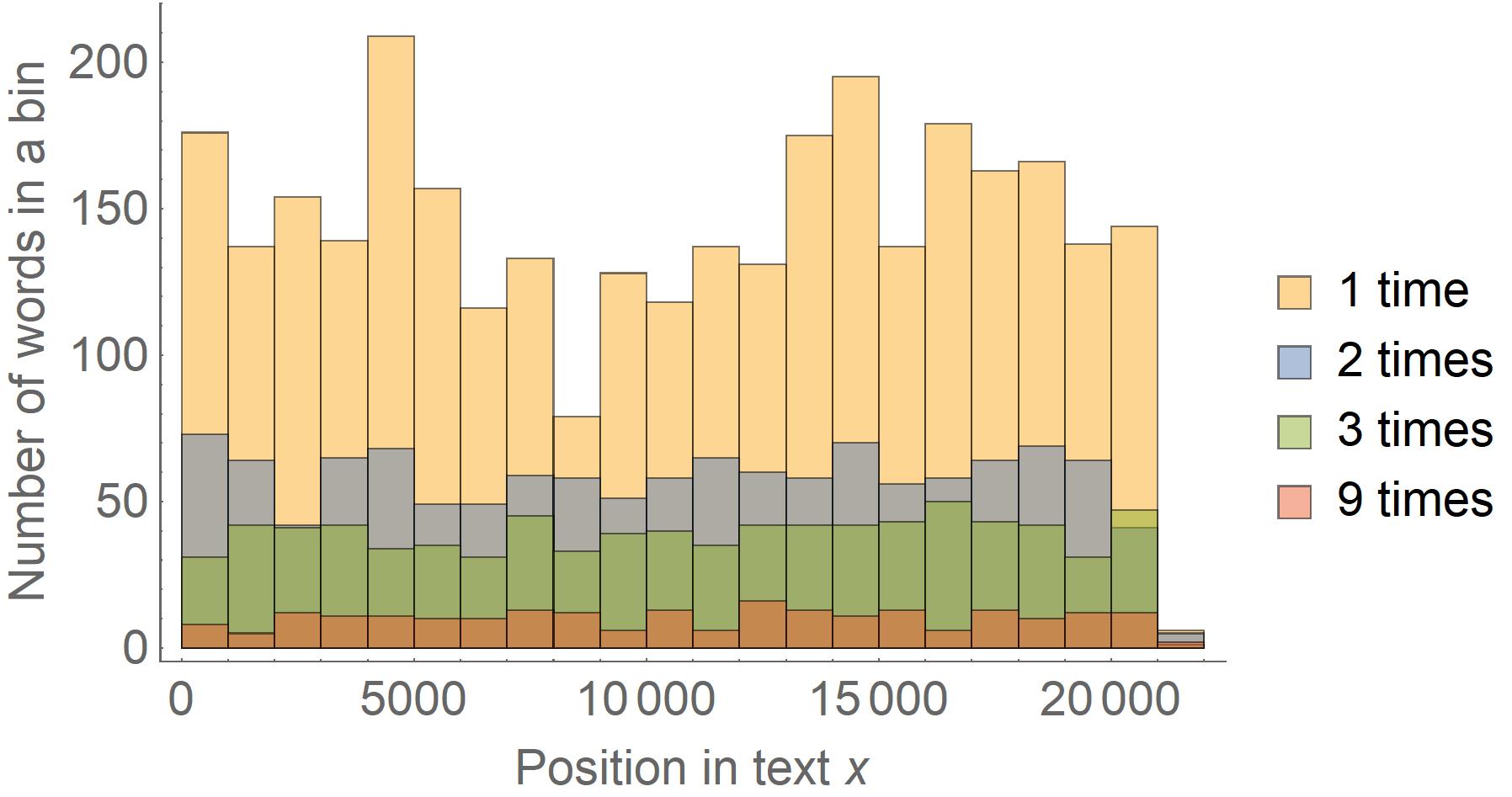}}
\subfigure[]{\includegraphics[width=0.42\textwidth]{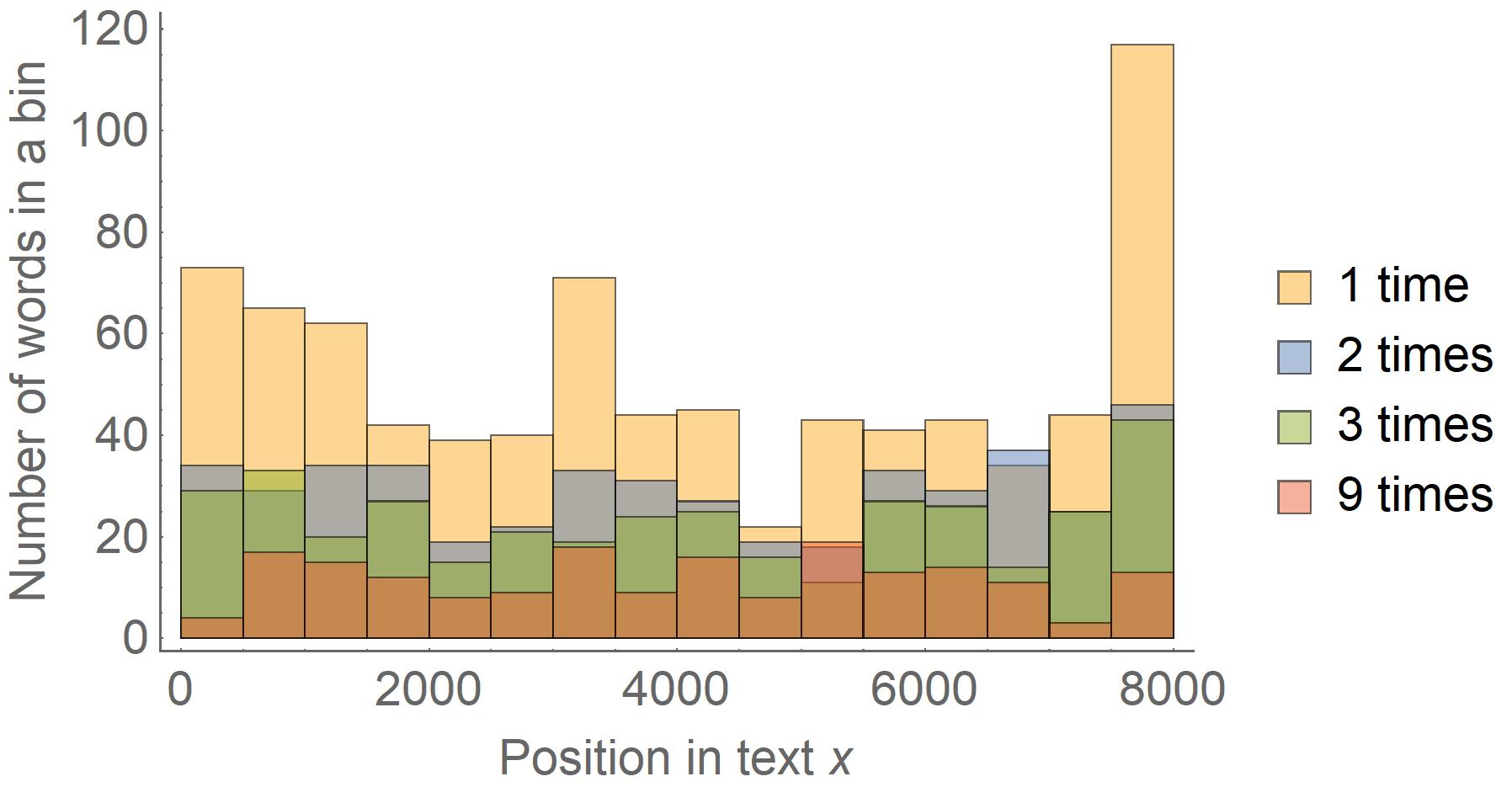}}\\
\subfigure[]{\includegraphics[width=0.23\textwidth]{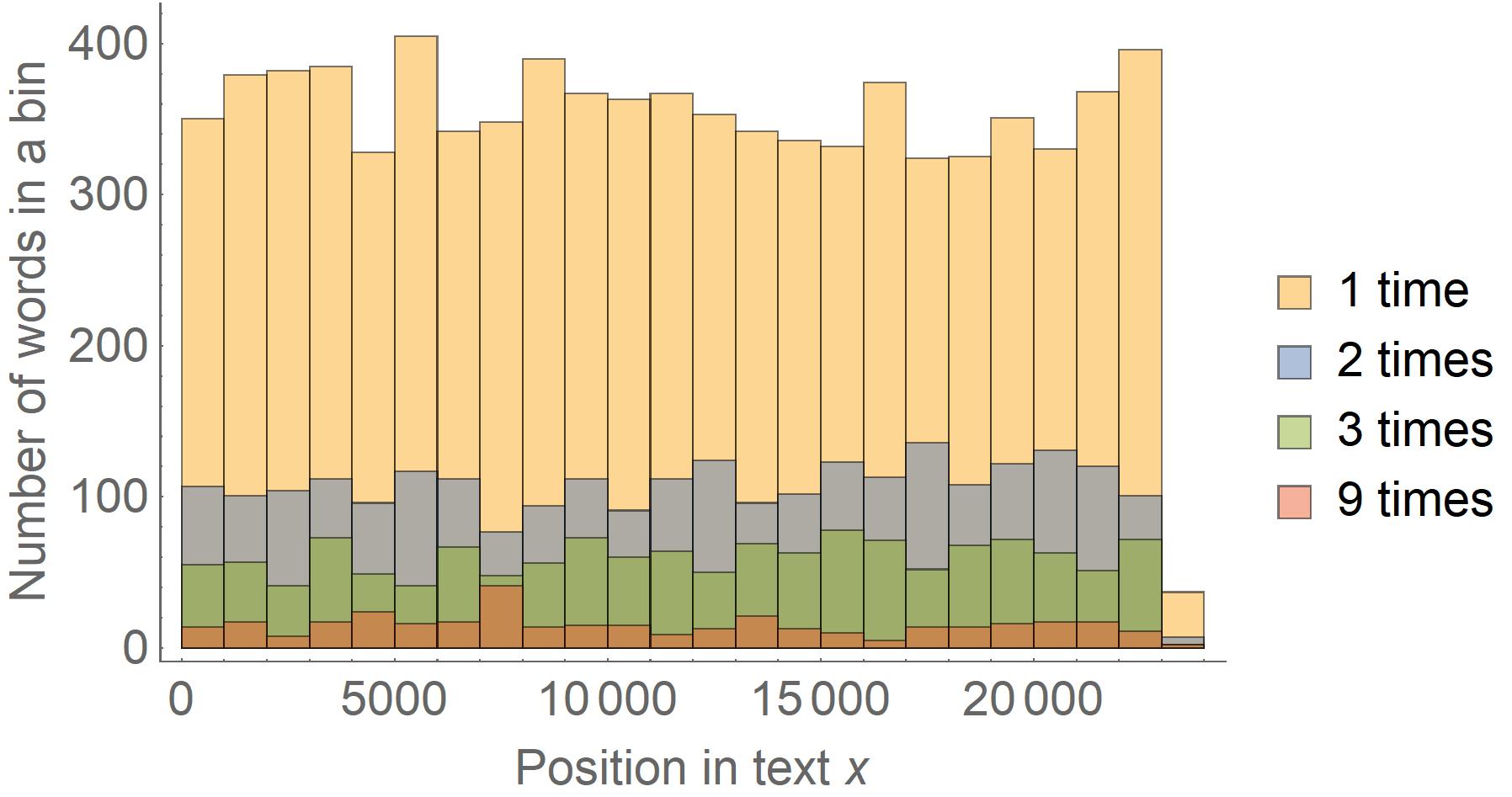}}
\subfigure[]{\includegraphics[width=0.23\textwidth]{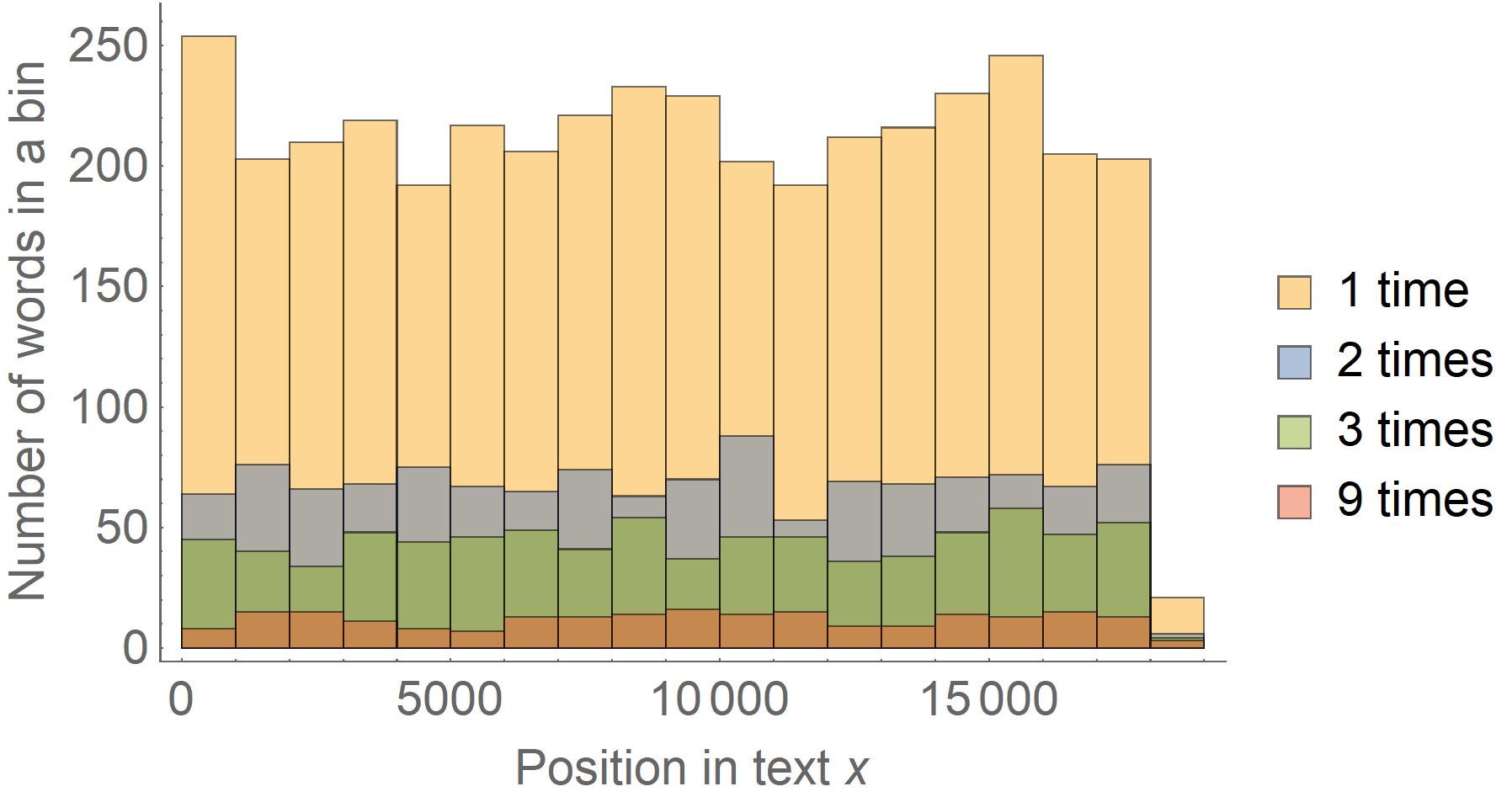}}
\subfigure[]{\includegraphics[width=0.23\textwidth]{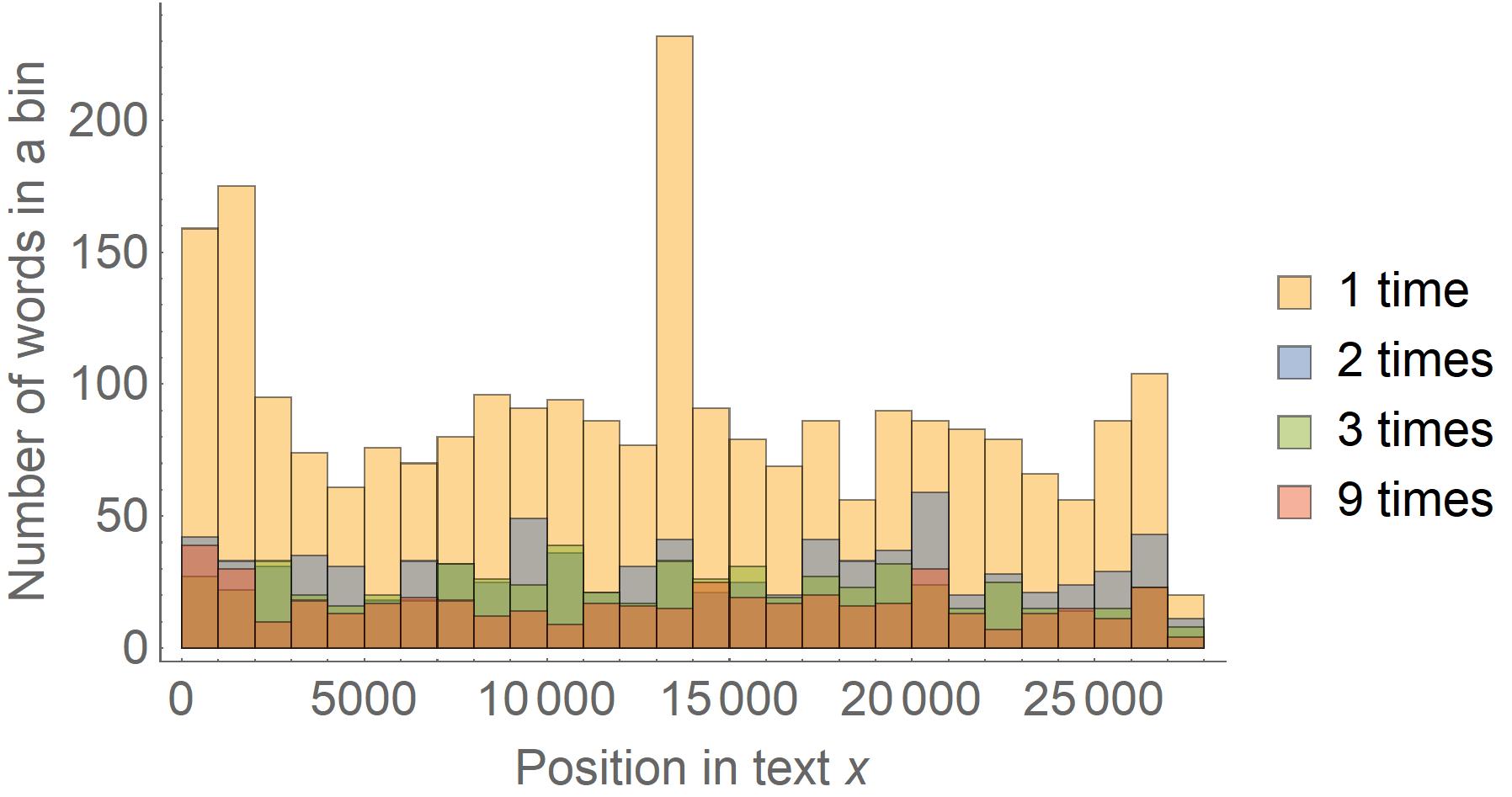}}
\subfigure[]{\includegraphics[width=0.23\textwidth]{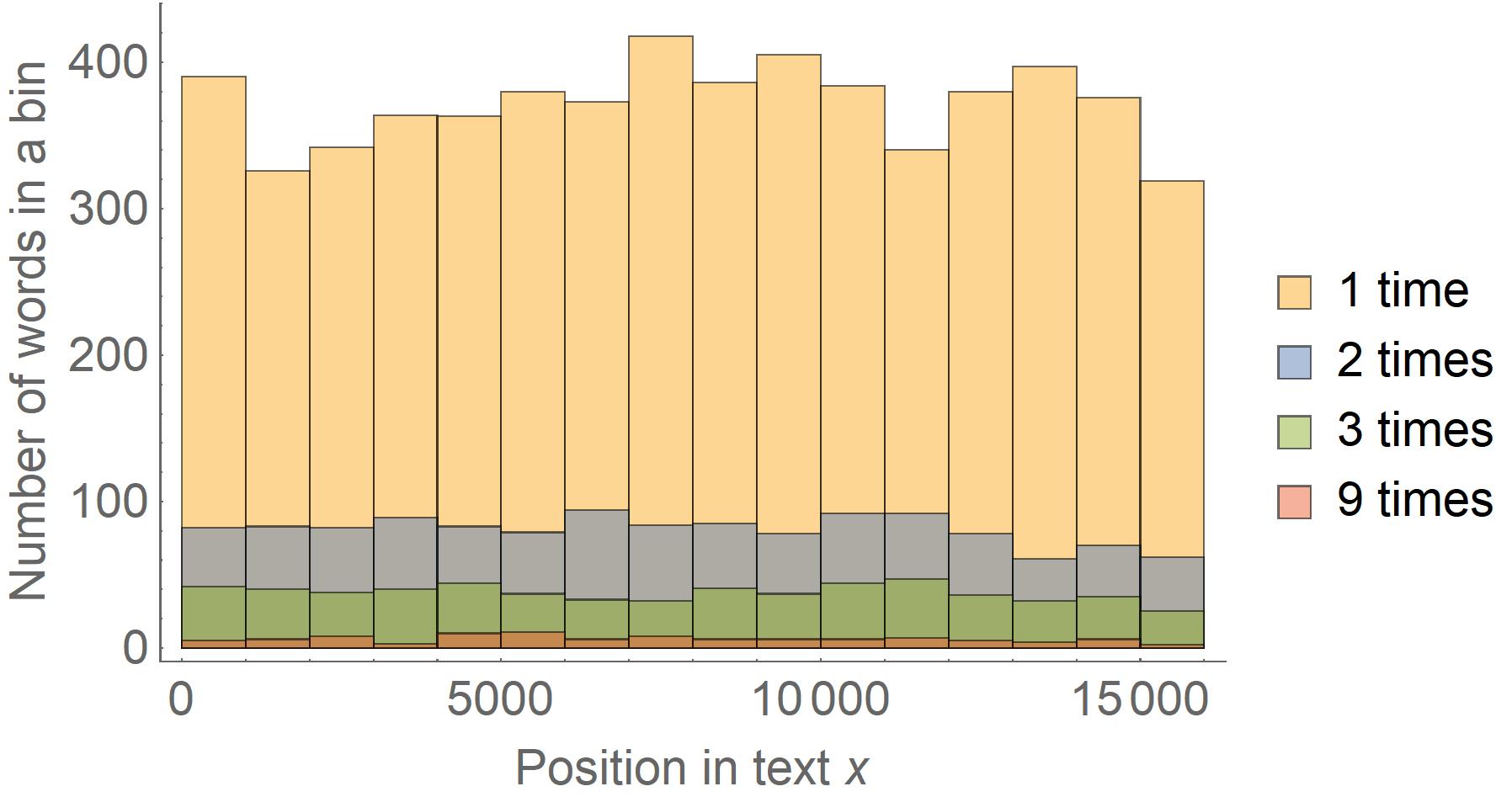}}
\caption{\label{fig:UniformDis}Histogram of all words appeared $k = 1, 2, 3, 9$ times. (a) the excerpt of $Moby Dick$, (b) $Cj2019$ and (c) Arabic, (d) Hindi, (e) Lao and (f) Polish texts in Leipzig Corpora. Independently of type of word, the words occurred the given times altogether seem to be distributed uniformly.} 
\end{figure*}

The increase of the word frequency is arithmetic so that we cannot apply the approach of the geometrically growing system developed by \cite{Chol-jun2022}. Of course, we cannot verify the trend for very rare words or hapax legomena because they appear only one or a few times in text. Meanwhile, the words of a given number of occurrences are altogether distributed uniformly throughout the text (Fig.~\ref{fig:UniformDis}). 
   
\begin{figure*}
\centering
\subfigure[]{\includegraphics[width=0.42\textwidth]{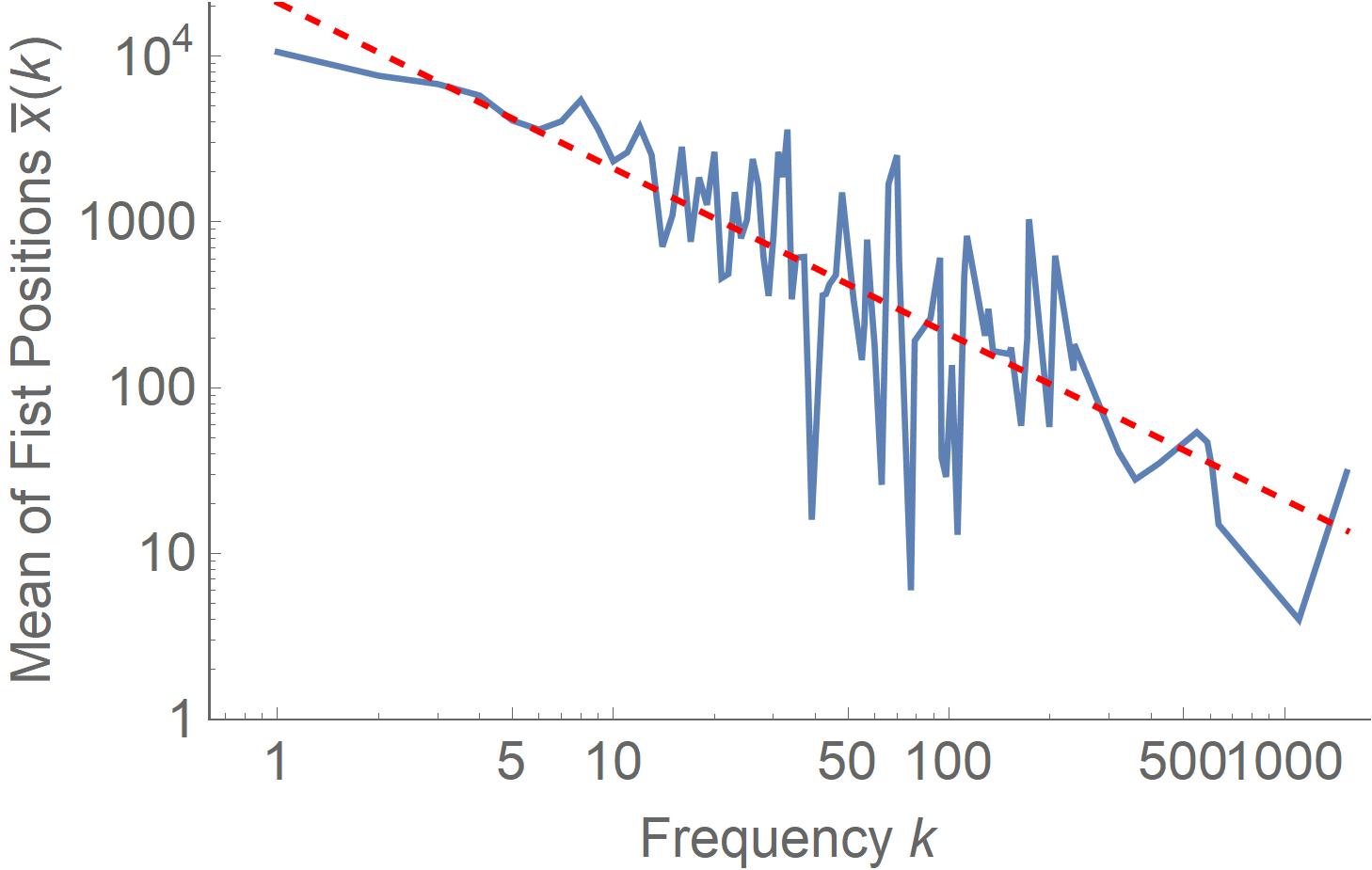}}
\subfigure[]{\includegraphics[width=0.42\textwidth]{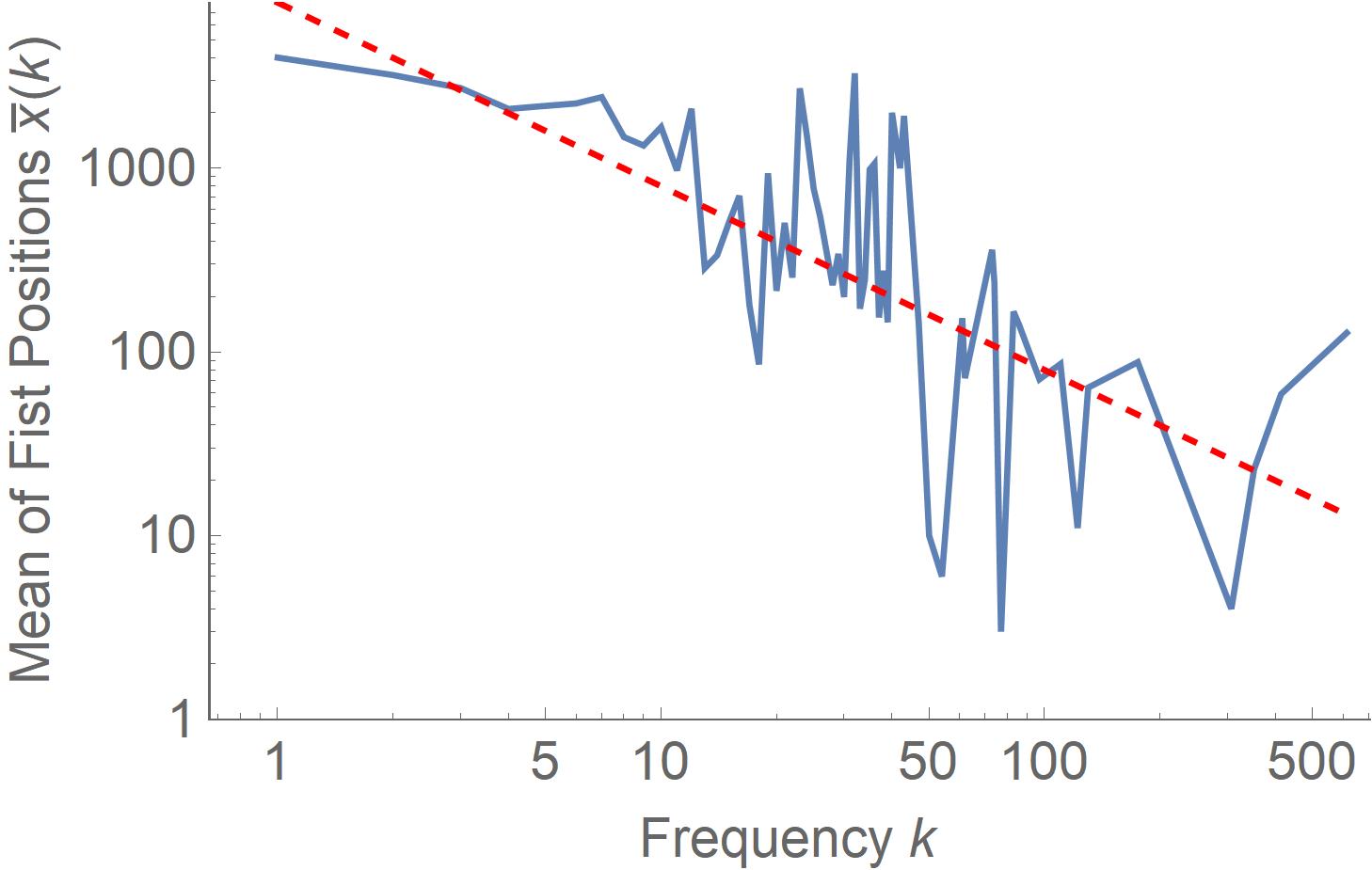}}\\ 
\subfigure[]{\includegraphics[width=0.23\textwidth]{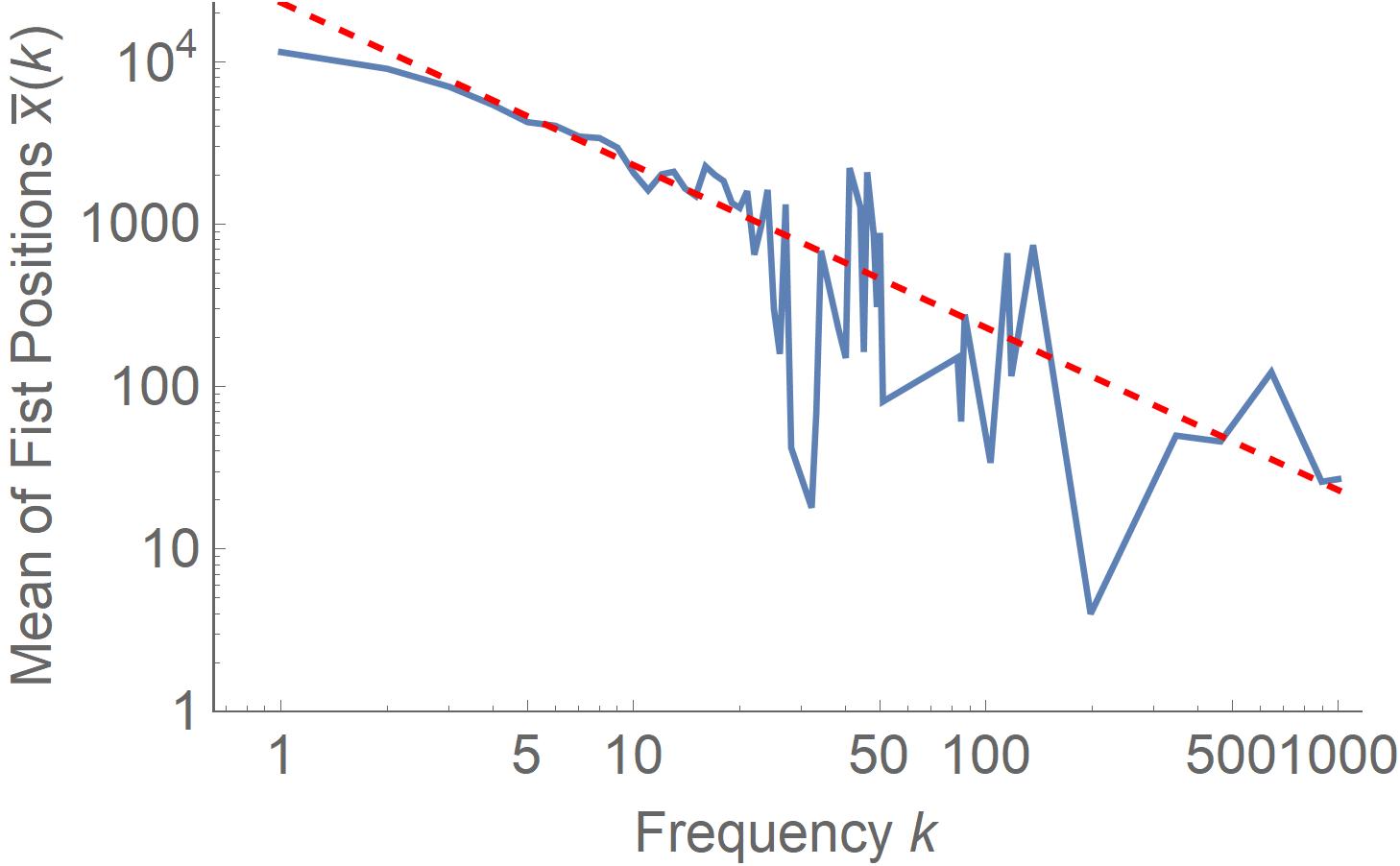}}
\subfigure[]{\includegraphics[width=0.23\textwidth]{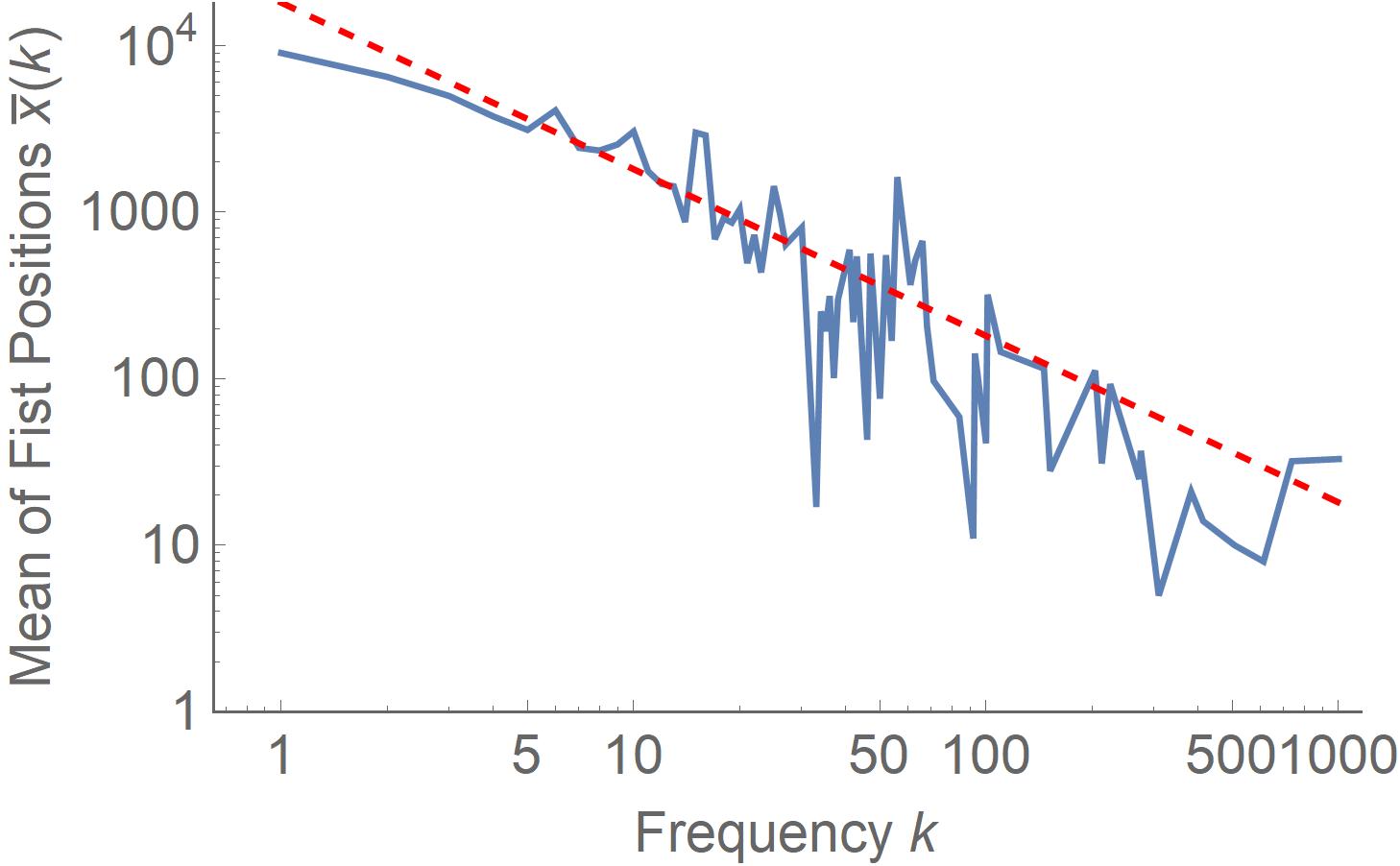}}
\subfigure[]{\includegraphics[width=0.23\textwidth]{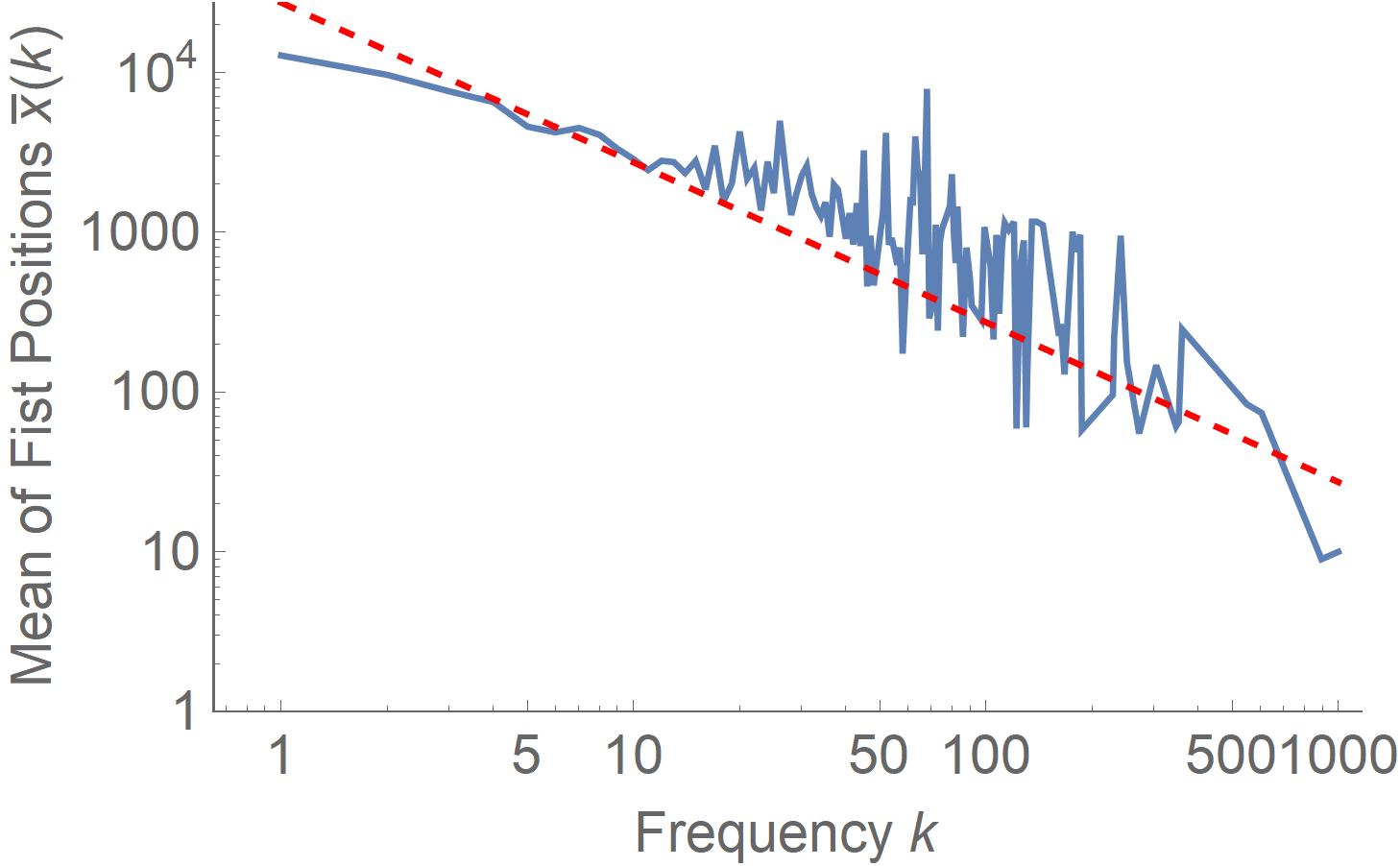}}
\subfigure[]{\includegraphics[width=0.23\textwidth]{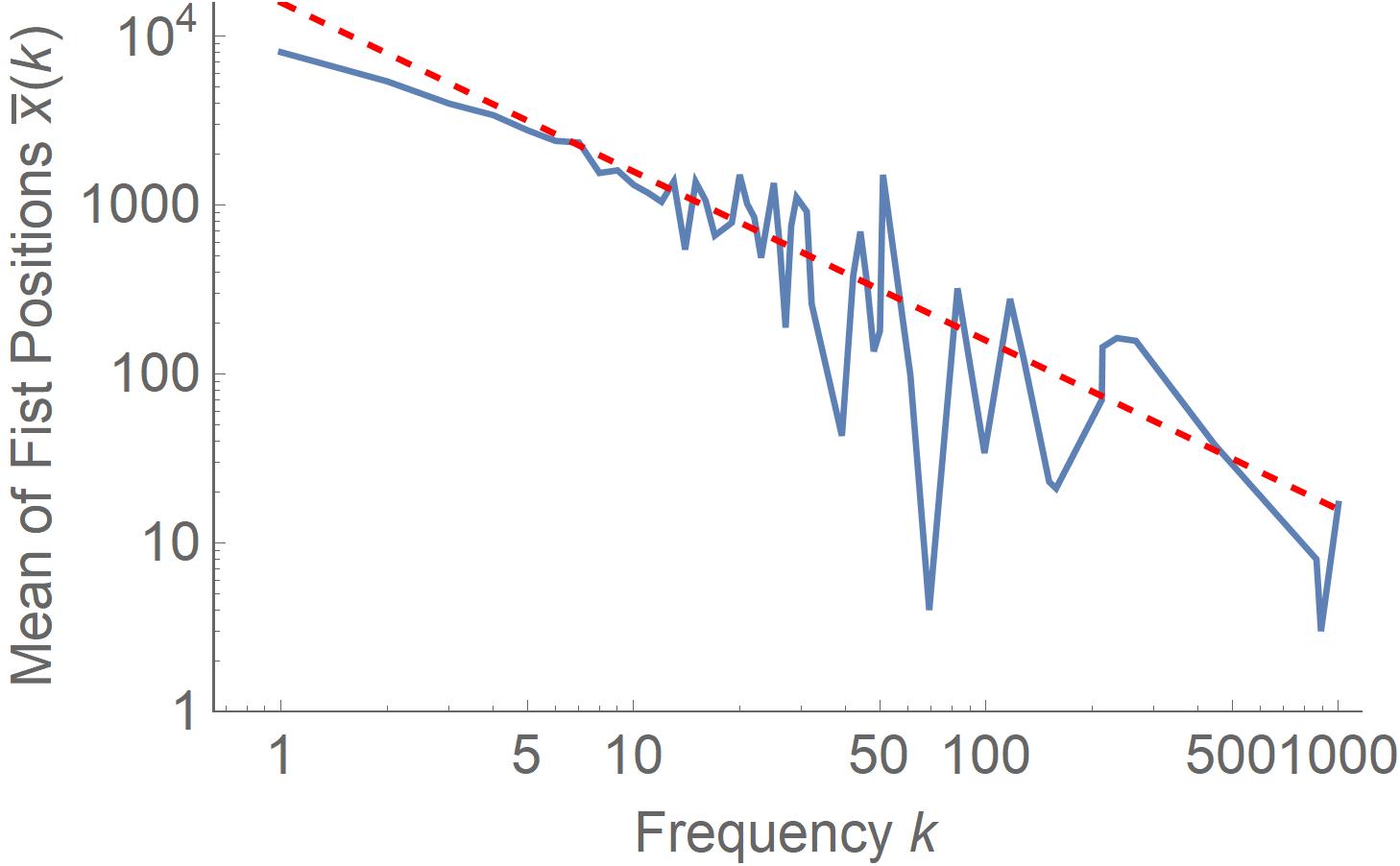}}
\caption{\label{fig:PoissonDist}The relation of the word frequency $k$ vs. the mean first positions $\bar{x}$ for the words of the same $k$. (a) the excerpt of $Moby Dick$ and (b) $Cj2019$ and (c) Arabic, (d) Hindi, (e) Lao and (f) Polish texts in Leipzig Corpora. The red dashed line implies $\bar{x}=\frac{X}{k}$. This relation seems coincident with data throughout all the frequencies.} 
\end{figure*}

We try to show that the uniformity can also be extended to the ``distribution'' of rare words or even hapax legomena. If a word appears uniformly, its rate could follow a Poisson distribution. Then, its first position in text or an instantaneous span  $x$ between its replicas should follow an exponential distribution \citep[e.g. see][]{Feller1968}: $p(x)=\lambda\exp(-\lambda x)$. Here $\lambda$ stands for the probability of occurrence of the word and is defined as $\lambda=\frac{k}{X}$, where $k$ stands for the word frequency and $X$ for the length of text or the tokens (later we will see that $p(x)$ is expressed in a binomial format, more exactly, which gets closer to the exponential distribution when $k$ or $\lambda$ get greater). In fact, uniform, Poisson and exponential distributions can be said to be three different faces of a cube. We can estimate the mean of $x$ as $\bar{x}=\int^{\infty}_{0}x\lambda\exp(-\lambda x)=\frac{1}{\lambda}$. Therefore, we could expect an inverse relation between $k$ and the mean of first positions $\bar{x}$. For the rare words, it seems impossible to get a statistical mean $\bar{x}$ of the word because of a deficiency of samples. Instead of the ``mean’’ for one word, we can get the mean for all the words of the same $k$. If the words are supposed to appear uniformly, we can expect a relation
\begin{linenomath} \begin{align}
\bar{x}=\frac{X}{k}\label{eq:xbar},
\end{align} \end{linenomath} 
We test this relation in the real texts. Figure~\ref{fig:PoissonDist} shows that the above inverse relation holds all over the frequencies approximately for sample texts. 

The uniform distribution seems to originate from the homogeneity of text. In fact, the distribution and statistics of words should not be changed by swapping parts of text so that the distribution of words can be supposed to be uniform throughout the text. However, considering that many non-Zipfian distributions appear homogeneously in a domain, it is not clear that the homogeneity might not be necessary for Zipf’s law.

On behalf of mathematical formalism, we adopt the following definition for the language process.
\begin{definition}
We would call such a process embodying the following axiom as property simply an L-process:
\end{definition}
\begin{axiom}
Every word appears uniformly in position during the process. In other words, the series $x_1(y), x_2(y), \dots, x_k(y)$ are distributed uniformly.
\end{axiom}

The L-process recalls us ``preferential attachment'' model, where the probability of a word occurrence is proportional to its past occurrences. Also in the L-process, a frequent word appears frequently and vice versa. Therefore, we can expect a similar result here. In addition, the aforementioned definition could make modeling easier and more intuitive than ``preferential attachment''. Our observation shows that language process might have originally the property of the L-process without a swap of parts. Hereafter, we consider the L-process in place of the language process. 

In the L-process, the first positions of words of the same frequency seem to follow the exponential distribution, according to the above logic:
\begin{linenomath} \begin{align}
p(x_1(y\vert_k))=\frac{k}{X}\exp(-\frac{k}{X}x_1(y\vert_k))\label{eq:px1yk},
\end{align} \end{linenomath} 
or simply 
\begin{linenomath} \begin{align}
p(x\vert_k)=\frac{k}{X}\exp(-\frac{k}{X}x)\label{eq:pxk},
\end{align} \end{linenomath} 
However, this is an approximation only at great $k$. The case of $k = 1$ is an obvious exception for that. According to our definition, words of $k = 1$ must be distributed uniformly (see Fig.~\ref{fig:UniformDis}, considering that the positions of words of $k = 1$ are just their first position), but not exponentially as the above equation. We can derive a distribution for that logically.

\begin{figure*}
\centering
\subfigure[]{\includegraphics[width=0.42\textwidth]{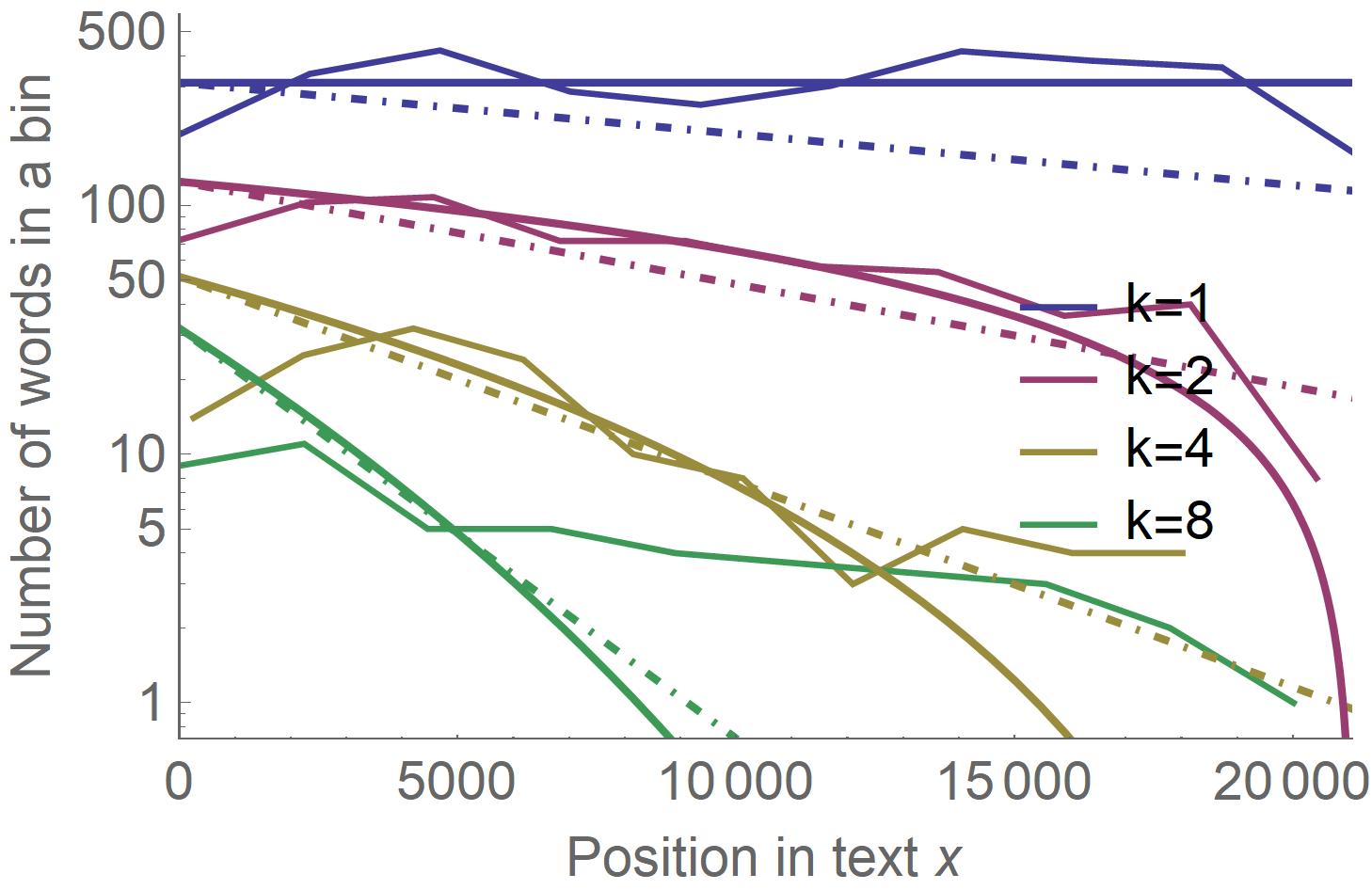}}
\subfigure[]{\includegraphics[width=0.42\textwidth]{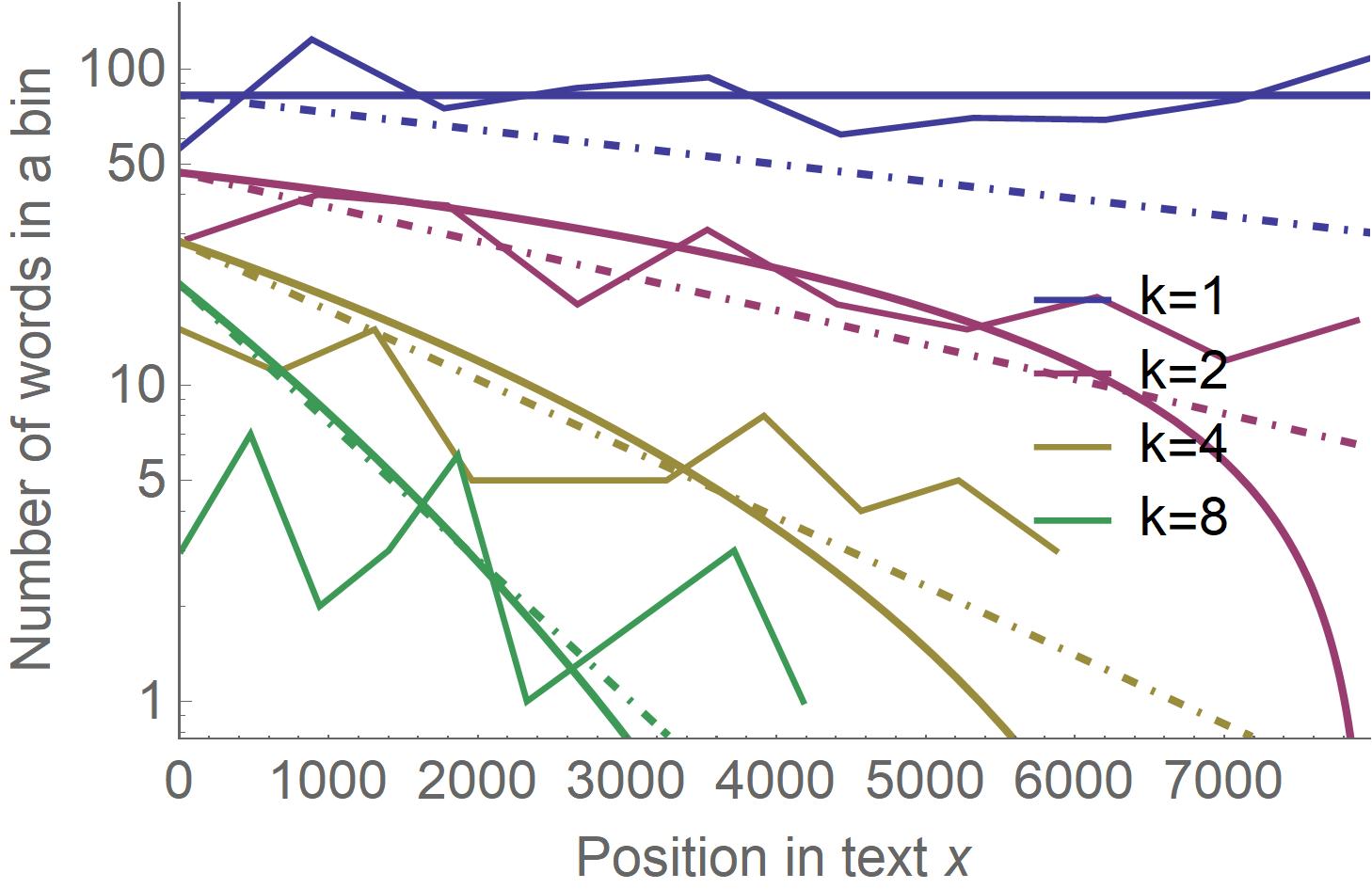}}   \\
\subfigure[]{\includegraphics[width=0.23\textwidth]{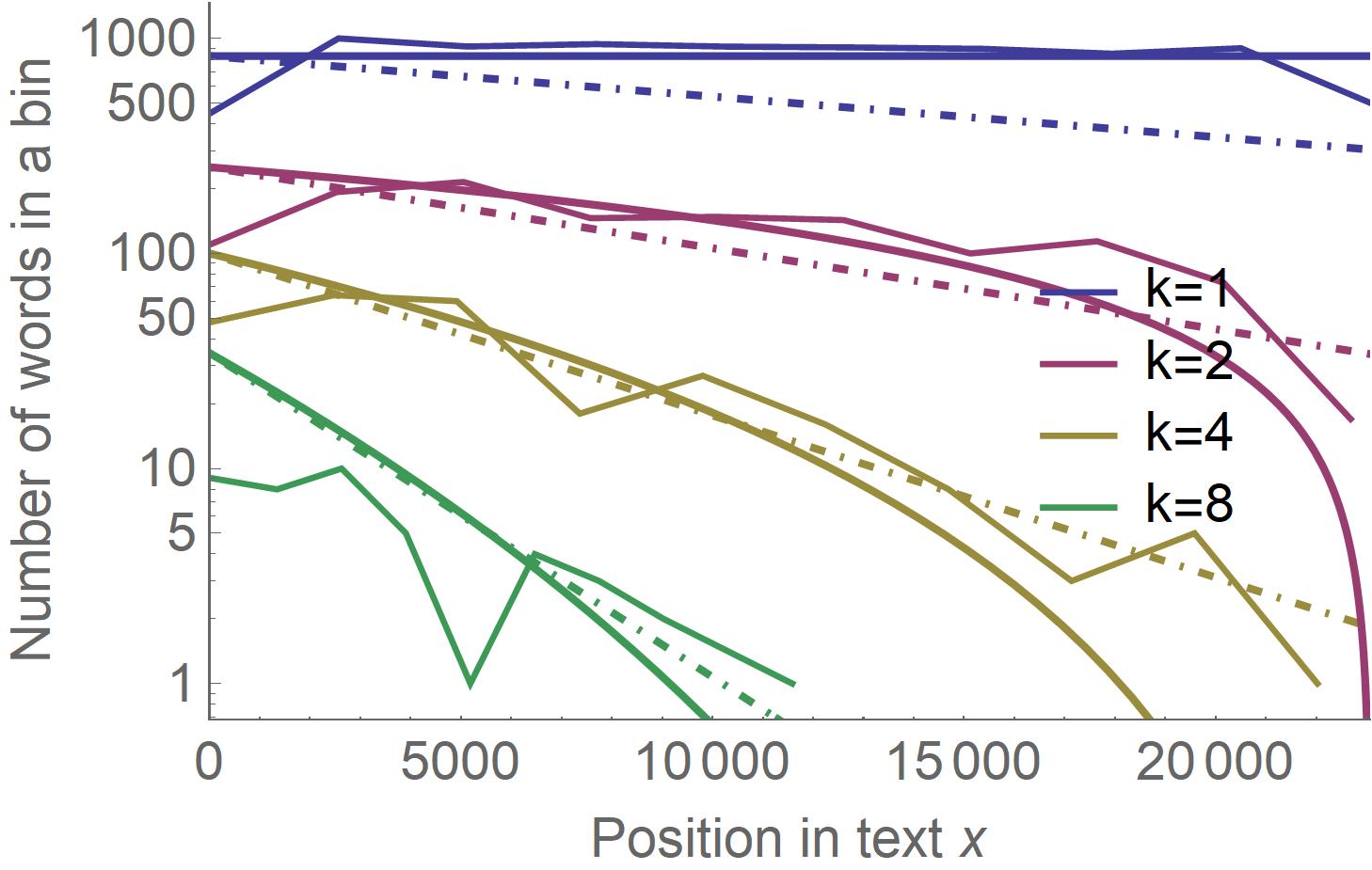}}
\subfigure[]{\includegraphics[width=0.23\textwidth]{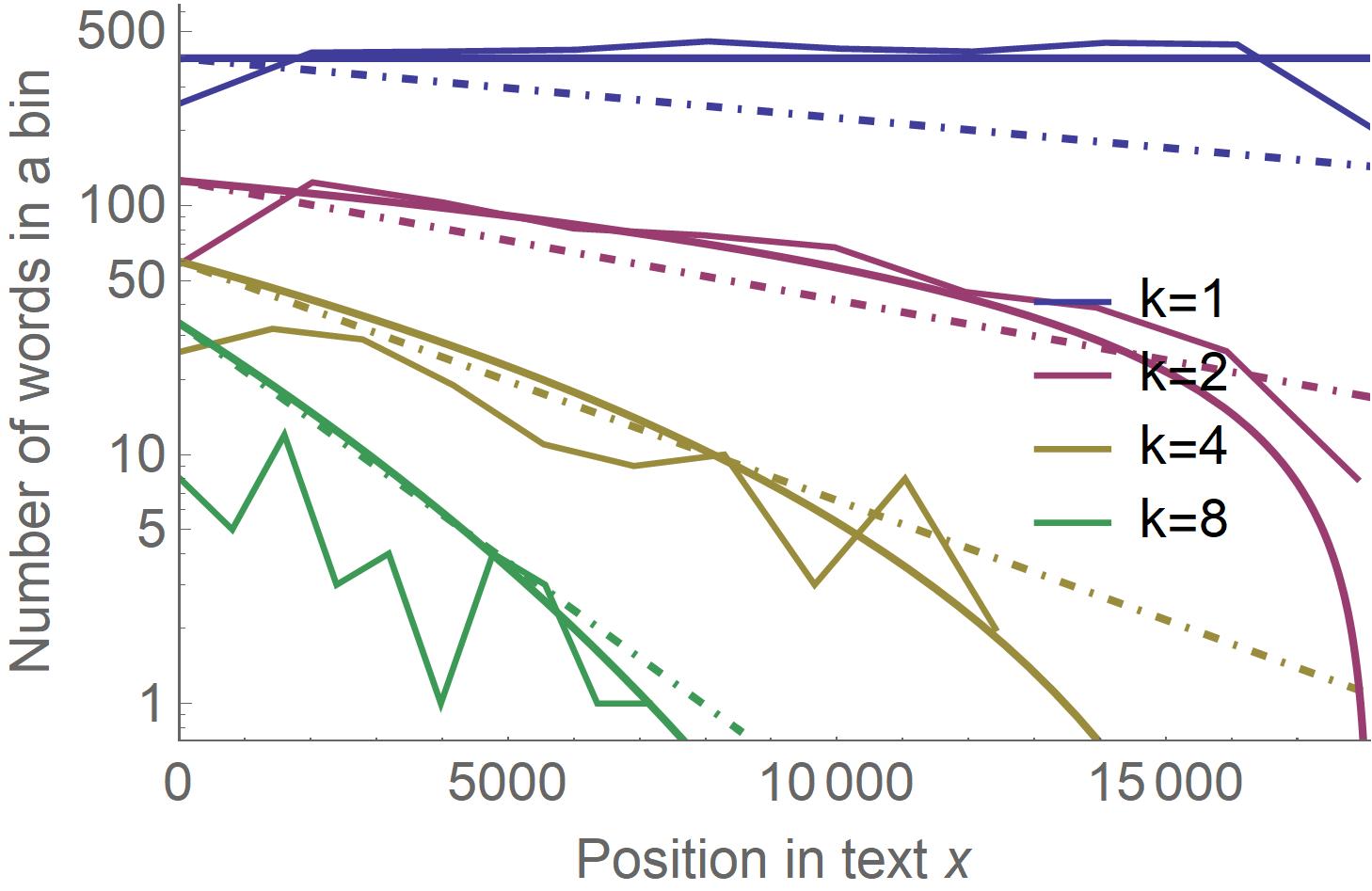}}   
\subfigure[]{\includegraphics[width=0.23\textwidth]{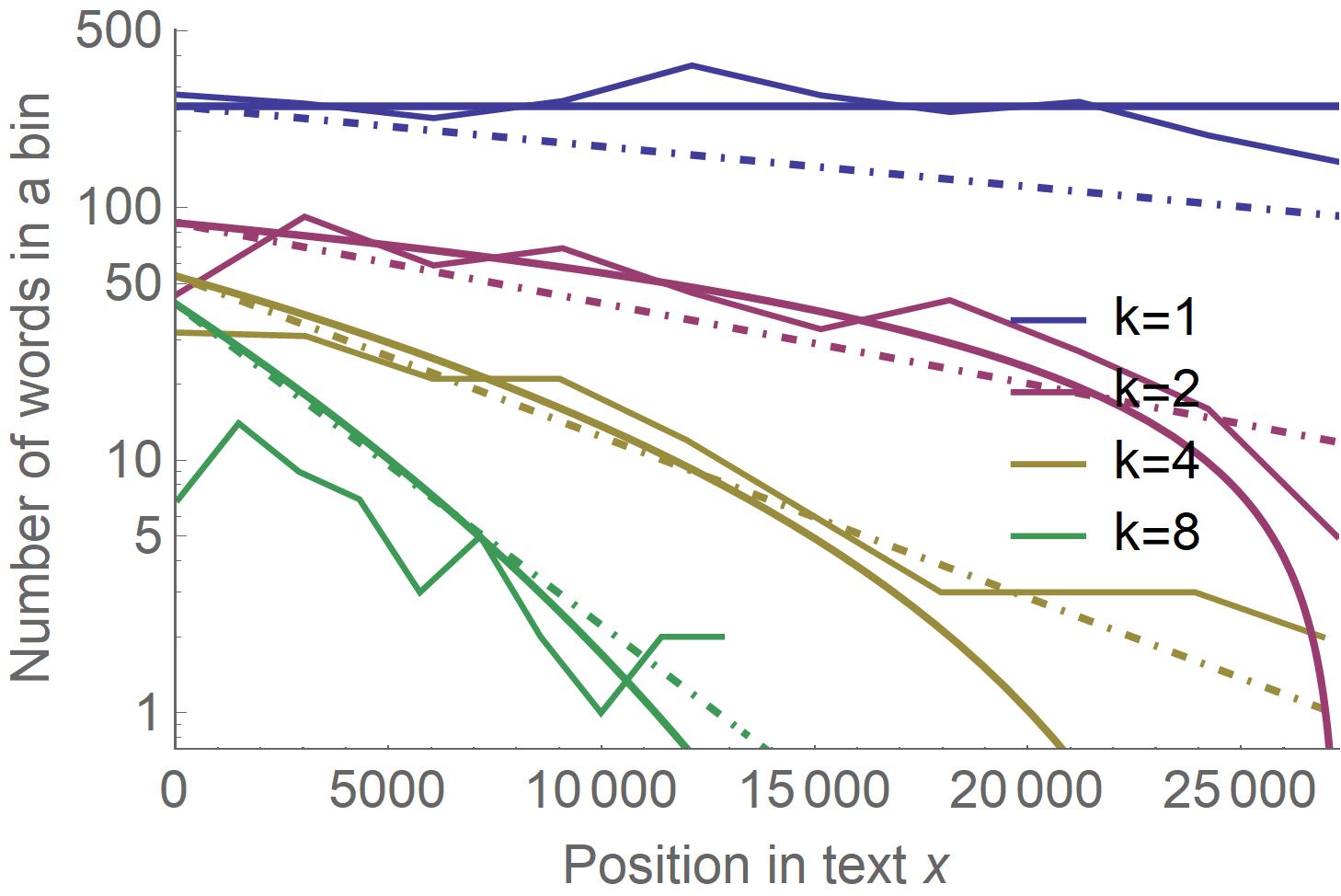}}
\subfigure[]{\includegraphics[width=0.23\textwidth]{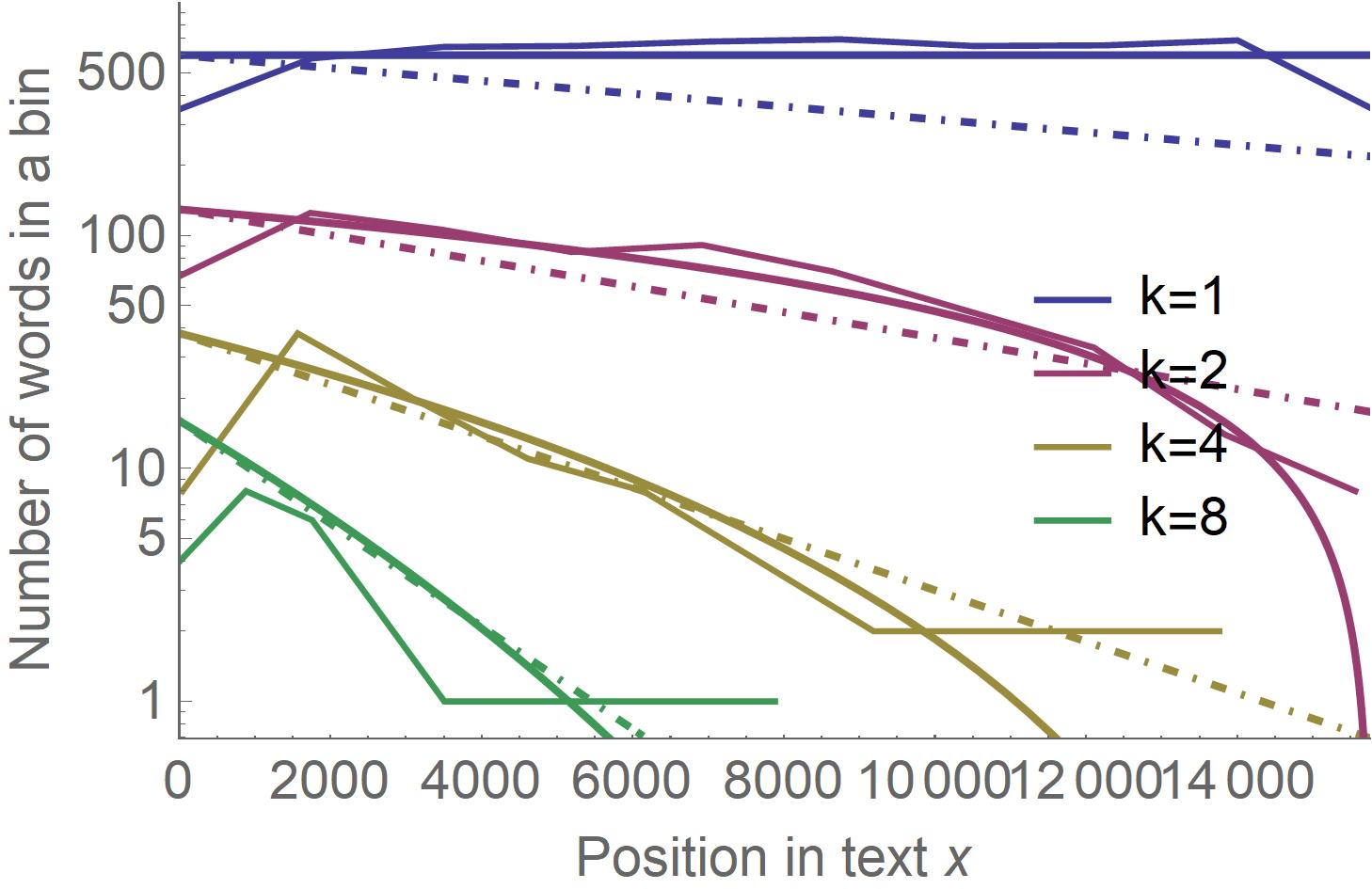}}   
\caption{\label{fig:ExpBinom}The frequency of firstly appeared words with given frequency $k = 1, 2, 4, 8$. (a) the excerpt of $Moby Dick$, (b) $Cj2019$ and (c) Arabic, (d) Hindi, (e) Lao and (f) Polish texts in Leipzig Corpora. The segmented curves stand for the frequency of first appeared words within a bin when dividing text by 10 bins equally. The solid lines stand for the binomial-type model while the dash-dotted lines for an exponential distribution. Obviously, for $k=1$, the binomial-type model is superior while for other $k$s two models of distribution are similar.} 
\end{figure*}

If in a text of length $X$ appear $k$ replicas of a word uniformly, the whole number of such configurations of the word is $\binom{X}{k}$, while the number of configurations that the word at first appears at $x$ is $\binom{X-x}{k-1}$, which can be simply derived by suggesting that after the first position $x$, $k-1$ replicas are distributed arbitrarily (uniformly). We can check these results by using formula $\binom{X}{k}=\binom{X-1}{k-1}+\binom{X-1}{k}$. Thereafter, we can obtain  
\begin{linenomath} \begin{align}
p(x_1(y\vert_k))=\frac{\binom{X-x_1(y\vert_k)}{k-1}}{\binom{X}{k}}\label{eq:px1yk1},
\end{align} \end{linenomath} 
or simply 
\begin{linenomath} \begin{align}
p(x\vert k)=\frac{\binom{X-x}{k-1}}{\binom{X}{k}}\label{eq:pxk1},
\end{align} \end{linenomath} 
Here we can conclude:
\begin{lemma}
In the L-process, the first position x of a word occurring k times has distribution in Eq.~\eqref{eq:pxk1}.
\end{lemma}
We can check that this formula can converge to the above exponential distribution at greater $k$. Figure~\ref{fig:ExpBinom} shows the exponential and binomial-type distribution of first positions of words for the given frequencies.

The above formula can be used to derive Zipf’s law which is shown in the following section.

\section{Derivation of Heaps' law}\label{sec:Heaps}

Most authors derived Heaps’ law from Zipf’s law \citep[e.g. see][]{Bernhardsson2011, Boytsov2017, van Leijenhorst2005, Lu2010}, adopting Zipf’s law as an axiomatic property of language processes without proof. However, as Zipf’s law seems subtler and less intuitive than Heaps’ law, a reverse derivation seems more desirable. Deriving Heaps’ law, the previous works take the mean spans of words as the first positions of them, but such an assumption is clearly improper. The first position is no more than an instance of span, but cannot represent the mean span.

We can intuitively expect that the rate of new words should be lowering as going to the end of text because ``already appeared'' word types are growing and they all in turn continue appearing in the later part of text. Assuming that the word appearance follows a uniform distribution, let’s formulate this process. 

\begin{figure*}
\centering
\subfigure[]{\includegraphics[width=0.42\textwidth]{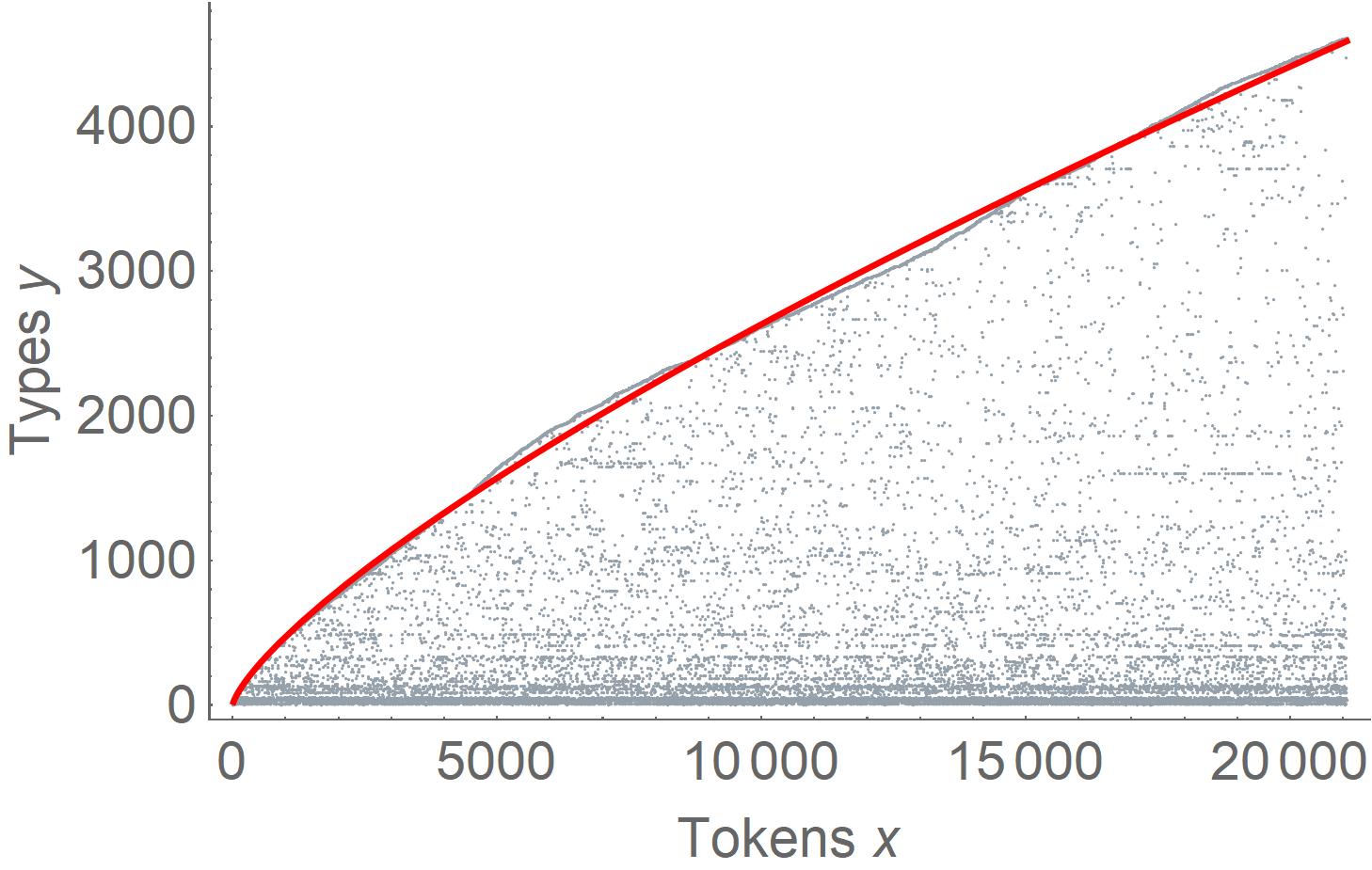}\label{fig:HeapsDiag_1}}
\subfigure[]{\includegraphics[width=0.42\textwidth]{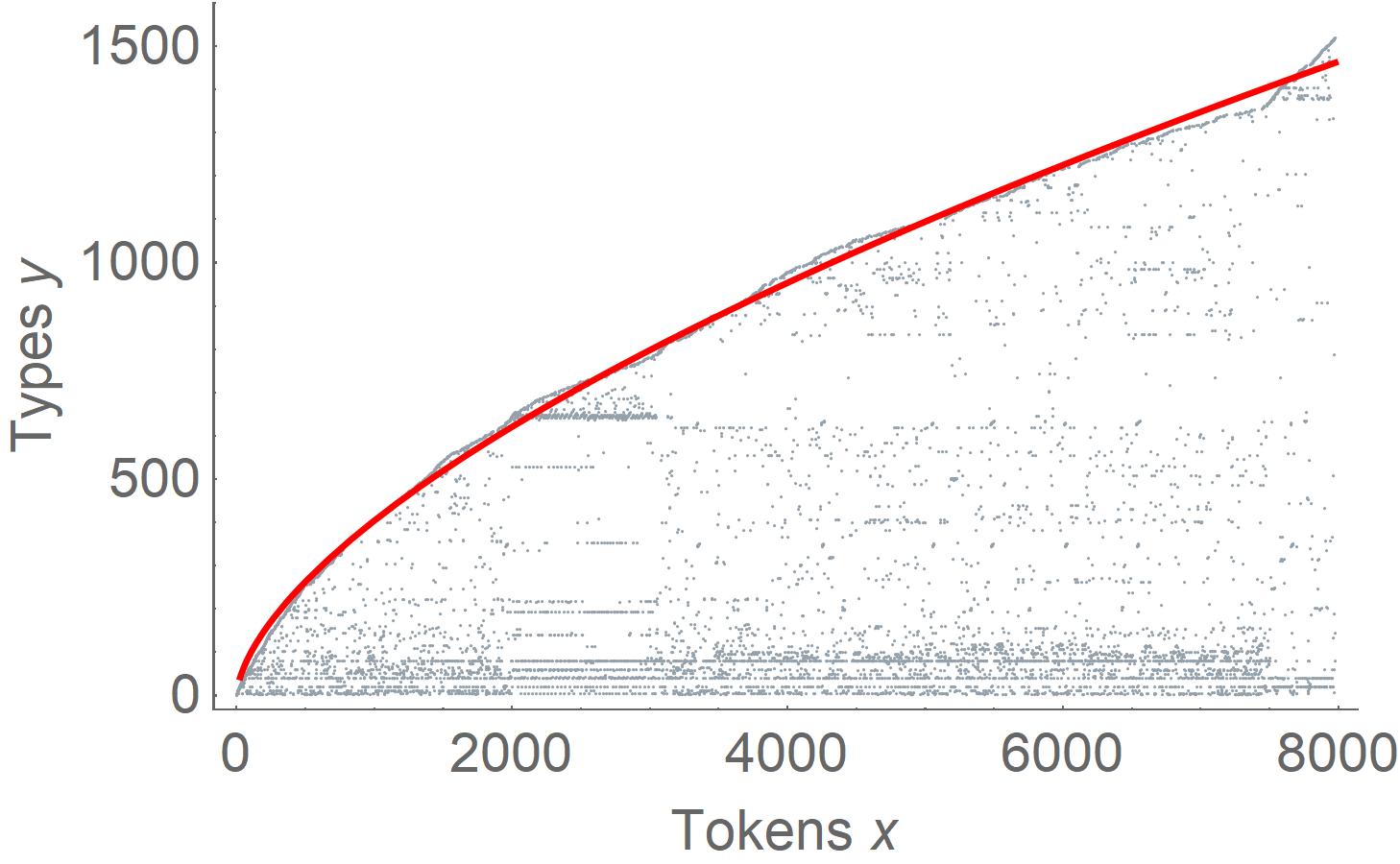}\label{fig:HeapsDiag_2}}\\
\subfigure[]{\includegraphics[width=0.23\textwidth]{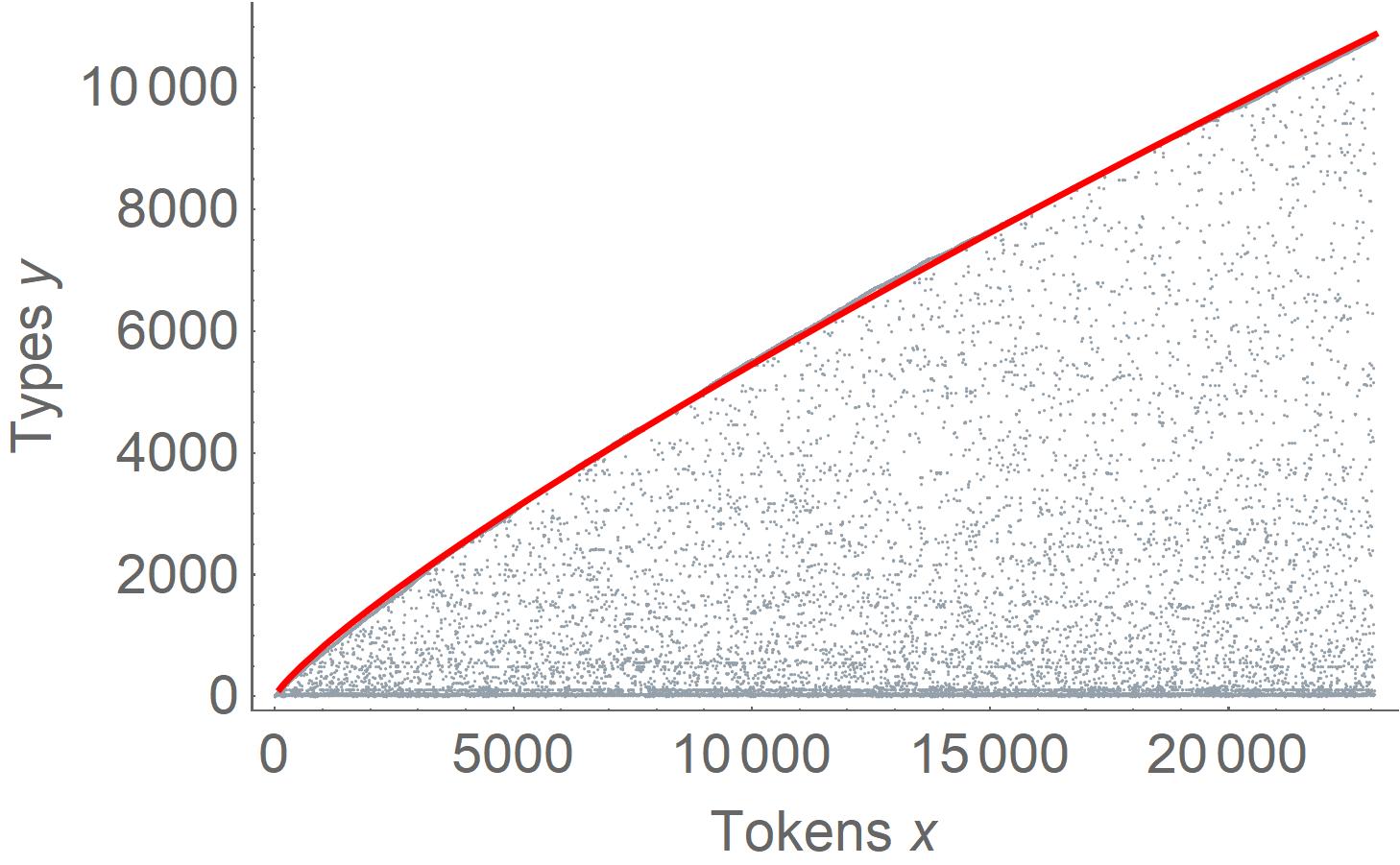}}
\subfigure[]{\includegraphics[width=0.23\textwidth]{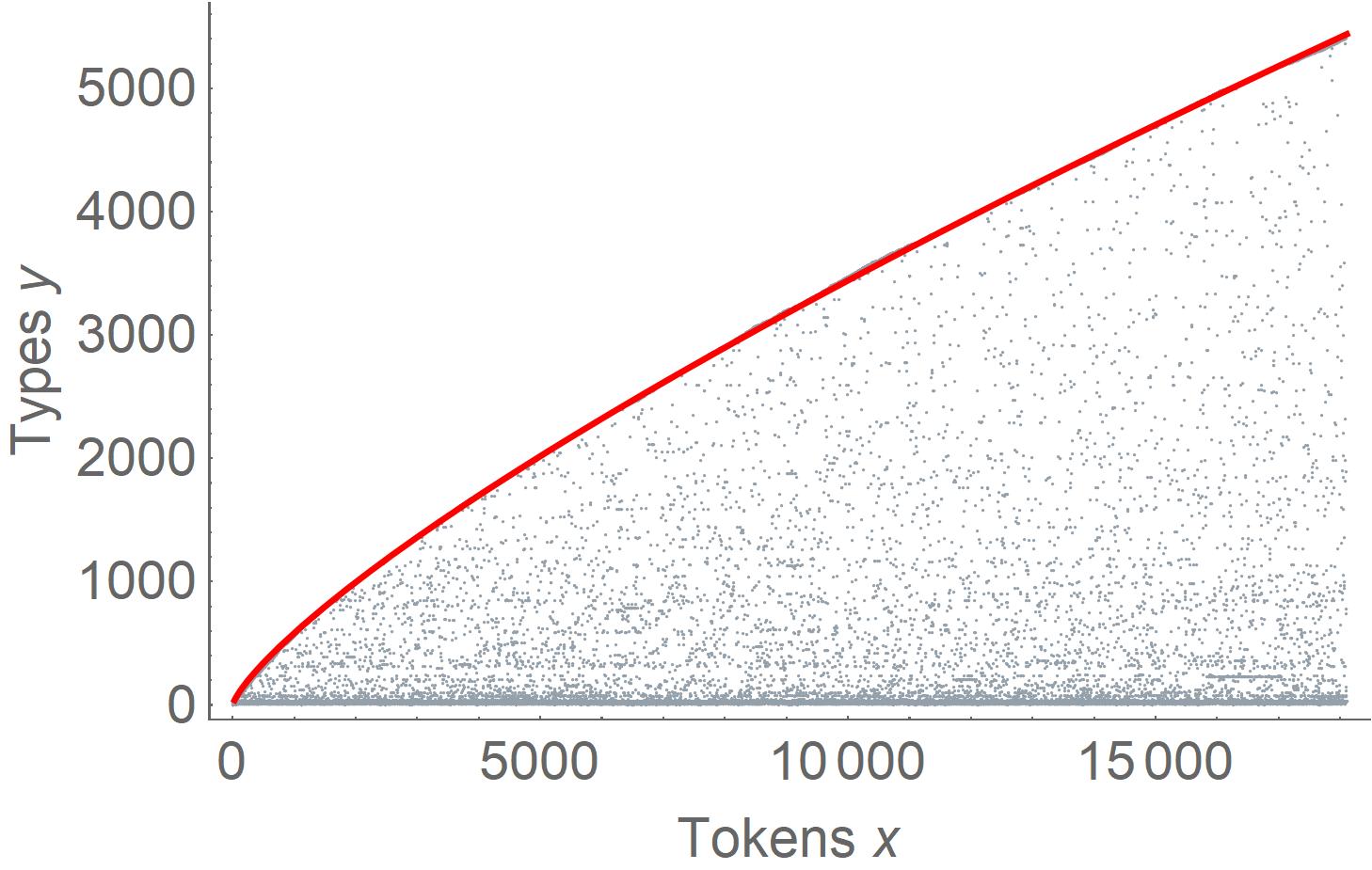}}
\subfigure[]{\includegraphics[width=0.23\textwidth]{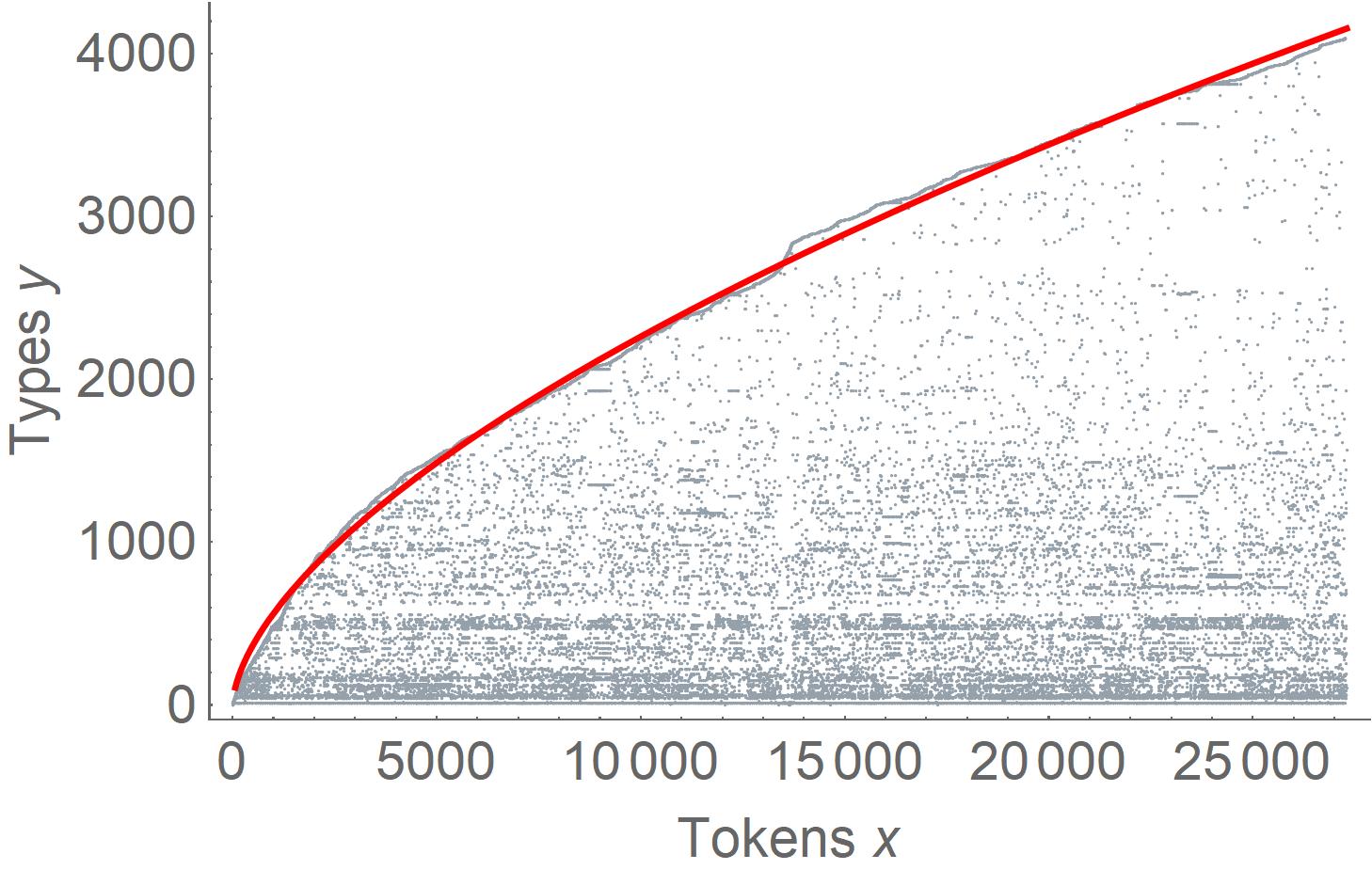}}
\subfigure[]{\includegraphics[width=0.23\textwidth]{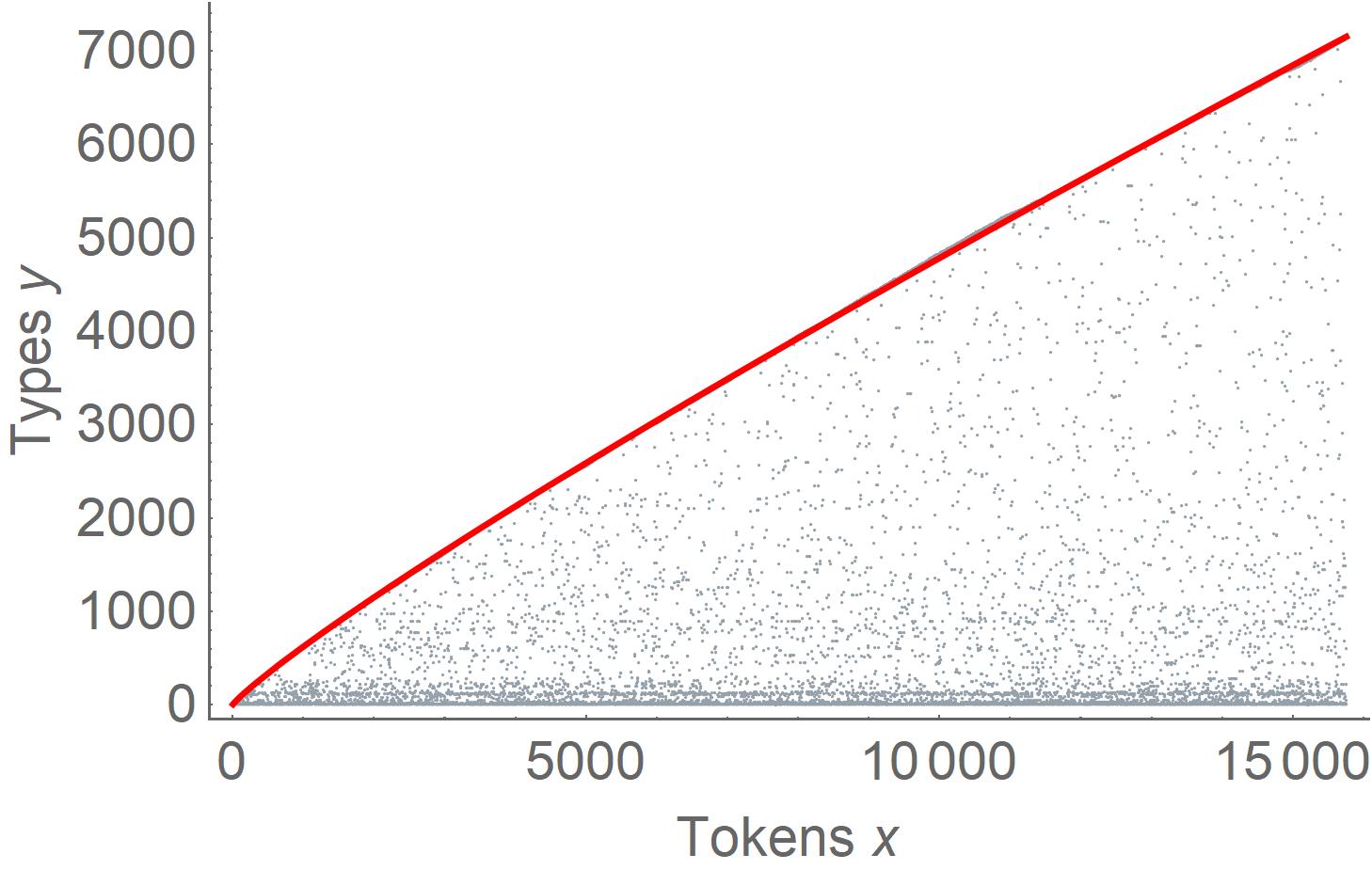}}
\caption{\label{fig:HeapsDiag}Heaps’ diagram. (a) the excerpt of $Moby Dick$, (b) $Cj2019$ and (c) Arabic, (d) Hindi, (e) Lao and (f) Polish texts in Leipzig Corpora. The abscissa $x$ stands for the position in text while the ordinate $y$ for the word types. The dots imply what type of word appears in what position in text. The red curve shows Heaps’ law with the best-fit parameters. Though in the first part (Introduction) and the last part (References) in $Cj2019$, the slope jumps due to many new words, Heaps’ law approximates the overall trend of word increase throughout the text.  } 
\end{figure*}

Figure~\ref{fig:HeapsDiag_1},~\ref{fig:HeapsDiag_2} show Heaps' law $y=Kx^{\beta}$ with best-fit parameters $K \sim 2.73$, $\beta\sim 0.75$ for the excerpt of $Moby Dick$ while $K\sim 5.59$ and $\beta \sim 0.62$ for text of $Cj2019$ despite local undulation.

If the length of text increases by $dx$, this increment is composed of $dy$ new words and replicas of already appeared (i.e., old) words. In our formalism, a function $x_1(y)$ or simply $x(y)$, i.e., the first position $x$ of the word of the type $y$, represents Heaps’ law.  We can represent the mean span $\bar{x}(y)$ of a word of type $y$ as the first position of the word $x(y)$ divided by a variable $\omega(y)$ that must be random: 
\begin{linenomath} \begin{align}
\bar{x}(y)=\frac{x(y)}{\omega(y)}
\label{eq:xbar_y},
\end{align} \end{linenomath} 

The rate of the word type $y$ within $dx$ is expected as $\frac{dx}{\bar{x}(y)}$. We can integrate such a word rate not from $y = 1$, i.e. the start of text, but from $y = y_i$ to the given $y$, where $y_i$ is a position whence Heaps' law seems to hold. In fact, the law does not hold from the start of text. 

Considering that the rate of new words + the rate of old words = 1, we can establish the following equation for the rate of new words:
\begin{linenomath} \begin{align}
\frac{d y}{d x}&=1-\sum_{y=y_i}^{y}\frac{1}{\bar{x}(y)}\label{eq:Heapseq}\\
&\approx 1-\int_{y_i}^{y}\frac{dy}{\bar{x}(y)}=1-\int_{y_i}^{y}\frac{\omega(y)dy}{x(y)}=1-\tilde{\omega}(y)\int_{y_i}^{y}\frac{dy}{x(y)}\\
&=1-\tilde{\omega}\int_{y_i}^{y}\frac{dy}{x(y)}\label{eq:Heapseq2},
\end{align} \end{linenomath}
where in the fourth equation $\omega(y)$ is extracted outside of integral into a variable $\tilde{\omega}(y)$ that also should be a function of $y$. In Eq.~\eqref{eq:Heapseq2} we approximate $\tilde{\omega}(y)$ simply by a constant $\tilde{\omega}$. The fact that $\tilde{\omega}(y)$ varies slowly along with $y$ will be shown by experiment later (see below Eq.~\ref{eq:ombarmicro}). Eq.~\eqref{eq:Heapseq2} is a kind of integro-differential equation.

We can adopt a general solution method of differential equation, giving a type of predefined solution function and determining its parameters. Then it can be easily shown that Heaps’ law $y=Kx^{\beta}$ is a solution of Eq.~\eqref{eq:Heapseq2} (see Appendix~\ref{app:HeapsDer}), if setting Heaps’ law as an ansatz and relating the macroscopic quantities $ K,\beta$ and the microscopic ones $y_i ,\tilde{\omega} $ as follows:
\begin{linenomath} \begin{align}
\tilde{\omega}&=1-\beta,\label{eq:omegabar} \\
\beta &=1-\tilde{\omega},\label{eq:beta} \\ 
y_i&=\left(K \beta^{\beta}\right)^{\frac{1}{1-\beta}},\label{eq:yi} \\
K&=\frac{y_i^{\tilde{\omega}}}{(1-\tilde{\omega})^{(1-\tilde{\omega})}},\label{eq:K}.
\end{align} \end{linenomath} 
The fact that the above obtained $K$ and $\beta$ are mere constants independent of $y$ can justify that Heaps' law can be derived for the L-process.

\begin{figure*}
\centering
\subfigure[]{\includegraphics[width=0.42\textwidth]{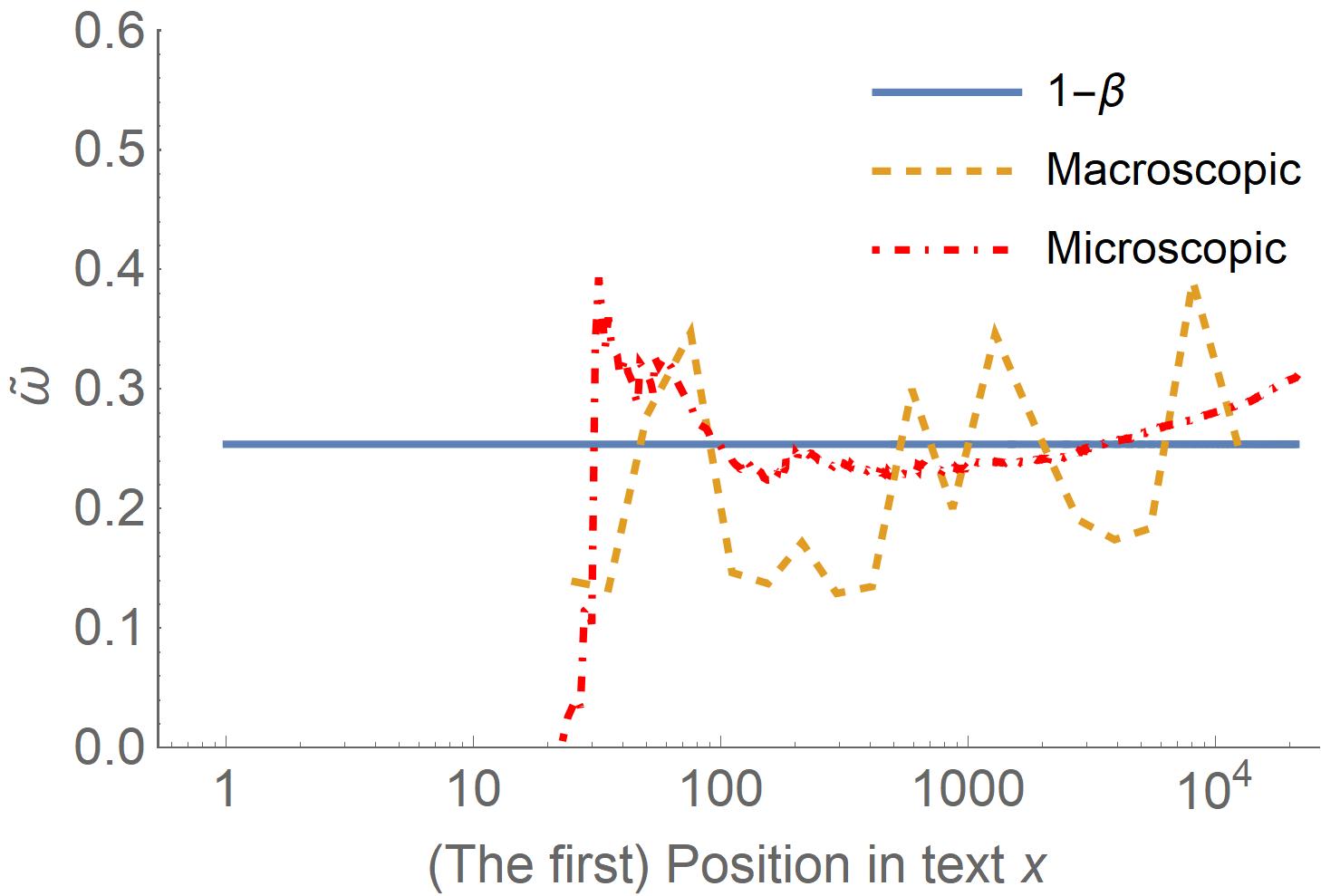}}
\subfigure[]{\includegraphics[width=0.42\textwidth]{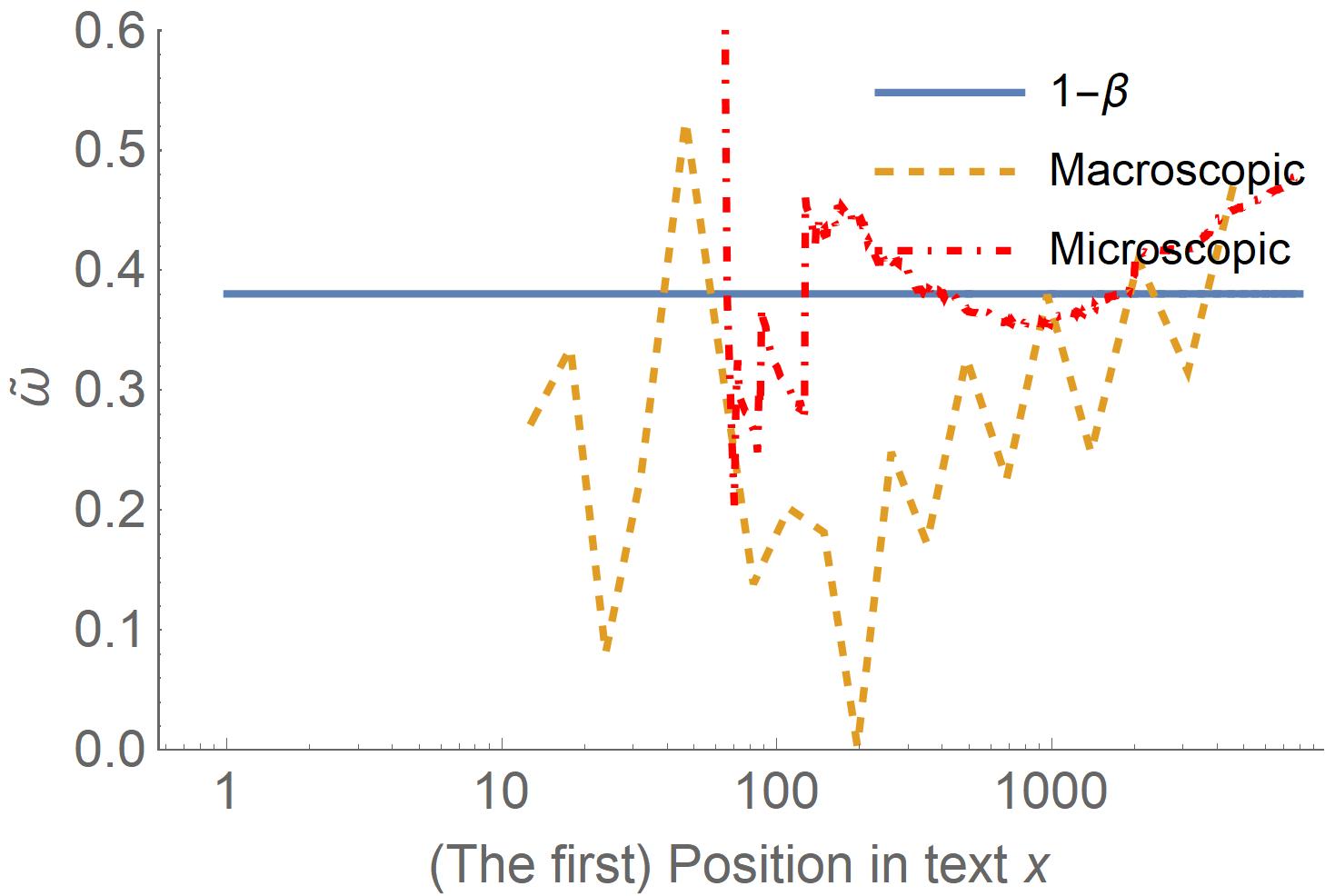}}\\
\subfigure[]{\includegraphics[width=0.23\textwidth]{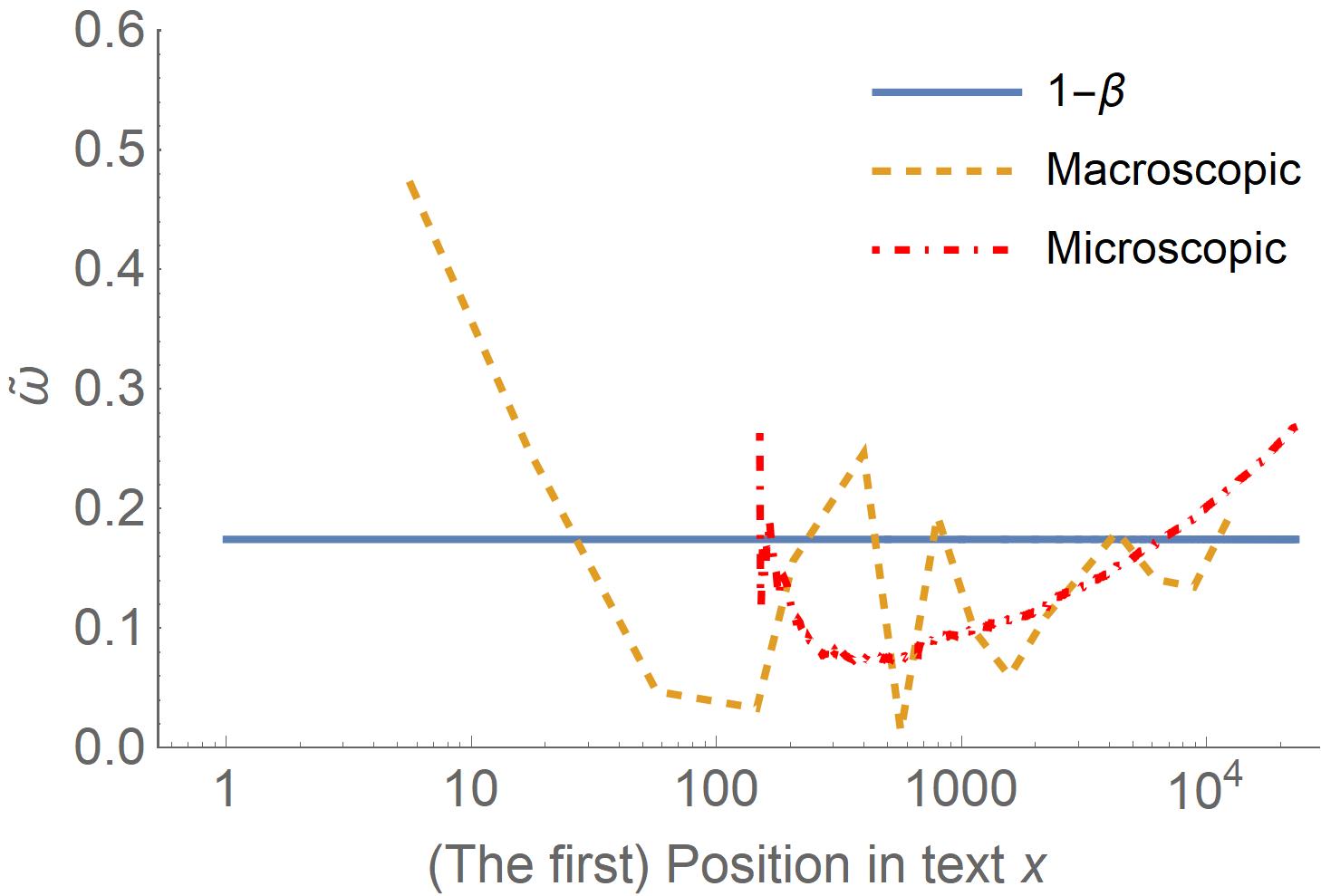}}
\subfigure[]{\includegraphics[width=0.23\textwidth]{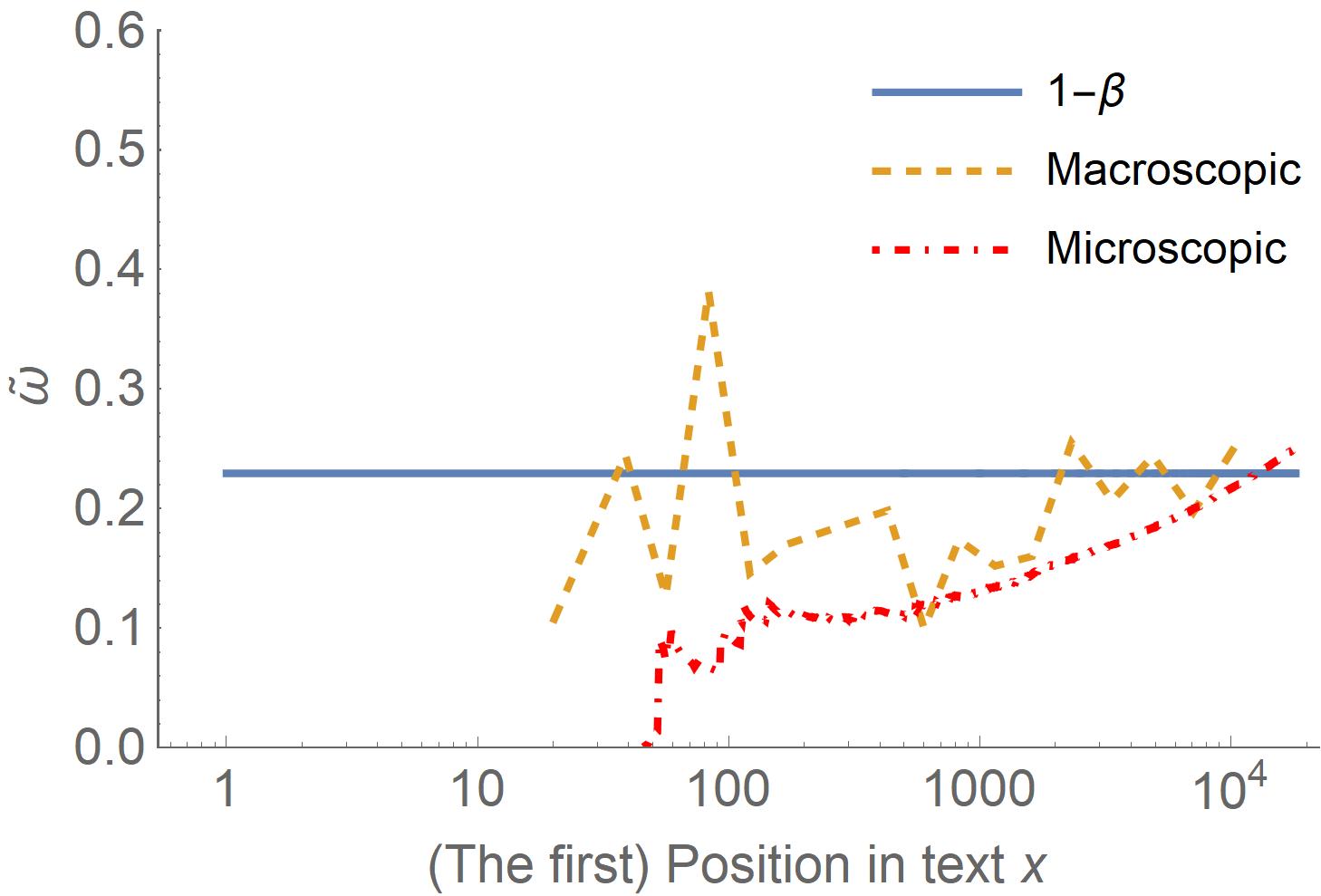}}
\subfigure[]{\includegraphics[width=0.23\textwidth]{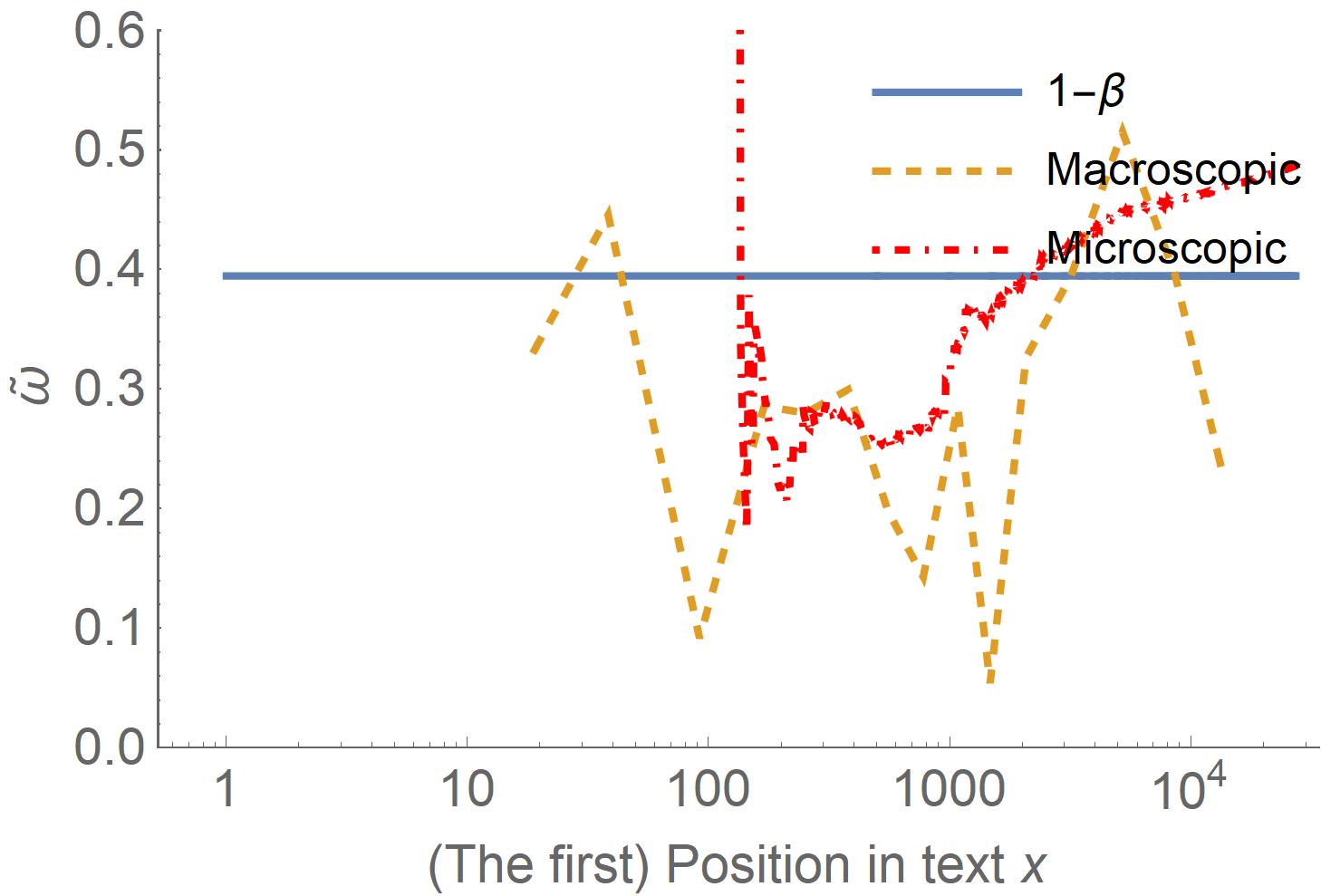}}
\subfigure[]{\includegraphics[width=0.23\textwidth]{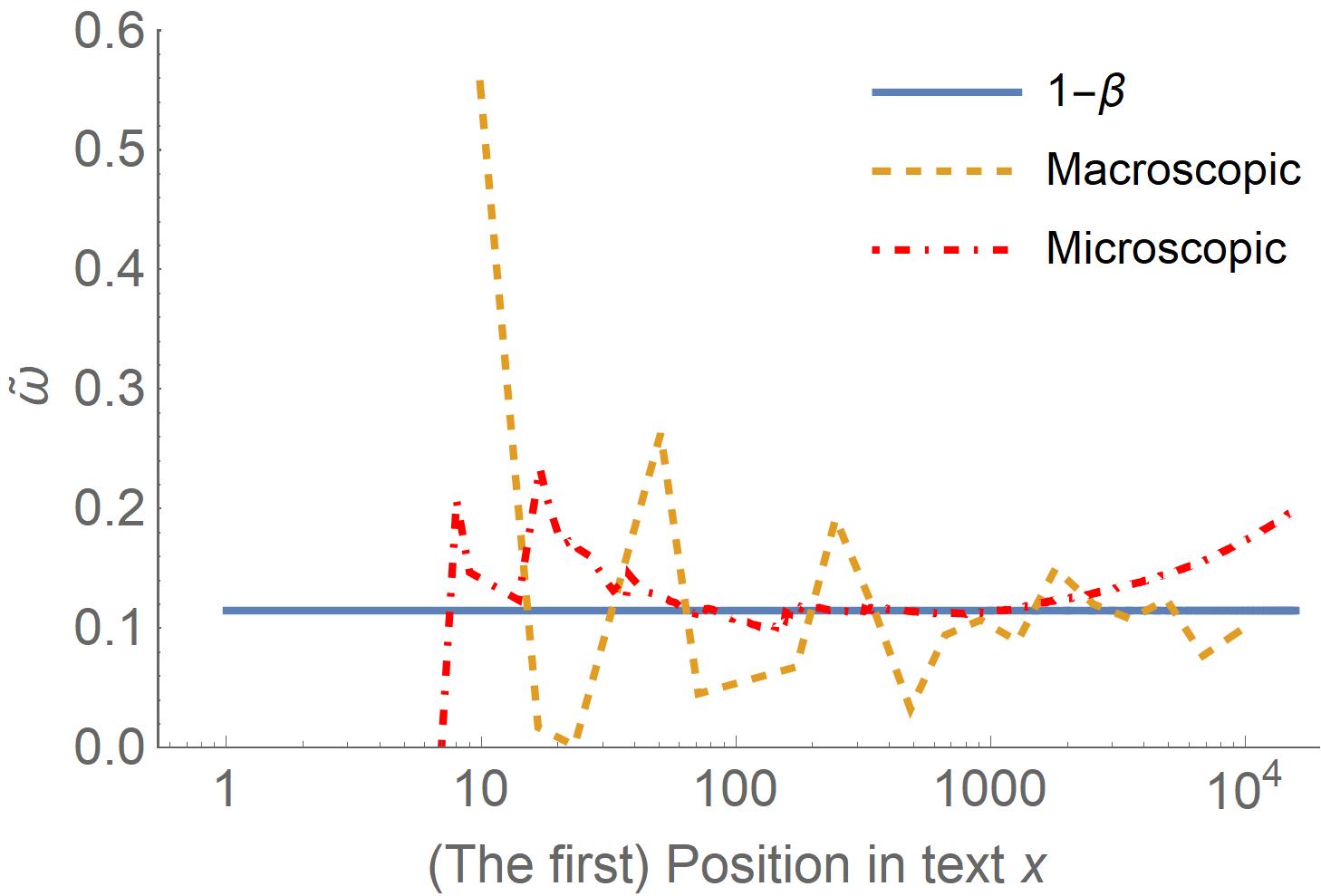}}
\caption{\label{fig:ombarvar}Variation of $\tilde{\omega}$ vs. $x$. (a) the excerpt of $Moby Dick$, (b) $Cj2019$ and (c) Arabic, (d) Hindi, (e) Lao and (f) Polish texts in Leipzig Corpora. Variation of $\tilde{\omega}$ vs. $x$ evaluated in three ways: $1-\beta$ (Eq.~\ref{eq:omegabar}, where $\beta$ is a best-fit constant throughout text), macroscopic (Eq.~\ref{eq:omegabar}, where $\beta$ is inferred within adjacent segments of the text), microscopic (Eq.~\ref{eq:ombarmicro}). The evaluations show consistently the constancy of $\tilde{\omega}$.} 
\end{figure*}

Here we assumed a constancy of $\tilde{\omega}$. We can observe change of $\tilde{\omega}$ in text. Here are given two expressions for $\tilde{\omega}$: one is macroscopic (Eq.~\ref{eq:omegabar}), while another is microscopic, derived from Eq.~\eqref{eq:xbar} and ~\eqref{eq:Heapseq2}:
\begin{linenomath} \begin{align}\label{eq:ombarmicro}
\tilde{\omega}(y)=\frac{\sum_{y=y_i}^{y}\frac{1}{\bar{x}(y)}}{\sum_{y=y_i}^{y}\frac{1}{x(y)}}=\frac{\sum_{y=y_i}^{y}\frac{k(y)}{X}}{\sum_{y=y_i}^{y}\frac{1}{x(y)}}
\end{align} \end{linenomath}
Substituting $y_i$ from Eq.~\eqref{eq:yi} into the above equation, we can compare both expressions Eq.~\eqref{eq:omegabar} and ~\eqref{eq:ombarmicro}, which is shown in Fig.~\ref{fig:ombarvar}. The $\tilde{\omega}$ (Eq.~\ref{eq:omegabar}) is evaluated piecewise from $\beta$ best-fitted to Heaps diagram. In other words, we divided Heaps diagram into bins and inferred $\beta$ in every bin. Figure~\ref{fig:ombarvar} shows that both expressions  give approximately consistent result and even consistent with the above-inferred value $\tilde{\omega}=1-\beta$ (i.e, here $\beta$ is considered as a constant throughout text, but not evaluated in every bin). Additionaly, for the excerpt of $Moby Dick$ $\tilde{\omega}\sim 0.25$ and $y_i\sim 22.1$ are inferred while for the text of $Cj2019$ $\tilde{\omega}\sim 0.38$ and $y_i\sim 42.3$. Of course, Heaps' exponent $\beta$ seems nearly unity before $y=y_i$.

In fact, the constancy of $\tilde{\omega}$ holds approximately in the L-process. As we can see, the numerator and denominator in Eq.~\eqref{eq:ombarmicro} grow slowly and their increasing rates get lower along with $y$ because $x$ and $\bar{x}$ grow with $y$. Therefore, we can consider $\bar{\omega}$ is almost constant throughout text.

Now we can claim the following, which leads to Heaps’ law.

\begin{proposition}
Supposing that $\tilde{\omega}$ is a constant, in the L-process, the first positions $x$ of words can be approximated by a power-law function of their type number $y$: $y=K x^{\beta}$.
\end{proposition}
 
Additionally, as we showed, $\tilde{\omega}$ can be approximately considered as a constant throughout the L-process.

\section{Derivation of Zipf's law}\label{sec:Zipf}
\subsection{Equation Method}\label{sec:ZipfEqu}

Before deriving Zipf’s law, let’s see its mathematical framework and drawbacks of the previous works. Here we focus on the probability density function (PDF) of the word frequency $p(k)=\frac{n(k)}{Y}$. Then the rank of the frequency $r(k)$ can be defined as
\begin{linenomath} \begin{align}\label{eq:rankdef}
r(k)=Y \int_k^\infty p(k)dk,
\end{align} \end{linenomath}
where $Y$ is the whole types in text. Then Zipf's law implies 
\begin{linenomath} \begin{align}\label{eq:Zipfslaw}
k(r)\sim r^{-\alpha},
\end{align} \end{linenomath}
which in turn can lead to another power-law relation for $p(k)$
\begin{linenomath} \begin{align}
p(k)\sim k^{-A}.
\end{align} \end{linenomath}
It can easily be shown that $\alpha$ and $A$ are related as 
\begin{linenomath} \begin{align}\label{eq:aArel}
\frac{1}{\alpha}=A-1.
\end{align} \end{linenomath}.

Zipf's law also announced that $\alpha\approx 1$ in common English text. But many authors \citep[e.g. see][]{Bernhardsson2011, Boytsov2017, van Leijenhorst2005, Lu2010} derived that Zipf's exponent $\alpha$ has a relation with Heaps' exponent $\beta$ as
\begin{linenomath} \begin{align}\label{eq:abrel}
\frac{1}{\alpha}=\beta, \ A=1+\beta
\end{align} \end{linenomath}

However, there is an obvious exception (see Appendix~\ref{app:zipfLawSamples})

\begin{property}\label{th:prop2}
In a dictionary-type text which consists only of new words, $\beta\rightarrow1$ and $A\rightarrow\infty$, which is far beyond $A = 1 + \beta$, because the number of words $k > 1$ are negligible. 
\end{property}
Therefore, Eq.~\eqref{eq:abrel} has not a generality. We have to develop a new formula. 

\begin{figure*}
\centering
\includegraphics[width=0.42\textwidth]{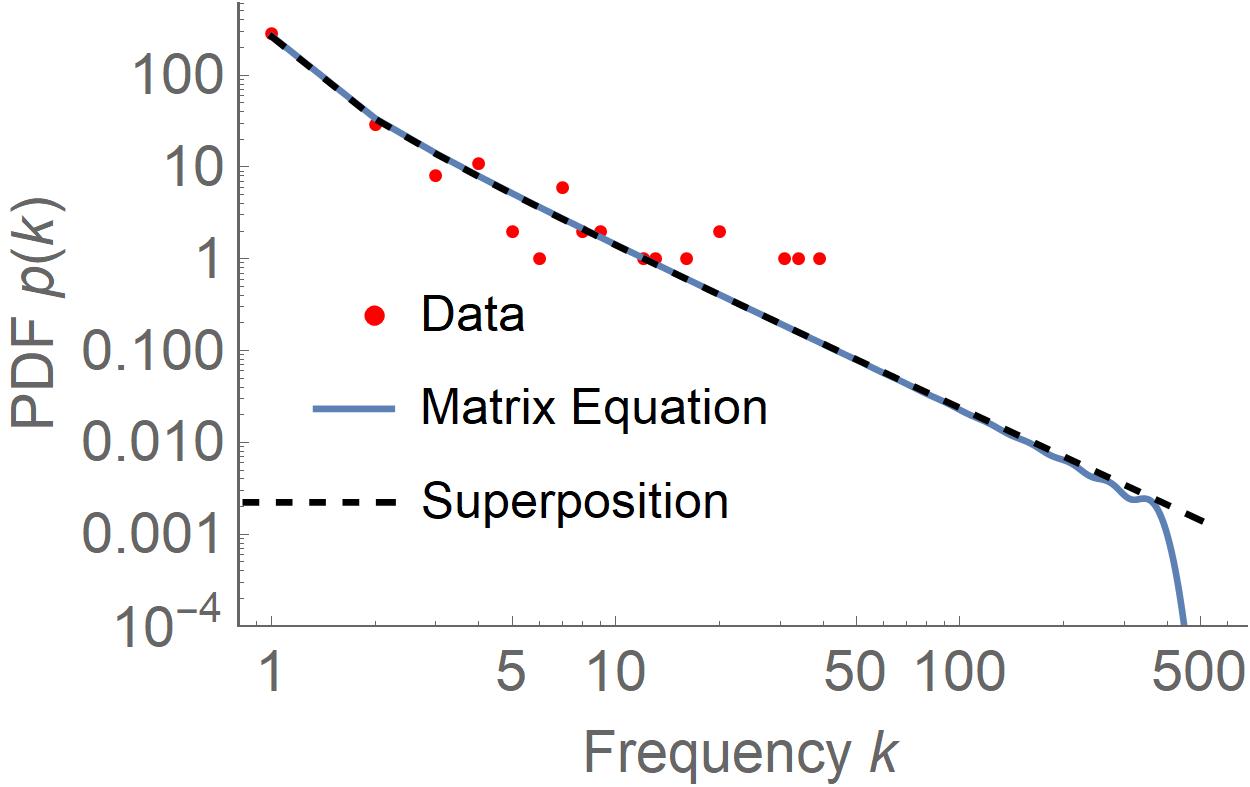}
\caption{\label{fig:matrEq} Comparison between the methods and data. The numerical solution for $p(k)$ in the matrix equation (solid) compared with data (dots) for the excerpt of $Moby Dick$ and the superposition model (dashed) seen later. The curves and data seem to be well consistent} 
\end{figure*}

Deriving Heap's law, we consider the formula for conditional probability $p(x)=\sum_{\bar{x}}p(\bar{x})p(x\vert{\bar{x}})$, where $\bar{x}=\frac{X}{k}$ and we can replace   $p(\bar{x})$ with $p(k)$ and $p(x\vert\bar{x})$ with $p(x\vert k)$. $p(x\vert k)$ can be defined by Eq.~\eqref{eq:pxk1}. Then the above conditional probability can be rewritten:
\begin{linenomath} \begin{align}
p(x)=\sum^{k_\text{max}}_{k=1}p(k)\frac{\binom{X-x}{k-1}}{\binom{X}{k}}\label{eq:px1},
\end{align} \end{linenomath}
where $k_\text{max}$ is the maximum of $k$ (word frequency). This is an equation we have to solve for $p(k)$. $p(x)$ can be obtained by $p(x)=\left\vert\frac{dy}{dx}\right\vert p(y)$ and Heaps' law $y=Kx^{\beta}$, assuming that $y$ follows a uniform distribution in $[1, Y]$ so that $p(y)=\frac{1}{Y}$ merely. Then we can express 
\begin{linenomath} \begin{align}
p(x)=\frac{K\beta}{Y}x^{\beta-1}.\label{eq:px2},
\end{align} \end{linenomath}
Eq.~\eqref{eq:px1} seems not to give an analytic solution. We can solve this equation numerically by using linear algebra, setting $k_\text{max} = X$. In relation to computational capacity limitation, we showed a result for small $X\approx$ a few hundreds. For an example, we analyzed an excerpt which consists of first 5 paragraphs of $Moby Dick$ with the length (or the tokens) $X = 685$. Figure~\ref{fig:matrEq} shows that the numerical solution is well consistent with data.

\subsection{Superposition Method}\label{sec:superMeth}
To get an analytic solution more easily and intuitively, now we propose a ``superposition'' model. We can express the rate (i.e., probability of occurrence) of a word by a ``leaf'' which has an area of unity. A word type can occur with a probability as a new token. If words appear one after one, ``a superposed area of appeared leaves'' implies a probability that the same word appears several times. The number of the leaves implies the number of tokens. Here we can find out the axiom of the L-process: all the leaves are equal area and their superposition patterns coincide with each other. 

\begin{figure*}
\centering
\subfigure[]{\includegraphics[width=0.21\textwidth]{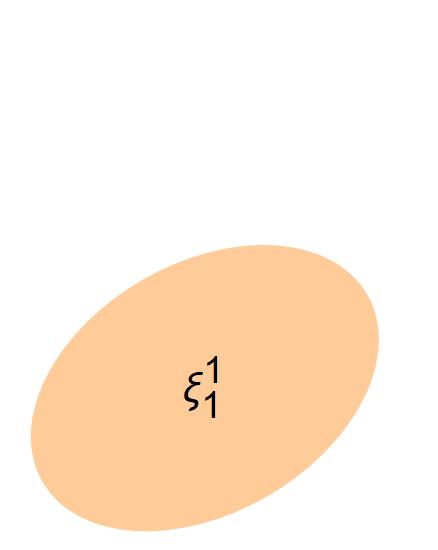}\label{fig:31leaf}}  
\subfigure[]{\includegraphics[width=0.3\textwidth]{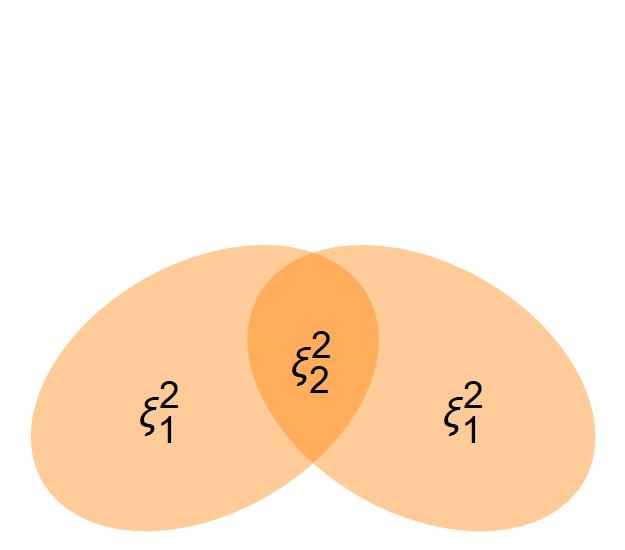}\label{fig:32leaf}}
\subfigure[]{\includegraphics[width=0.3\textwidth]{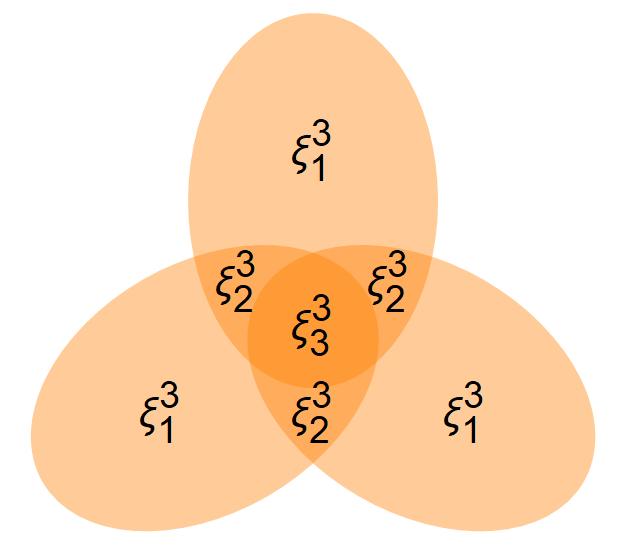}\label{fig:33leaf}}
\caption{\label{fig:3leaf} Superposition of the leaves. $\xi^X_k$ stands for the size of a part overlaid $k$ times by $X$ leaves. Such a part has $\binom{X}{k}$ replicas.} 
\end{figure*}

The area of every leaf is unity. $X$ can express the number of the leaves. If $X$ leaves are overlapped, we denote a $k$-fold overlaid part or its area by $\xi^X_k$. Figure~\ref{fig:3leaf} shows that such a part has altogether $\binom{X}{k}$ replicas (compare with the above-mentioned $p(x\vert k)$ to check). Therefore, the types (or the number) of $k$-times appeared words are equal to $Y_k=\binom{X}{k}\xi^X_k$. We define $Y=\sum_{k=1}^XY_k=\sum_{k=1}^X\binom{X}{k}\xi^X_k$, where $Y$ is the total area covered by the leaves or the whole word types in text. Then Heaps' law $y=Kx^{\beta}$ can be transformed into $Y=KX^{\beta}$. 

In Fig.~\ref{fig:3leaf} it can easily be known that 
\begin{linenomath} \begin{align}\label{eq:xidiv}
\xi^{X-1}_{k-1}=\xi^X_{k-1}+\xi^X_k, 
\end{align} \end{linenomath}
which is a key formula to relate the different frequencies. We can suppose that the word types, i.e. Heaps' law, affect the frequency distribution or Zipf's law. Adopting Heaps' law, we obtain $\xi^X_1=X^{\beta}-(X-1)^{\beta}$ from Fig.~\ref{fig:3leaf}. According to the above recursive relation, 
\begin{linenomath} \begin{align}\label{eq:xidiv1}
\xi^X_k=\xi^{X-1}_{k-1}-\xi^X_{k-1}=\sum _{j=0}^{k-1} (-1)^{j+k-1} \binom{k-1}{j} \left((X-j)^{\beta }-(X-j-1)^{\beta}\right).
\end{align} \end{linenomath}

It should be noted that $Y_k\rightarrow\infty$ for $X\rightarrow\infty$. The quantity of interest is not $Y_k$s, but the ratio $\frac{Y_k}{Y_1}$, i.e., the probability of the frequency, which we evaluate now. From the above equations, we obtain
\begin{linenomath} \begin{align}\label{eq:Ykrat}
\frac{Y_k}{Y_1}=\frac{\binom{X}{k}\sum _{j=0}^{k-1} (-1)^{j+k-1} \binom{k-1}{j} \left((X-j)^{\beta }-(X-j-1)^{\beta}\right)}{X\left(X^{\beta}-(X-1)^{\beta}\right)}, 
\end{align} \end{linenomath}
which can be factorized for convenience as follows:
\begin{linenomath} \begin{align}\label{eq:Ykratdiv}
\frac{Y_k}{Y_1}=\frac{\binom{X}{k}}{X^k}\frac{X^{k-1}\sum _{j=0}^{k-1} (-1)^{j+k-1} \binom{k-1}{j} \left((X-j)^{\beta }-(X-j-1)^{\beta}\right)}{\left(X^{\beta}-(X-1)^{\beta}\right)}, 
\end{align} \end{linenomath}

Let's consider the limit for $X\rightarrow\infty$, which holds for almost texts. This is an important assumption to make the derivation much simpler. Then, the first term tends as 
\begin{linenomath} \begin{align}\label{eq:Ykratdiv1term}
\lim_{X\rightarrow\infty}\frac{1}{X^k}\binom{X}{k}=\frac{1}{k!}.
\end{align} \end{linenomath}
Changing $X$ into a new variable $t=\frac{1}{X}$, the limit $X\rightarrow\infty$ is replaced with $t\rightarrow0$. Then we expand the ratio $\frac{(1-jt)^{\beta}-(1-(j+1)t)^{\beta}}{\left(1-(1-t)^{\beta}\right)}$ into a Taylor series with respect to $t$. Taking into account that $\sum _{j=0}^{k-1} (-1)^{j+k-1} \binom{k-1}{j}j^{\eta} =0$ if $\eta<k-1$, the second factor in Eq.~\eqref{eq:Ykratdiv} can be simplified as  
\begin{linenomath} \begin{align}\label{eq:Ykratdiv2term}
\frac{\sum _{j=0}^{k-1} (-1)^{j+k-1} \binom{k-1}{j} \left((1-j t)^{\beta}-(1-(j+1) t)^{\beta}\right)}{t^{k-1} \left(1-(1-t)^{\beta}\right)}=\prod _{j=1}^{k-1} (j-\beta). 
\end{align} \end{linenomath}
For example of $\beta=\frac{1}{2}$, the above formula gives $(-1)^{k-1}\frac{(2k-3)!!}{2^{k-1}}$. Eventually, we obtain the ratio as  
\begin{linenomath} \begin{align}\label{eq:discreterat}
\frac{Y_k}{Y_1}=\frac{\prod _{j=1}^{k-1} (j-\beta)}{k!},
\end{align} \end{linenomath}
which is the probability of discrete frequency $k$.

Considering $\sum_{k=1}^\infty\frac{Y_k}{Y_1}=\frac{1}{\beta}$ and $\prod _{j=1}^{k-1} (j-\beta)=\frac{\Gamma(k-\beta)}{\Gamma(1-\beta)}$, we can obtain the probability or PDF of the frequency in a discrete version of  Eq.~\eqref{eq:discreterat}:
\begin{linenomath} \begin{align}\label{eq:discreteP}
p(k)=\beta\frac{\prod _{j=1}^{k-1} (j-\beta)}{k!},
\end{align} \end{linenomath}
from which $n(k)=Yp(k)$, and in a continuous version:
\begin{linenomath} \begin{align}\label{eq:pYS}
p_{\text{YS}}(k)=\frac{\beta\Gamma(k-\beta)}{\Gamma(k+1)\Gamma(1-\beta)},
\end{align} \end{linenomath}
which we will call Yongsun distribution. Let's evaluate the asymptotic log-log slope of Yongsun function:
\begin{linenomath} \begin{align}\label{eq:AYS}
A_{\text{YS}}=-\frac{d(\log p_{\text{YS}}(k))}{d(\log k)}=\left\lbrace \begin{aligned}
 &1+\beta,\quad& k\rightarrow \infty\\
 &1-\gamma-\left.\frac{\Gamma^{\prime}(z)}{\Gamma(z)}\right|_{z=1-\beta},\quad& k\rightarrow 1
\end{aligned} \right.,
\end{align} \end{linenomath}
where $\gamma\approx0.577216$ is a constant called Euler's gamma. From Eq.~\eqref{eq:aArel} Zipf's exponent can be evaluated:
\begin{linenomath} \begin{align}\label{eq:aYS}
\alpha_{\text{YS}}=-\frac{d(\log k(r))}{d(\log r)}=\left\lbrace \begin{aligned}
 &\frac{1}{\beta},\quad& k\rightarrow \infty\\
 &\frac{1}{-\gamma-\left.\frac{\Gamma^{\prime}(z)}{\Gamma(z)}\right|_{z=1-\beta}},\quad& k\rightarrow 1
\end{aligned} \right..
\end{align} \end{linenomath}

In Fig.~\ref{fig:matrEq}, comparing this function with the above numerically obtained solution, we can claim that $p_{YS}(k)$ is consistent with the solution of the above matrix equation. As we can see, the exponents $\alpha$ and $A$ for frequent $(k \rightarrow\infty)$ words are coincident with previous results (Eqs.~\ref{eq:aArel} and ~\ref{eq:abrel}) and, more importantly, we obtained an asymptotic power-law behavior of $p(k)$. We can claim such a result, which leads to Zipf's law:

\begin{proposition}
In the L-process with infinite length, the distribution of word frequency has asymptotic power-law behavior. 
\end{proposition}

Note that $\left.\frac{\Gamma^{\prime}(z)}{\Gamma(z)}\right|_{z=1-\beta}$ is negative so that $\alpha_{\text{YS}}$ for $k\rightarrow 1$ is positive. The latter decreases monotonically with $\beta$ so that, when $\beta$ lowers in later part of text (see Heaps diagram in Fig.~\ref{fig:HeapsDiag}, Zipf's exponent $\alpha$ for $k\rightarrow 1$ increases. The above formula can explain property~\ref{th:prop2} excellently. In fact,
\begin{linenomath} \begin{align}
\lim_{\beta-1}\left.\frac{\Gamma^{\prime}(z)}{\Gamma(z)}\right|_{z=1-\beta}=-\infty
\end{align} \end{linenomath}
so that 
\begin{linenomath} \begin{align}
\lim_{\beta-1}A=\infty
\end{align} \end{linenomath}
which reproduces property~\ref{th:prop2}. In other words, the number of words $k > 1$ are negligible in $\beta \rightarrow 1$. 

\begin{figure*}
\centering
\subfigure[]{\includegraphics[width=0.42\textwidth]{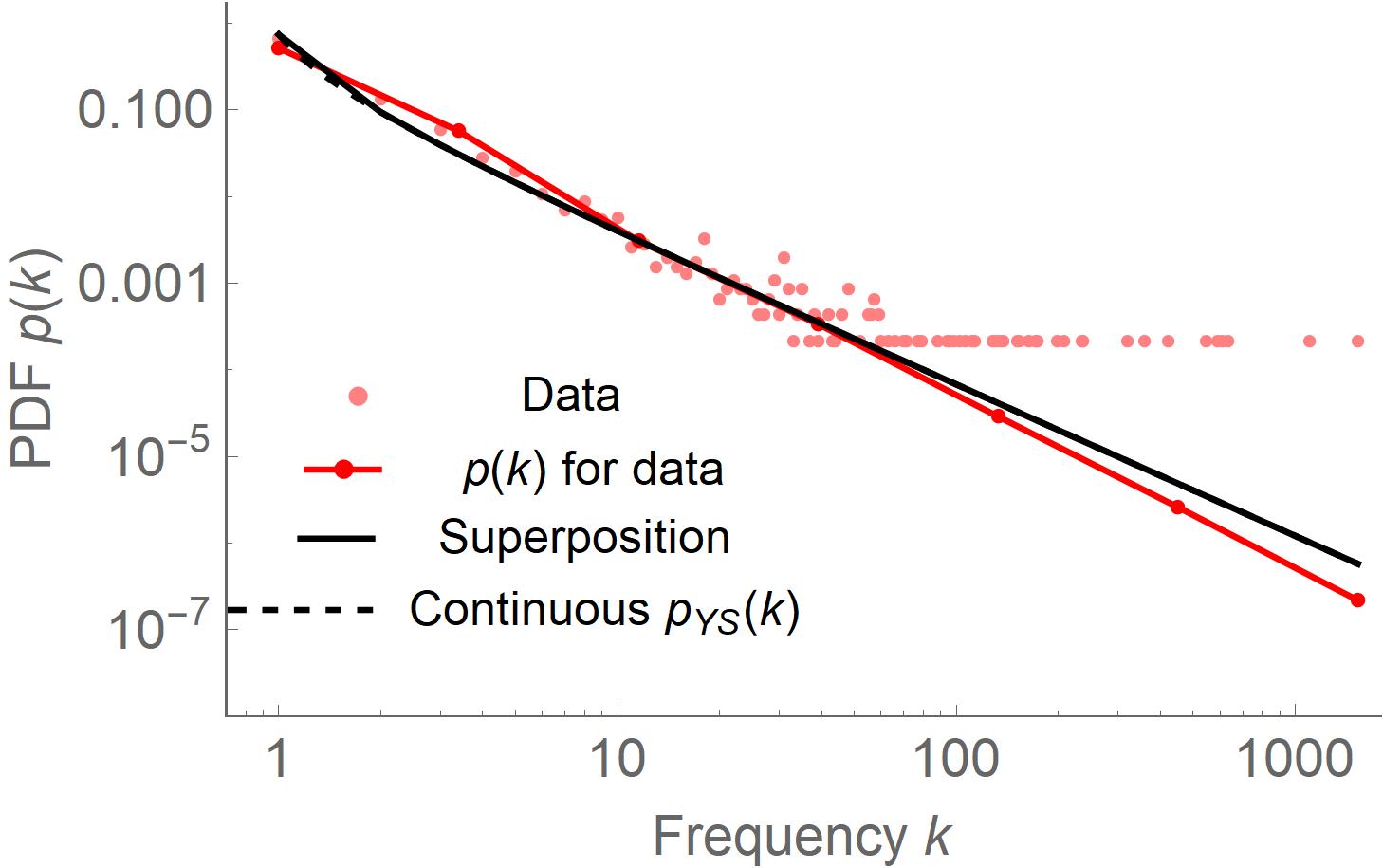}}
\subfigure[]{\includegraphics[width=0.42\textwidth]{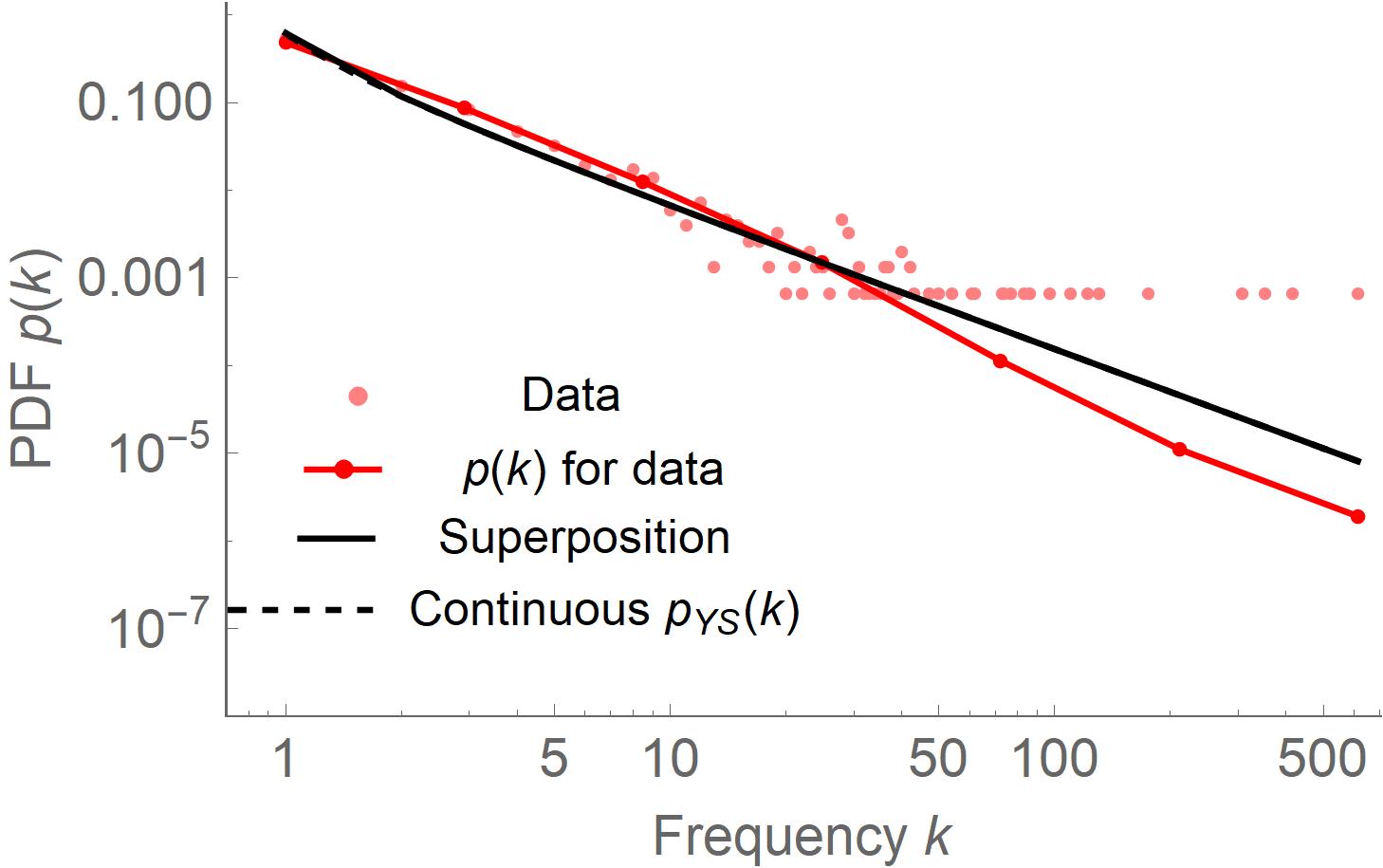}}\\
\subfigure[]{\includegraphics[width=0.23\textwidth]{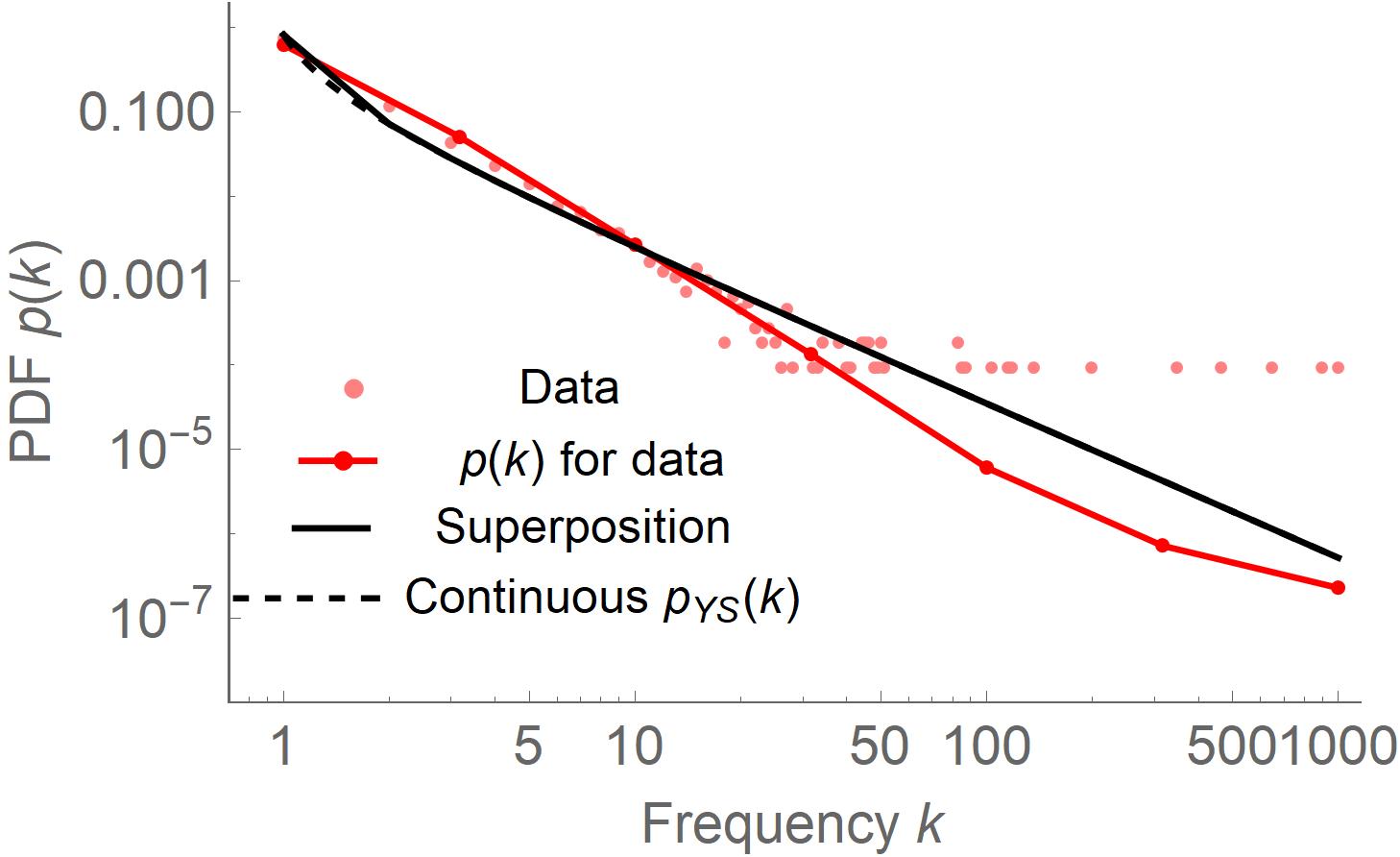}}
\subfigure[]{\includegraphics[width=0.23\textwidth]{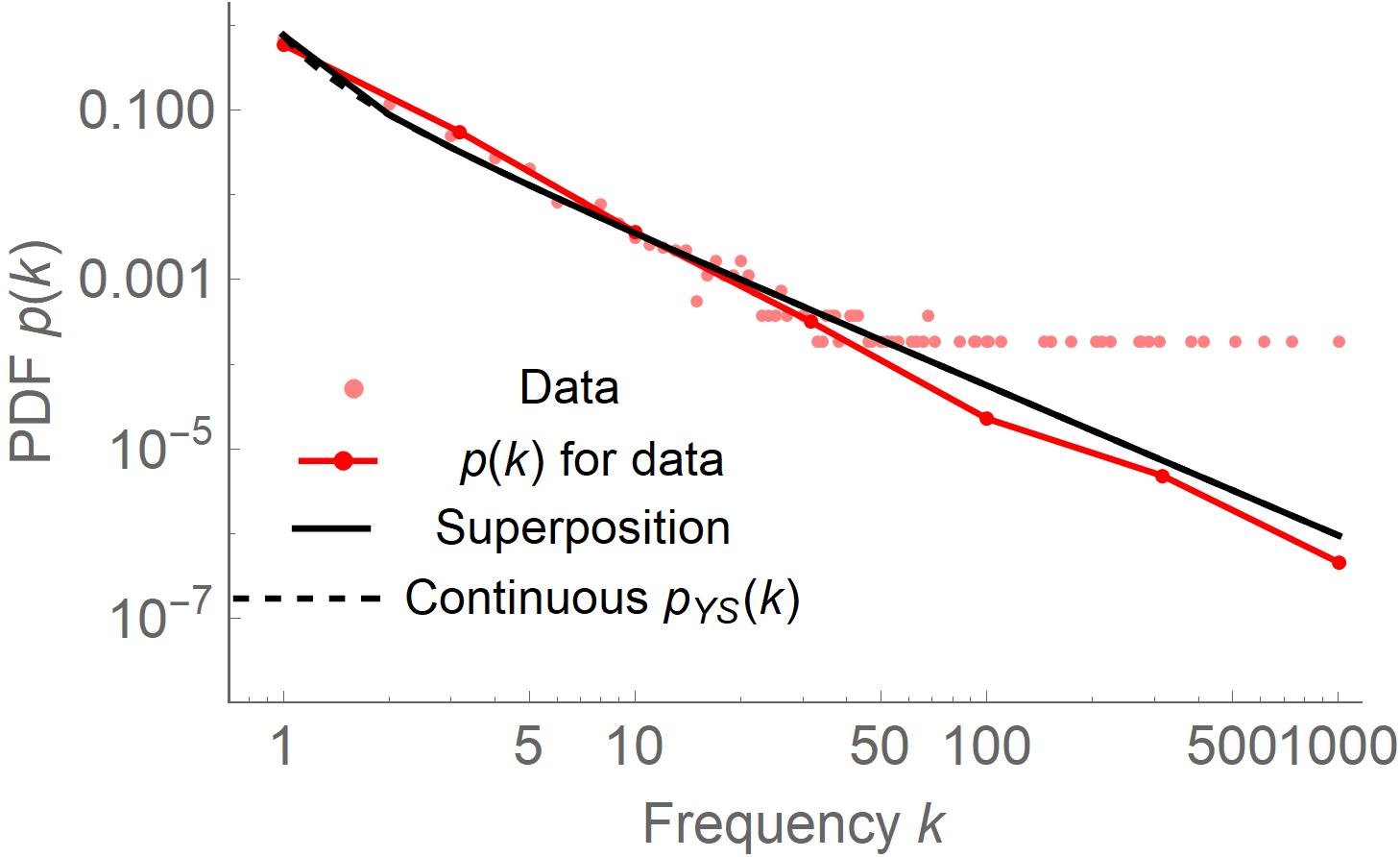}}
\subfigure[]{\includegraphics[width=0.23\textwidth]{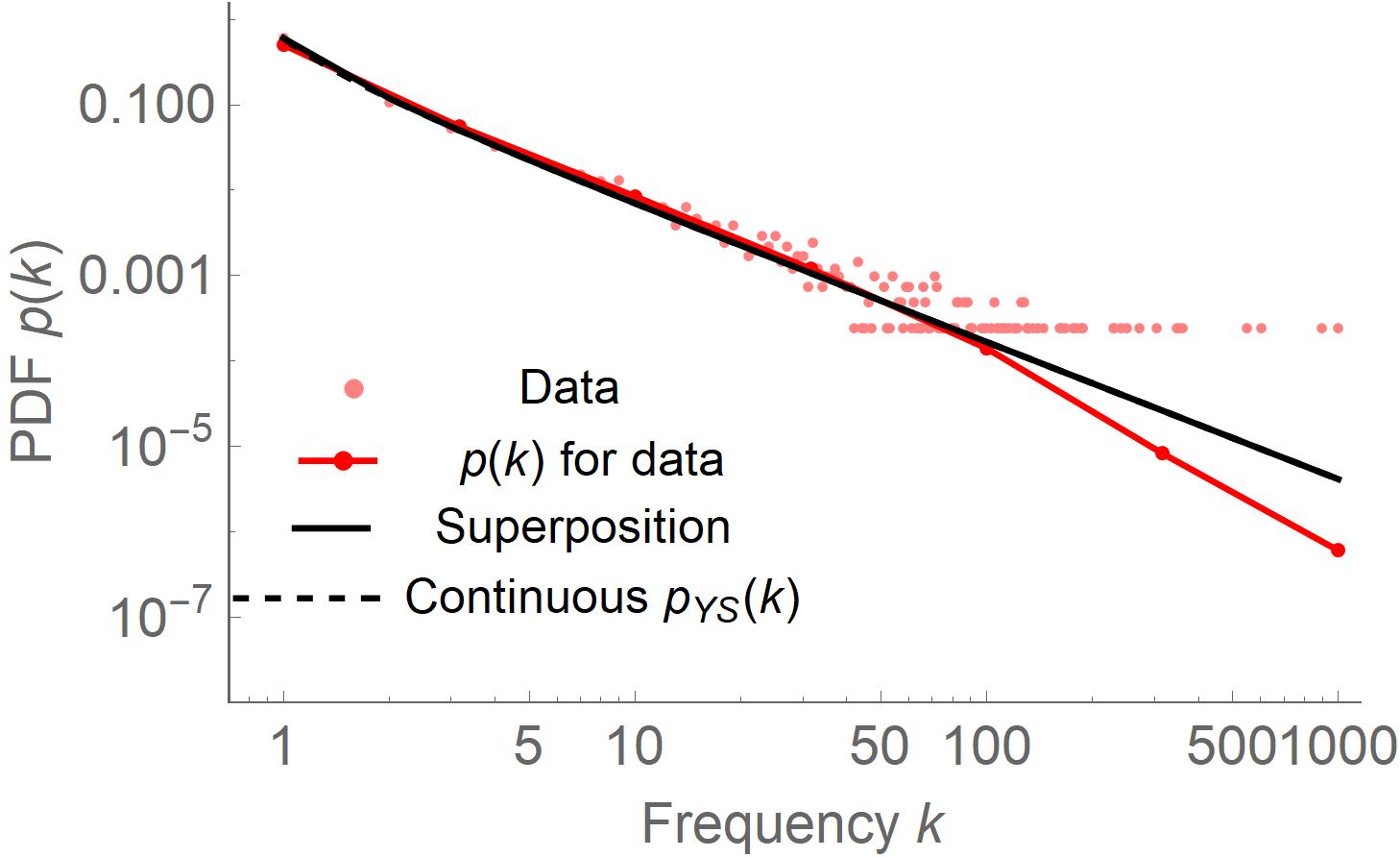}}
\subfigure[]{\includegraphics[width=0.23\textwidth]{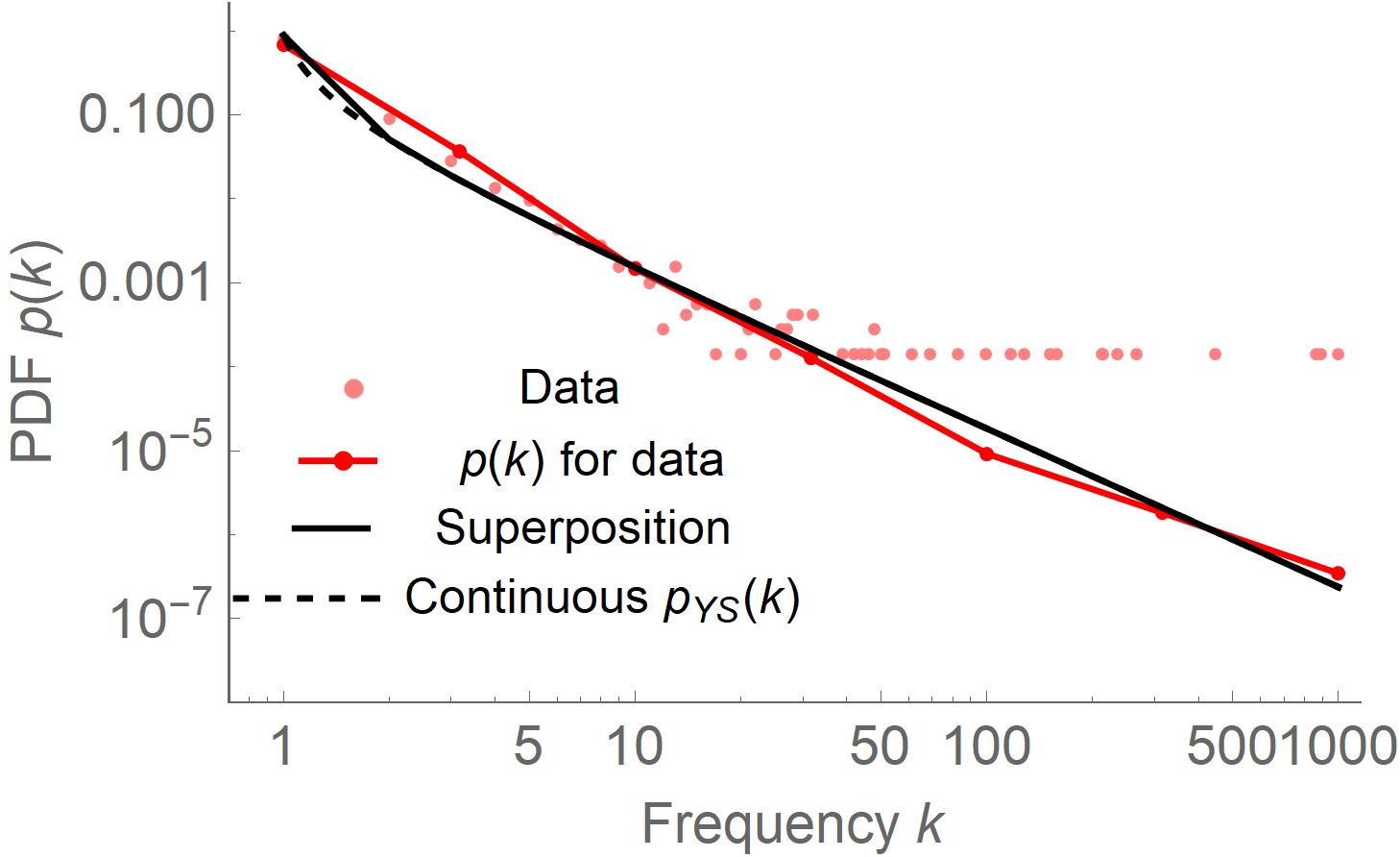}}
\caption{\label{fig:pkdiag} $p(k)$ diagrams. (a) the excerpt of $Moby Dick$, (b) $Cj2019$ and (c) Arabic, (d) Hindi, (e) Lao and (f) Polish texts in Leipzig Corpora. Data are pictured in two formats: data points themselves and PDF. ``Superposition'' stands for the discrete version of the model while ``Continuous $p_{YS}(k)$'' for the continuous version.} 
\end{figure*}

\begin{figure*}
\centering
\subfigure[]{\includegraphics[width=0.42\textwidth]{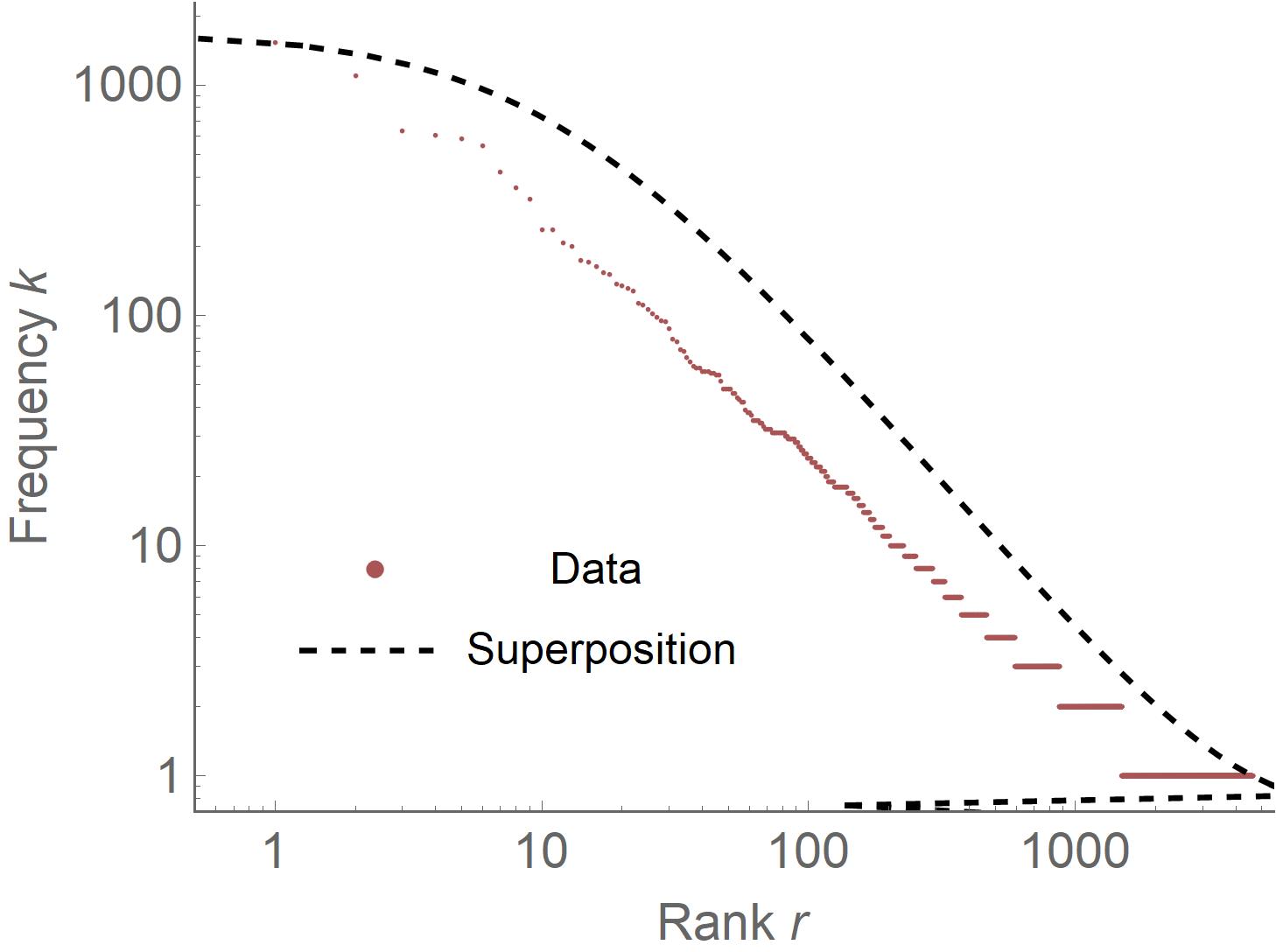}}
\subfigure[]{\includegraphics[width=0.42\textwidth]{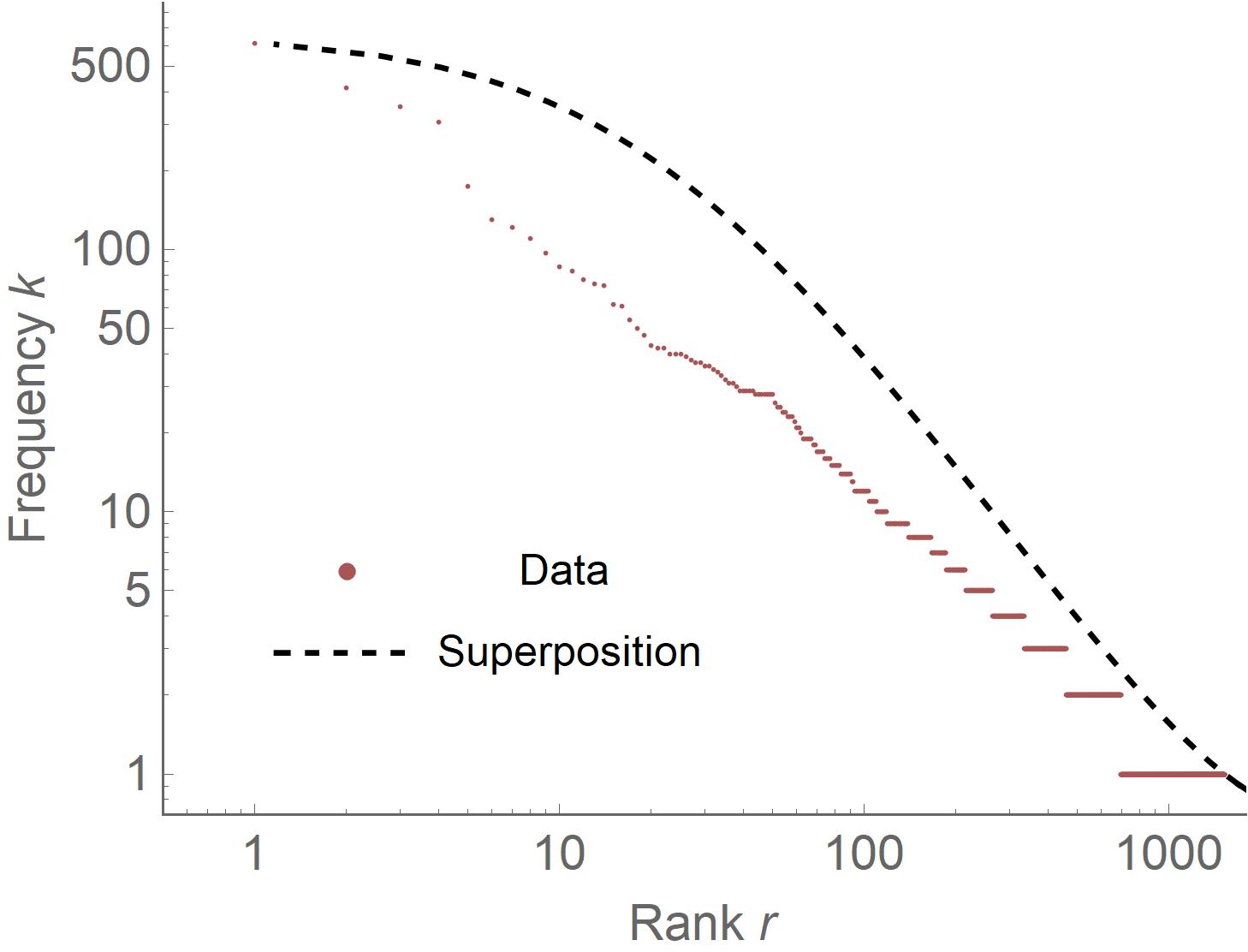}}\\
\subfigure[]{\includegraphics[width=0.23\textwidth]{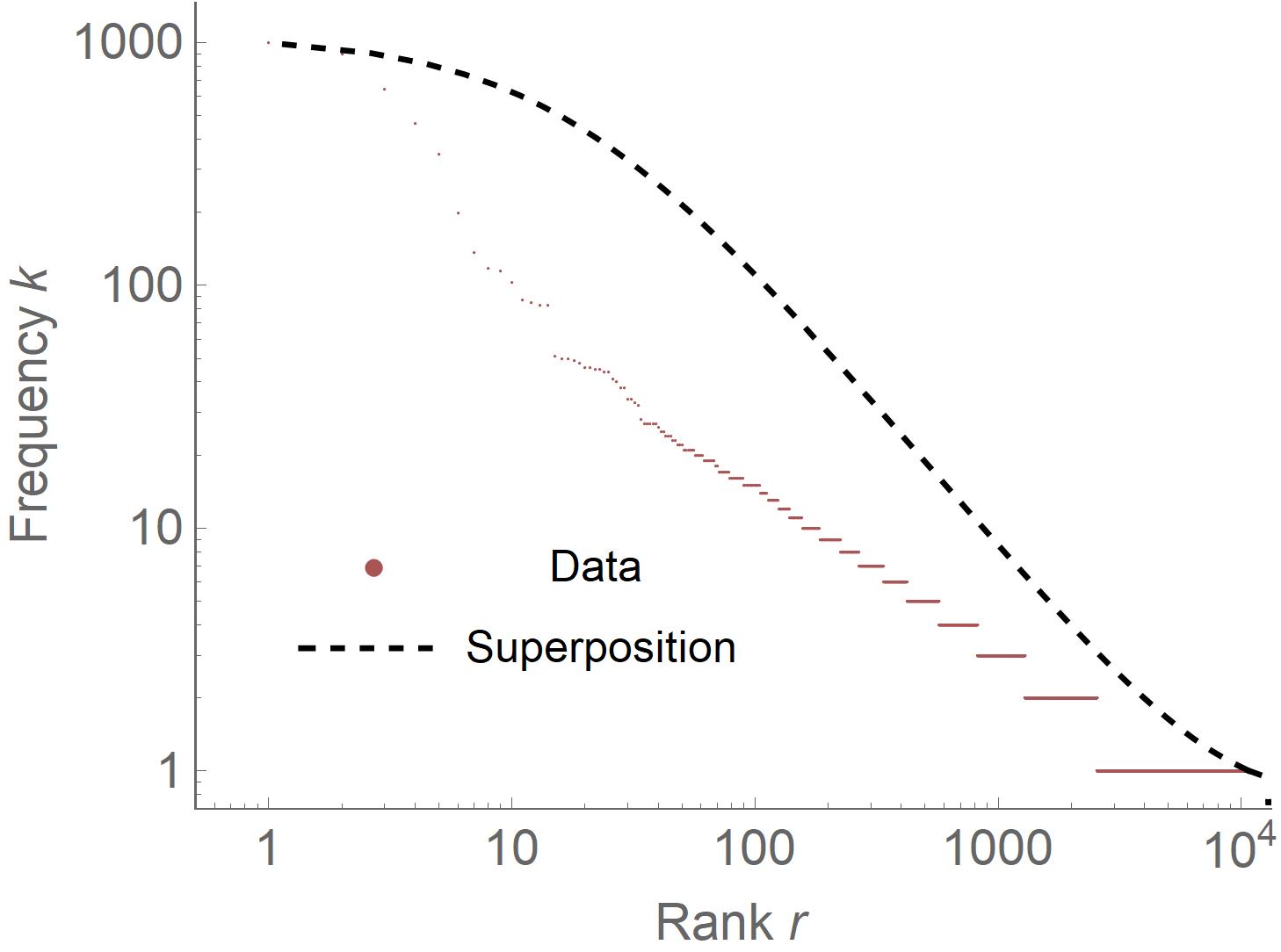}}
\subfigure[]{\includegraphics[width=0.23\textwidth]{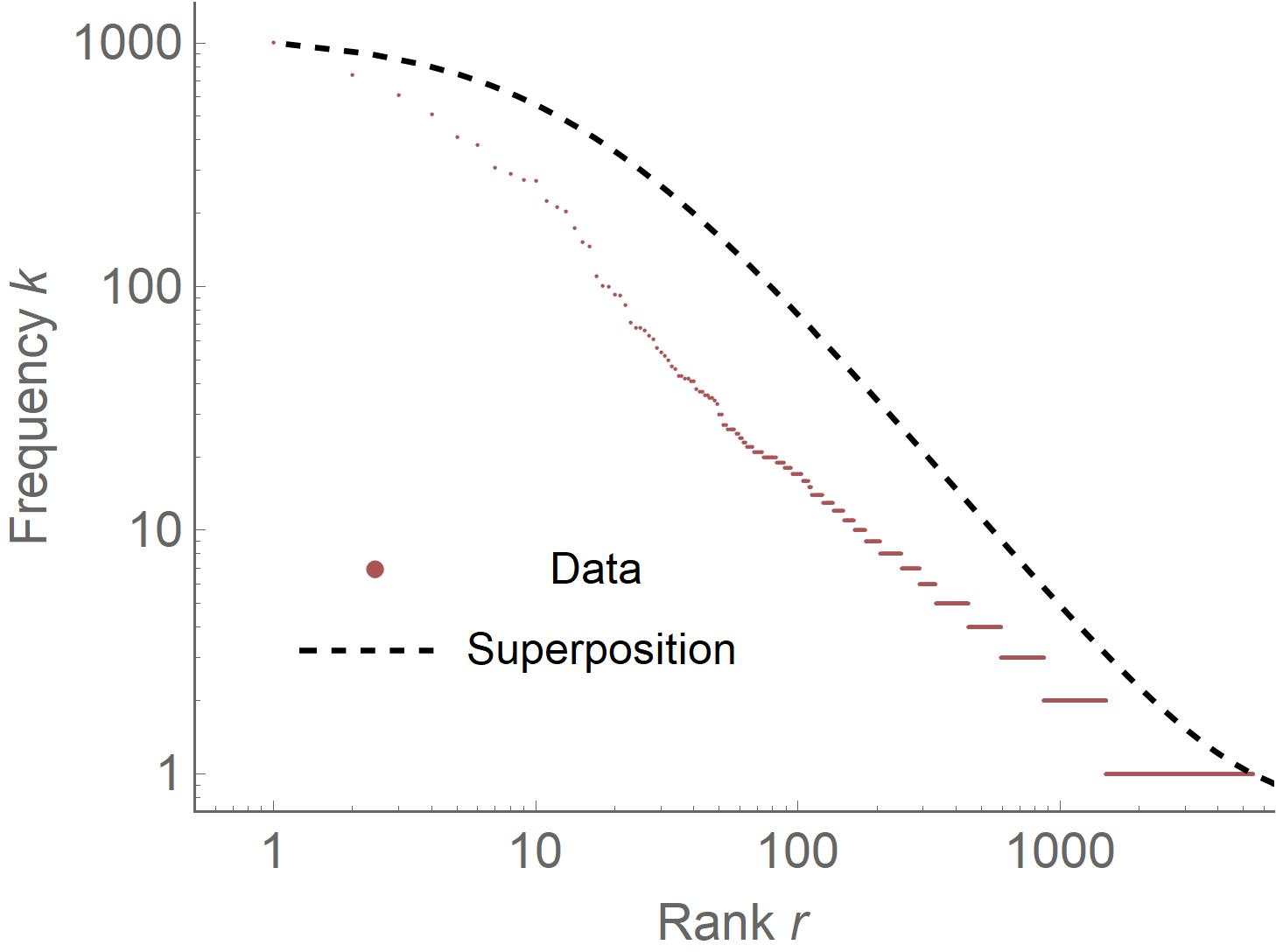}}
\subfigure[]{\includegraphics[width=0.23\textwidth]{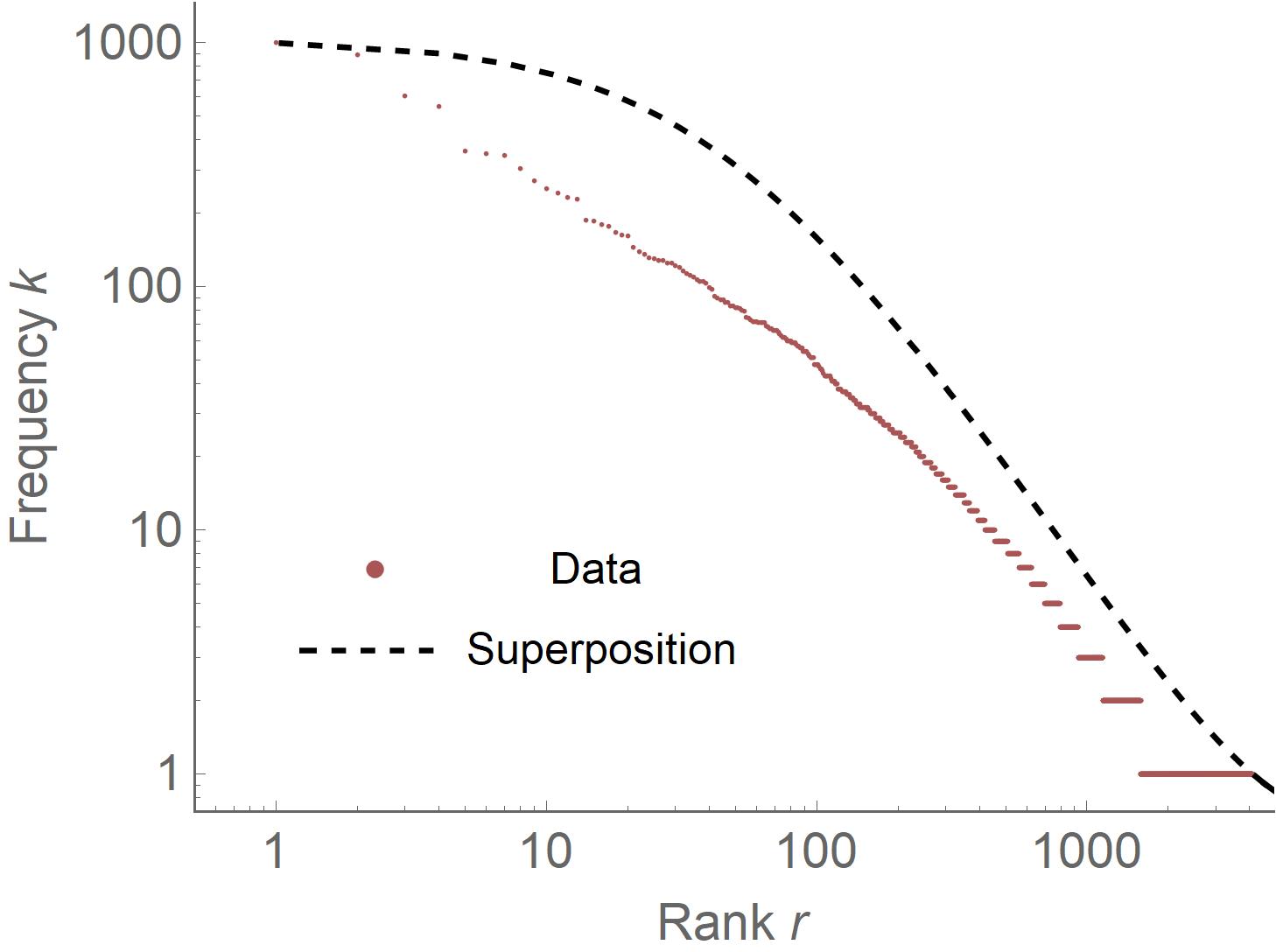}}
\subfigure[]{\includegraphics[width=0.23\textwidth]{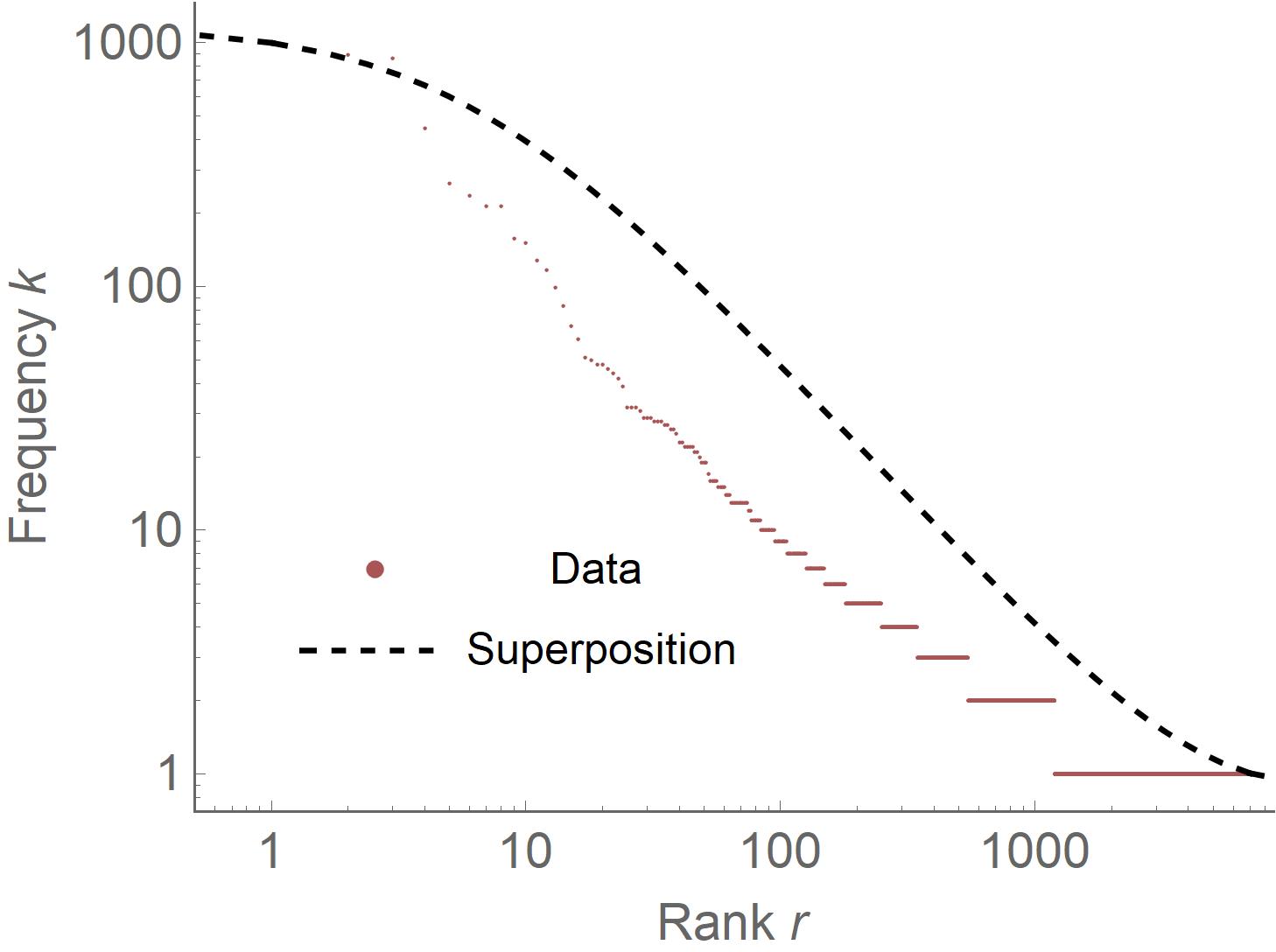}}
\caption{\label{fig:rkdiag}$r(k)$ diagrams. (a) the excerpt of $Moby Dick$, (b) $Cj2019$ and (c) Arabic, (d) Hindi, (e) Lao and (f) Polish texts in Leipzig Corpora. ``Superposition'' stands for the continuous version of the model. } 
\end{figure*}

Figure~\ref{fig:pkdiag} and ~\ref{fig:rkdiag} shows that the ``superposition of the leaves'' model fits well data both in $p(k)$ and Zipf's diagrams. For the normalization, we evaluate the rank in Fig.~\ref{fig:rkdiag} not by Eq.~\eqref{eq:rankdef} but by 
\begin{linenomath} \begin{align}\label{eq:rankdefnorm}
r(k)=\frac{Y}{\int_1^{k_\text{max}} p(k)dk} \int_k^{k_\text{max}} p(k)dk+1,
\end{align} \end{linenomath}
which can give $r\rightarrow 1$ for $k\rightarrow k_\text{max}$ and $r\rightarrow Y$ for $k\rightarrow 1$, where $k_\text{max}$ is the maximum frequency among the words appearing in the text. This formula in turn seems to deform Zipf’s diagram from the original pure power-law into a convex profile such as Zipf-Mandelbrot law in the limit $r\rightarrow 1$ for $k\rightarrow k_\text{max}$:
\begin{linenomath} \begin{align}
k(r)=\frac{C}{(r+B)^{\alpha}}
\end{align} \end{linenomath} 
In fact, let's define 
\begin{linenomath} \begin{align}
p(k)=Mk^{-A}\quad \text{and}\quad N= \int_1^{k_\text{max}} p(k)dk,
\end{align} \end{linenomath}
then we obtain $r(k)=\frac{MY}{N(A-1)}k^{1-A}+1$. Comparing with Zipf-Mandelbrot law, we can define parameters:
\begin{linenomath} \begin{align}
C=\left(\frac{MY}{(A-1)N}\right)^{\frac{1}{\beta}},\quad B=-1.
\end{align} \end{linenomath}
Figure~\ref{fig:pkdiag} shows that our model explains the statistics of all sample texts in various languages and genres very well, especially $p(k)$. In Zipf's diagram, Fig.~\ref{fig:rkdiag} for $r(k)$, data and the model show a discrepancy, but it is natural because when deriving model we use an assumption $X \rightarrow\infty$ so that a normalization factor $N=\int_1^{k_\text{max}} p(k)dk$ lifts the model curve. However, coincidence in $p(k)$ is excellent and a trend in Zipf's diagram is also well explained. From this logic we could explain a flattened Zipf's diagram for most frequent words. $r(k)$ model could be more developed.

\cite{Yu2018} analyzed texts in 50 languages and observed a 3-segment shape in Zipf’s law. They highlighted downward bending of Zipfian curve and claimed that this property could be explained by a cognitive model that, however, seems a mere replica of ``preferential attachment'' in mathematics. This downward bending can be explained in our theory (dashed line in Fig.~\ref{fig:rkdiag}). This property seems to originate from abnormal behavior of YongSun distribution (Eq.~\ref{eq:pYS}) for $k\rightarrow 1$. In other words, the rare words are more than expected by a pure power-law distribution. An upside bending in Fig.~\ref{fig:rkdiag} seems to be related to the superior limit of integration $k_{max}$ or, maybe, the additive term + 1 in Eq.~\eqref{eq:rankdefnorm}.

Amd previous works \citep[e.g. see][]{Yu2018, Natale2018} showed that Zipf's exponent (in absolute value) increases for rare words of rank greater than $\sim10^4$. TThis could be explained by that the previous works studied the corpora or very long texts while in this paper merely short texts are considered so that Heaps exponent can be constant. In fact, most novels contain $10^4$ or $10^6$ words \footnote{For example, see the homepage ``How many words should my novel be?'' at \url{https://www.reddit.com/r/writing/comments/78dwch/how_many_words_should_be_my_novel_be/}}. We can think that those words are common for most novels. Therefore, in a corpus, which can be considered to consist of the novels, the rare words of rank greater than $10^4$ are contained less, which implies that Heaps' exponent $\beta$ lowers for $y>10^4$ and, from Eq.~\eqref{eq:aYS}, Zipf's exponent $\alpha$ in corpora could increase for rank greater than $\sim10^4$. 

\section{More Properties}\label{sec:MoreProp}
Besides the aforementioned properties, we found more, analyzing texts. 
 
\begin{property} \label{th:prop3}
Zipf’s law appears not only in long text or corpora, but also in a short text which includes a few paragraphs or even in a list composed of incomplete sentences such as titles. 
\end{property}

As mentioned in Sec.~\ref{sec:intro}, previous works showed that Zipf’s law appears in various types of language process such as children’s utterance or schizophrenic speech. We have already seen a very short excerpt of $Moby Dick$. For other examples, see Appendix ~\ref{app:zipfLawSamples}. Very diverse phenomena including non-regular, non-complete English and non-English texts, n-grams involve Zipf’ law, which let us consider the law as a stable or even ``absolute’’ property of the language process. However, there is no ``absolute'' frequency of the words except a few grammatical ones. This might imply that Zipf’s law itself can never be an axiom. Some models introduced in Sec.~\ref{sec:intro} need a series of sophisticated operations to give the law, which cannot explain such a ``stability’’ of Zipf’s law in the language process.

\begin{property} \label{th:prop4}
Among the parts of sentences, the frequency lowers from the subject to the object, then to the adverbial, and at last to the predicate, neglecting the attributive as independent part. 
\end{property}

This property seems to be purely linguistic. In fact, the hero of a story might have the most frequency and the story would be woven with various actions of him. A clear point of Zipf’s or Mandelbrot’s optimization model is a trend to maximize the information in language processes. During inspection, we can be sure that almost words in text never repeat except a few words. But those a few words show Zipf’s law, where even non-iterative words are involved so that extending power-law distribution even to the frequency of 1 or 2. The verbs or the adjectives as the predicative have the main responsibility of such a maximization of information.

More properties can be found in previous works and by yourselves.

\section{Conclusion}
In this paper, searching for a basic property wherefrom Heaps’ and Zipf’s laws could be derived, we declared a uniform distribution of words in the language process (or, typically, the L-process) by experiment. The property can give a provable and reproducible basis of the mathematical modeling of the language process. In fact, starting from the uniformity, we can derive Heaps’ and Zipf’s. Old words keeping on appearing in the later part of text, so the rate of new words decreases, which implies Heaps’ law. The ``superposition model,'' where areas of all the leaves are equal and their superposition is also equal so that it coincides with the uniformity of words, gives an asymptotic power-law distribution of word frequency, i.e., Zipf’s law.

In this paper, deriving the laws is carried on in a semi-analytical way. Some assumptions are given not by rigorous proof but by intuition in observation. The uniformity of words might be an example, which, however, we can take as an axiom. In deriving Heaps’ law, we assume a parameter as constant, which has only qualitative interpretation. It is probable that the ``constant’’ could vary. Most of texts that we considered here have relatively short lengths. It should be studied how the ``constant'' can vary in long texts such as corpora, how it affects to the resultant Zip’s law and if it has a systematic trend in change throughout text. This might be resolved within the frame of the L-process. In deriving Zipf’s law, we make modeling easier and simpler by using a geometrical model called a ``superposition model'' in place of a rigorous ``equation method,'' which has no mathematical proof. Such picking up properties makes the model more intuitive, while lowering the mathematical rigor of the logic. Fortunately, the model gives results excellently consistent with the analysis of data, which is justified by the diverse sample texts. It should be noted that in deriving Zipf’s law there is not any best-fitting of data used but we can determine Zipf’s diagram in a deductive way. It should be more discussed to add more mathematical rigor to the theory.

We could expect that those laws in other examples belong to the L-process or arithmetically growing system. The laws seem to be less relevant to philological or linguistic aspects or differences between languages. Recent applications of Zipf’s law to evaluate the quality of a produced text by a neural network or machine learning is based on the assumption that the law is internal property of natural language. However, according to the discussion, the law could appear not only in language but also in other phenomena based on the L-process. And word frequency could be deviated from the law temporally even in text, for example, in a change of context. Therefore, this should be discussed more.



\section*{Conflict of interest}
The author has no conflicts to disclose.

\section*{Data availability}
Data used in this paper are available at the website addresses indicated or by corresponding with the author.

\newpage
\numberwithin{equation}{section}
\numberwithin{figure}{section}

{\LARGE SI Appendix to ``Proper Interpretation of Heaps' and Zipf's Laws''}
\appendix
\section{Solution of Eq.~\eqref{eq:Heapseq2}}\label{app:HeapsDer}

As mentioned in Sec.~\ref{sec:Heaps}, the increment of the length of text, $dx$, is composed of $dy$ new words and replicas of already appeared words. Based on the assumption of the uniformity of the word distribution in the L-process, for the old words, we can integrate a word rate $\frac{dx}{\bar{x}(y)}$, where ${\bar{x}(y)}$ is the mean span of the word type $y$, for the word type from $y = y_i$ to the given $y$, where $y_i$ is a position whence Heaps’ law seems to hold. Introducing a constant $\tilde{\omega}$ that can introduce $x(y)$ in place of ${\bar{x}(y)}$, we can obtain the following integro-differential equation:

\begin{linenomath} \begin{align*}\qquad\qquad\qquad
\frac{d y}{d x}&=1-\tilde{\omega}\int_{y_i}^{y}\frac{dy}{x(y)}, \qquad\qquad\qquad \eqref{eq:Heapseq2}
\end{align*} \end{linenomath}

To solve Eq.~\eqref{eq:Heapseq2}, we substitute $y=Kx^{\beta}$ into the equation. Taking into account that
\begin{linenomath} \begin{align}
x&=\left(\frac{y}{K}\right)^{1/\beta },\label{eq:xfromy} \\
\frac{dy}{dx}&=K \beta x^{(\beta-1)}, \label{eq:dy2dx}
\end{align} \end{linenomath} 
we can rewrite Eq.~\eqref{eq:Heapseq2} as follows:
\begin{linenomath} \begin{align}
K \beta x^{(\beta-1)}
&=1-\tilde{\omega} \int_{y_i}^{y} \frac{dy}{x(y)}
=1-\bar{\omega } \int_{y_i}^{y} \left(\frac{K}{y}\right)^{1/\beta }dy
=1-\left.\frac{\tilde{\omega} K^{1/\beta} y^{(1-\frac{1}{\beta})}}{1-\frac{1}{\beta}}\right|^y_{y_i}\nonumber\\
&=1-\frac{\tilde{\omega} K^{1/\beta} y^{(1-\frac{1}{\beta})}}{1-\frac{1}{\beta}}+\frac{\bar{\omega } K^{1/\beta } y_i^{(1-\frac{1}{\beta})}}{1-\frac{1}{\beta}}
=1-\frac{\tilde{\omega} K x^{(\beta -1)}}{1-\frac{1}{\beta}}+\frac{\tilde{\omega} K^{1/\beta} y_i^{(1-\frac{1}{\beta})}}{1-\frac{1}{\beta}}.
\end{align} \end{linenomath} 

Equating $x$ terms and constants, respectively, on both sides, simple relations between the parameters are obtained:
\begin{linenomath} \begin{align}
\beta =1-\tilde{\omega}, \\
\tilde{\omega}=1-\beta, \\
\tilde{\omega} K^{1/\beta} y_i^{(1-\frac{1}{\beta })}
=\frac{1}{\beta}-1\Longrightarrow 
K=\left(\frac{\left(\frac{1}{\beta}-1\right) y_i^{(\frac{1}{\beta}-1)}}{\tilde{\omega}}\right)^{\beta}
=\left(\frac{\left(\frac{1}{\beta}-1\right) y_i^{(\frac{1}{\beta}-1)}}{1-\beta}\right)^{\beta}
=\left(\frac{1}{\beta}\right)^{\beta} y_i^{(1-\beta)}
=\frac{y_i^{\tilde{\omega}}}{\beta^{\beta}}, \\
y_i=\left(K \beta^{\beta}\right)^{\frac{1}{\tilde{\omega}}}
=\left(K \beta^{\beta}\right)^{\frac{1}{1-\beta}}
=K^{\frac{1}{1-\beta}}\beta^{\frac{\beta}{1-\beta}}.
\end{align} \end{linenomath} 
Those relations ensure that Heaps' law $y=Kx^{\beta}$ is a solution of Eq.~\eqref{eq:Heapseq2}.

\section{Zipf's law in various texts}\label{app:zipfLawSamples}
\begin{figure*}
\centering
\subfigure[]{\includegraphics[width=0.3\textwidth]{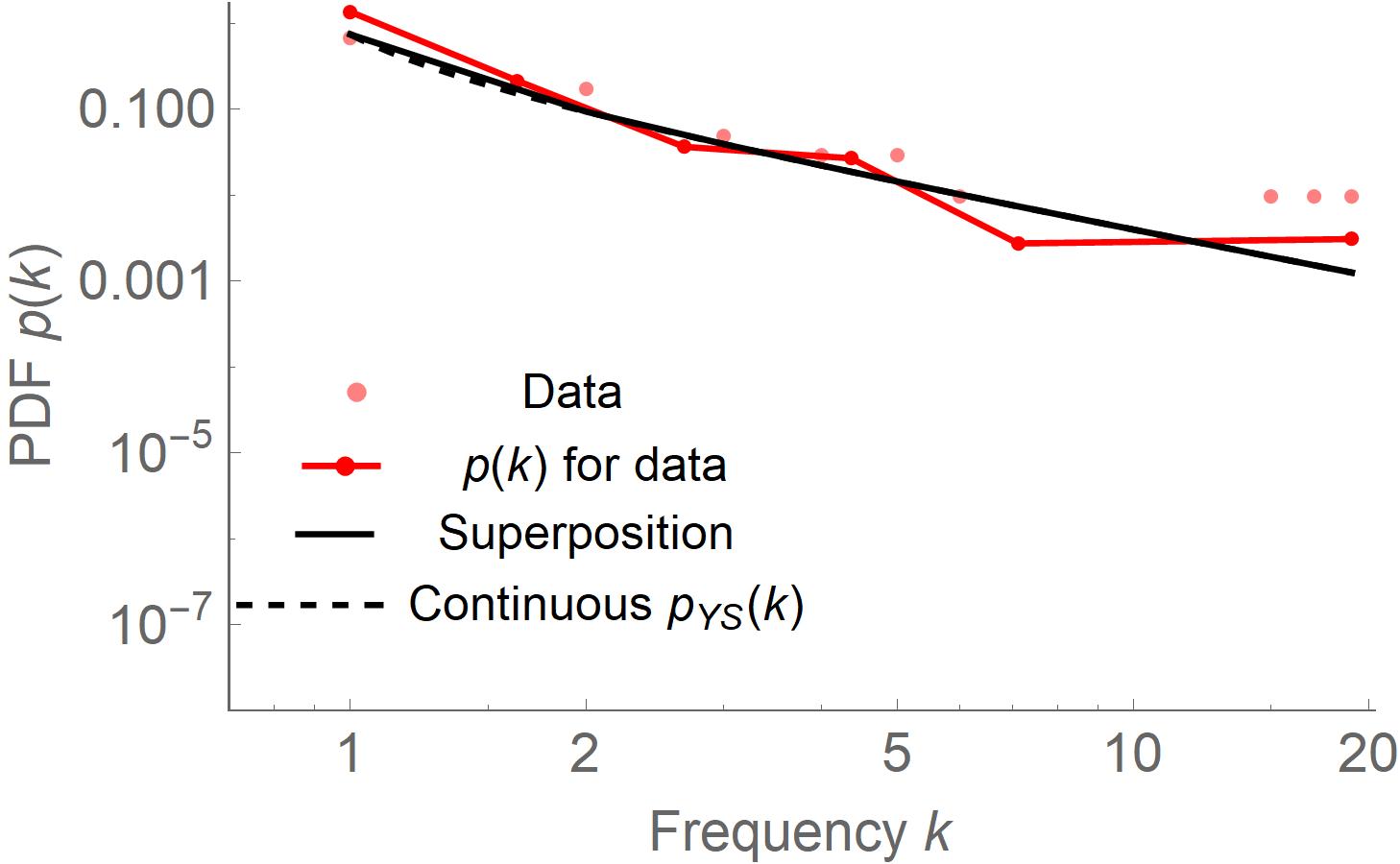}\label{fig:kfdiagram_RS}}  
\subfigure[]{\includegraphics[width=0.3\textwidth]{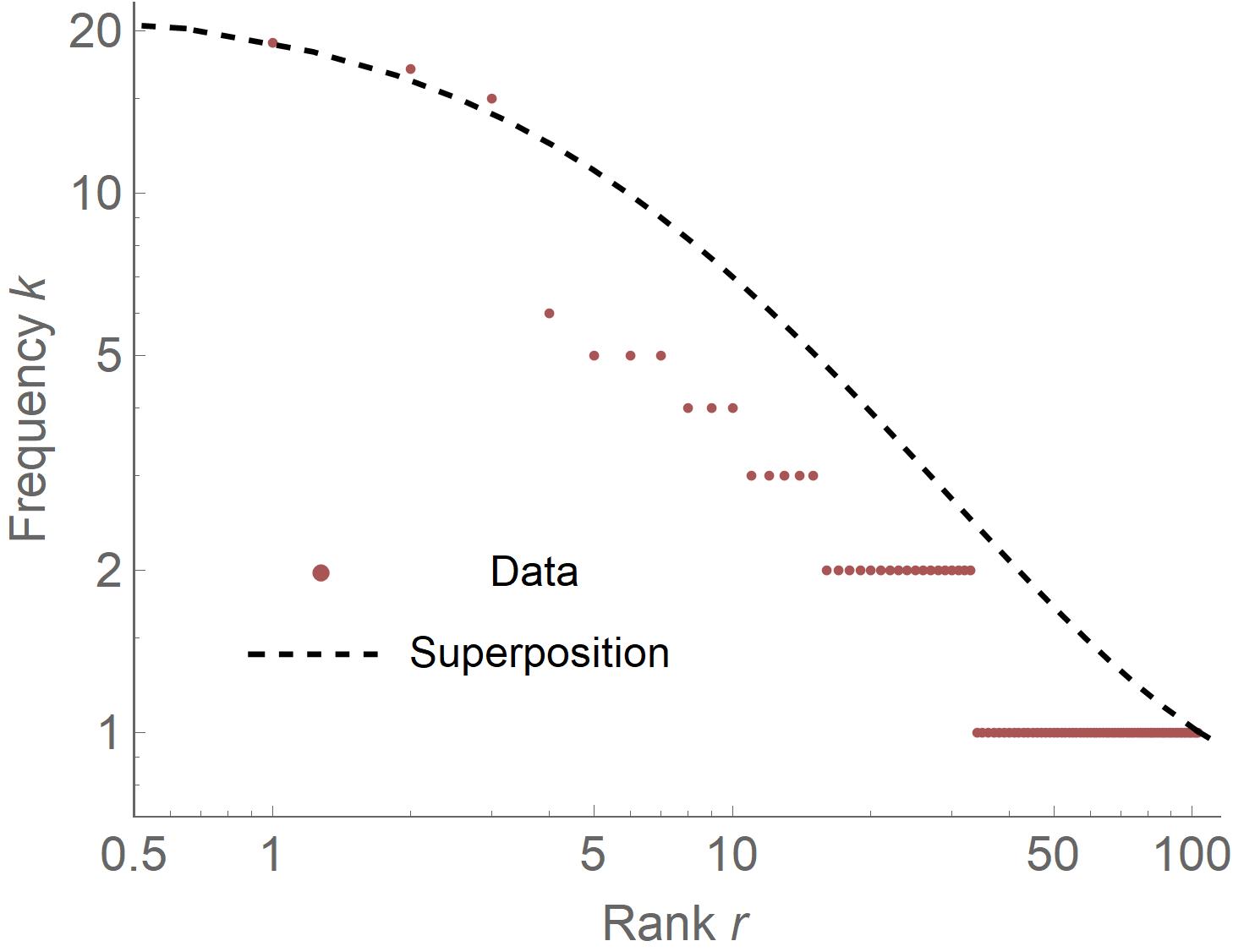}\label{fig:ZipfDiag_RS}}\caption{\label{fig:rusTxt} Interpretation of a short Russian text. (a) probability of word frequency $p(k)$, (b) Zipf's diagram $r(k)$} 
\end{figure*}

We took an experiment with a short excerpt of Russian text \footnote{We took the first 4 paragraphs as data in Татьяна Эсмантова, ``Русский язык: 5 элементов'' уровень А1, Санкт-Петербург, ``Златоуст'', 2013, 250 стp.,  ``Я учусь учиться'' 
}. We can find that Zipf’s and Heaps’ laws hold even for the non-English and very short text. Figure~\ref{fig:rusTxt} shows the result. The data is compared with the superposition model in $p(k)$ and Zipf's diagram.

\begin{figure*}
\centering
\subfigure[]{\includegraphics[width=0.3\textwidth]{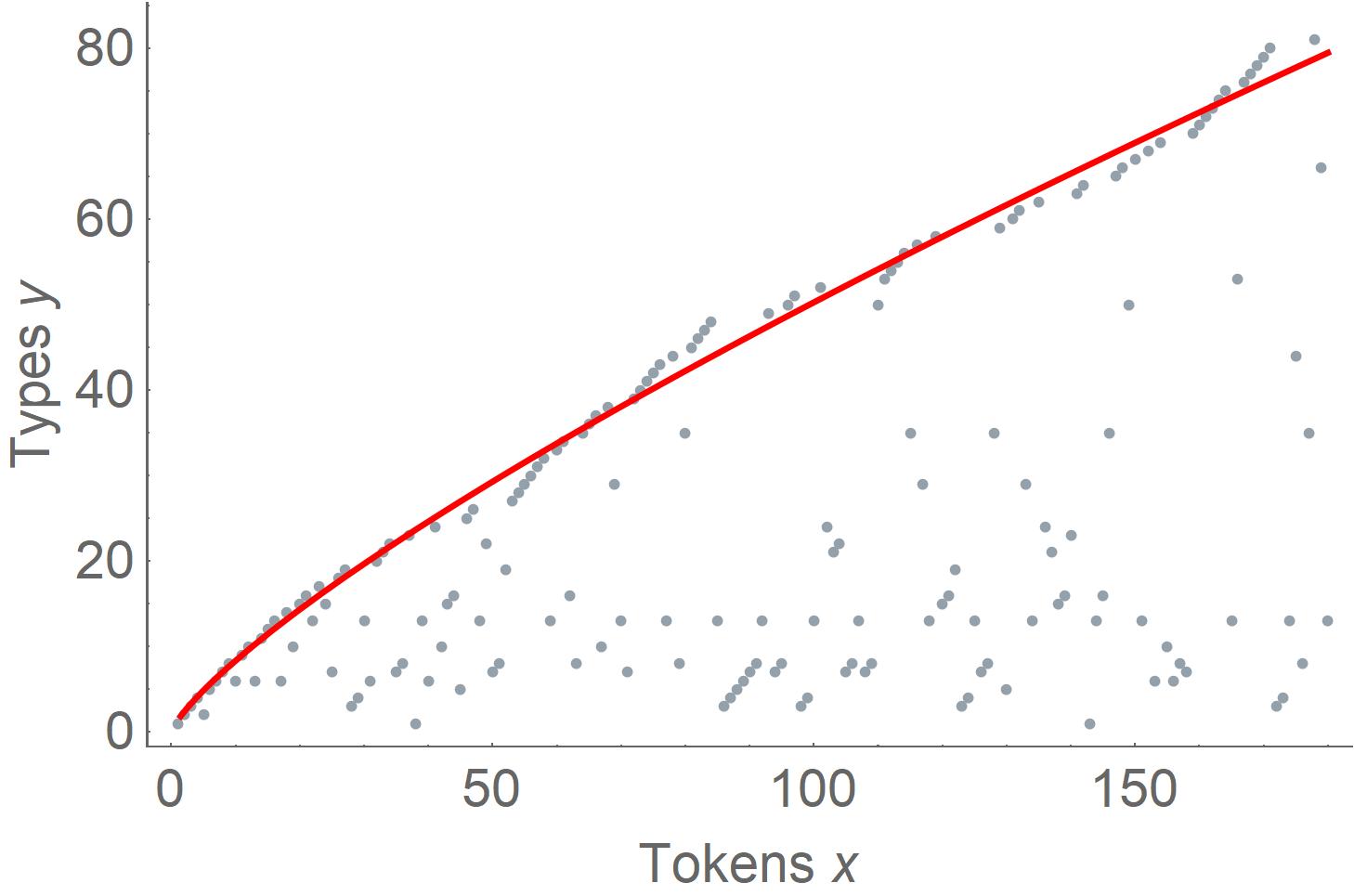}\label{fig:HeapsDiag_MC}}  
\subfigure[]{\includegraphics[width=0.3\textwidth]{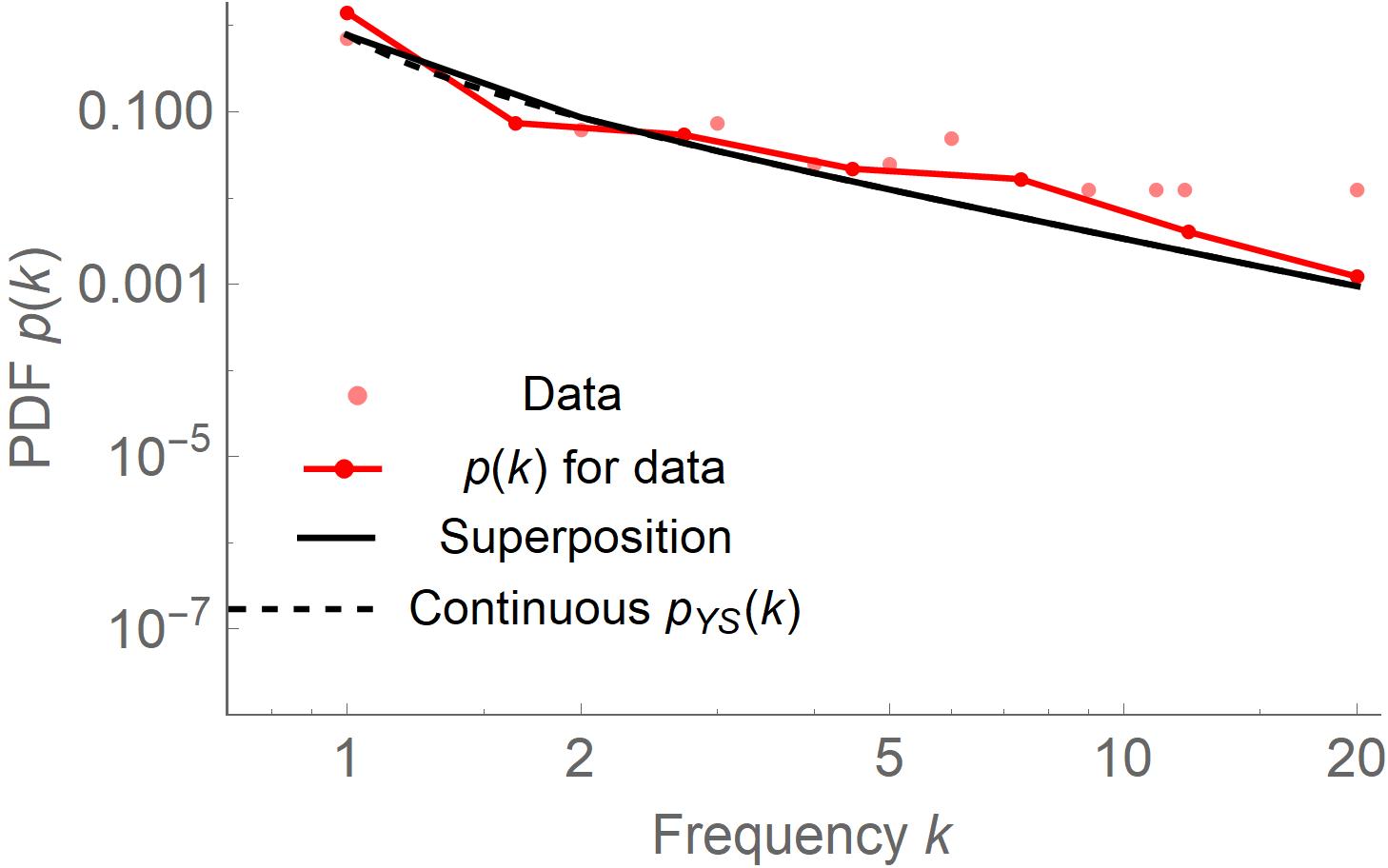}\label{fig:kfdiagram_MC}} 
\subfigure[]{\includegraphics[width=0.3\textwidth]{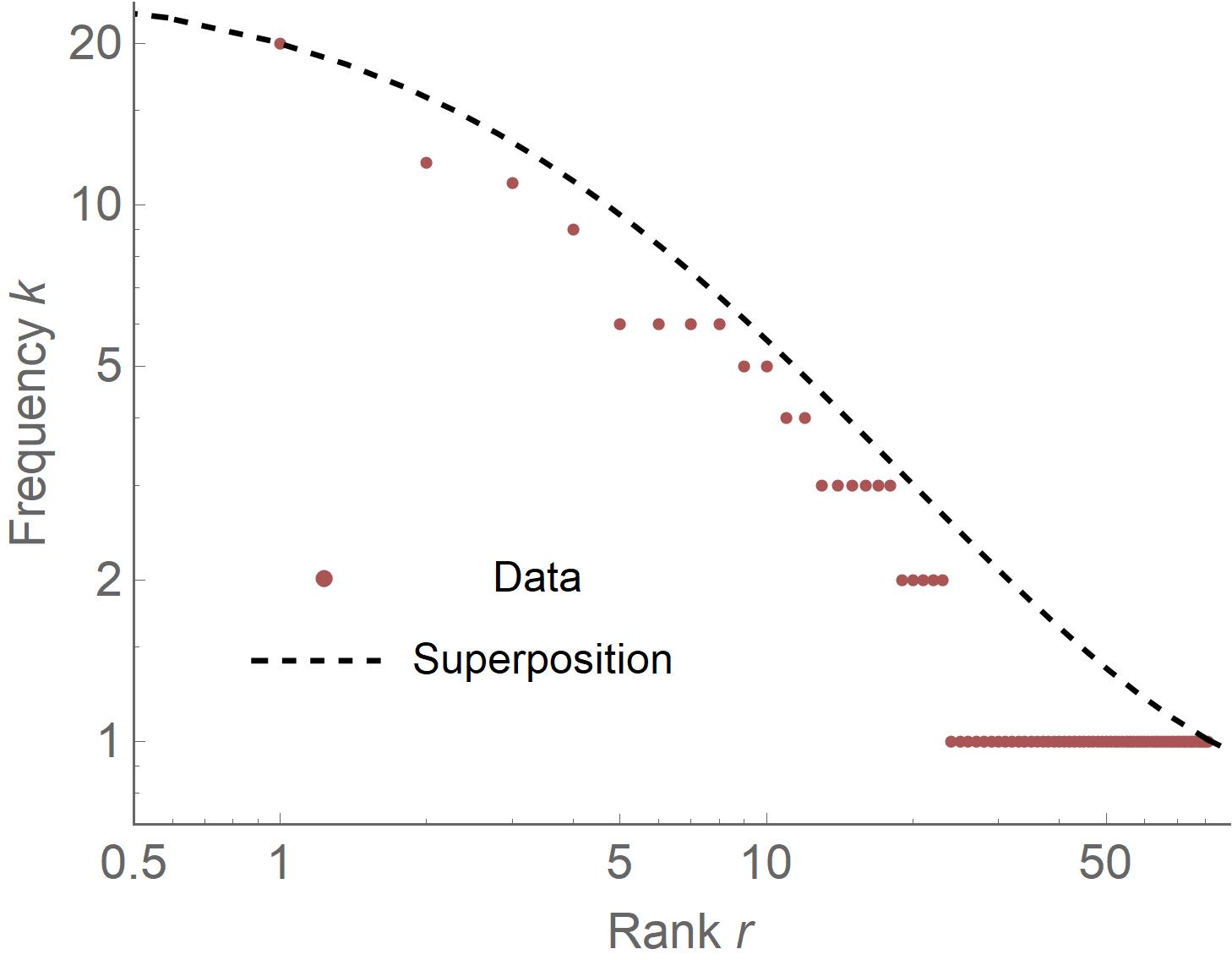}\label{fig:ZipfDiag_MC}}
\caption{\label{fig:listCorTitle} Interpretation of a list of more correlated titles. (a) Heaps' diagram (b) probability of word frequency (c) Zipf's diagram} 
\end{figure*}

\begin{figure*}
\centering
\subfigure[]{\includegraphics[width=0.3\textwidth]{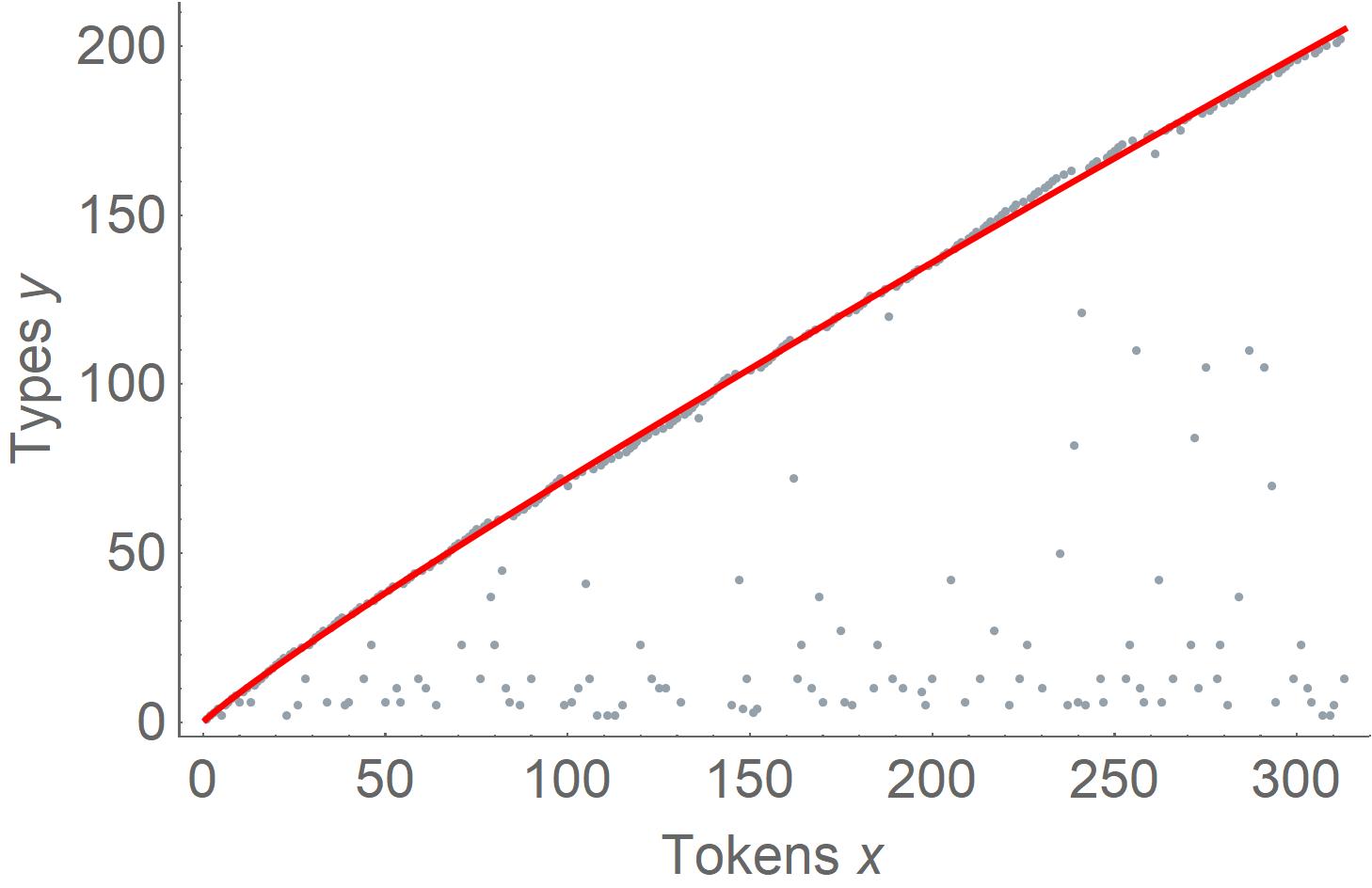}\label{fig:HeapsDiag_LC}}  
\subfigure[]{\includegraphics[width=0.3\textwidth]{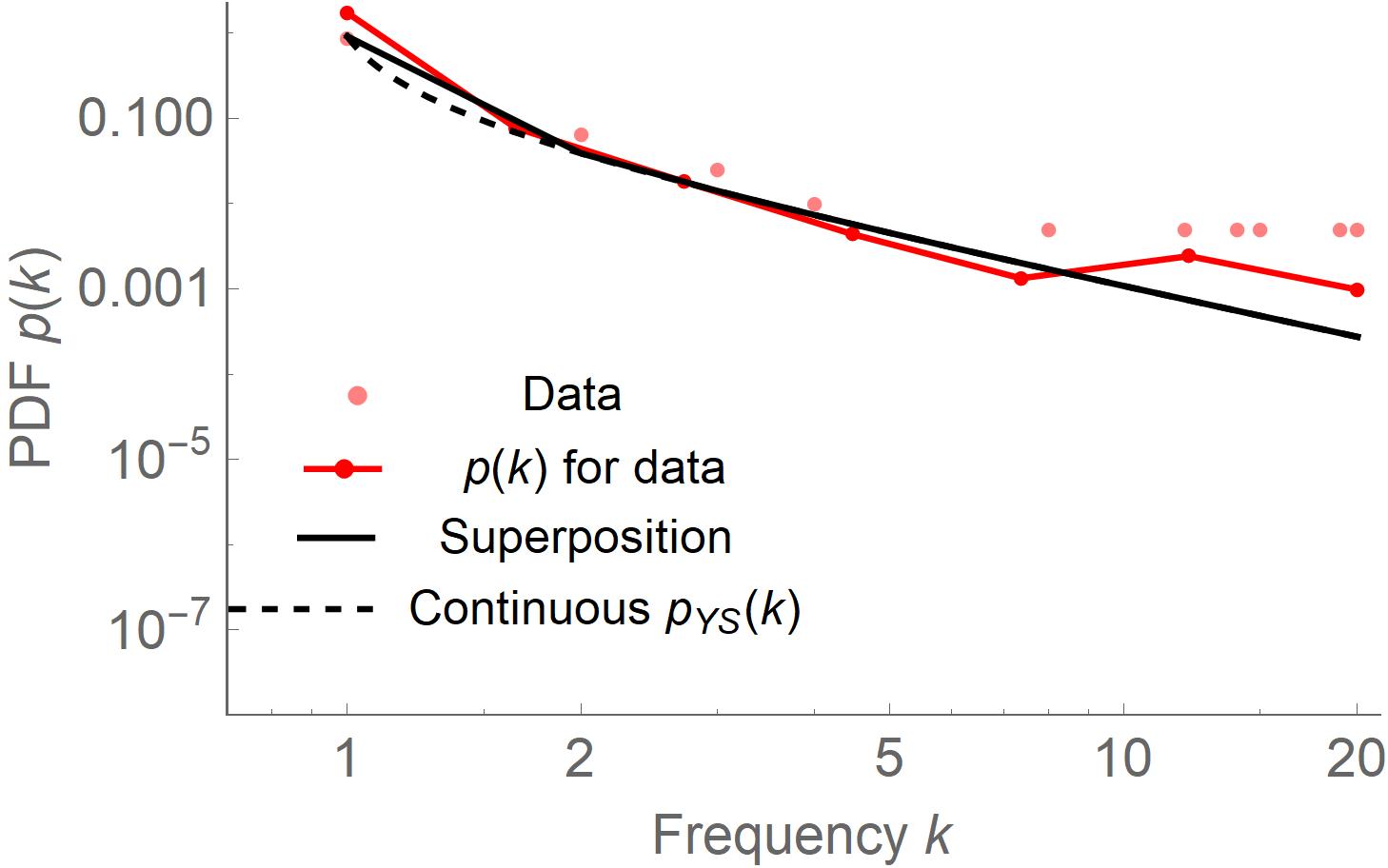}\label{fig:kfdiagram_LC}} 
\subfigure[]{\includegraphics[width=0.3\textwidth]{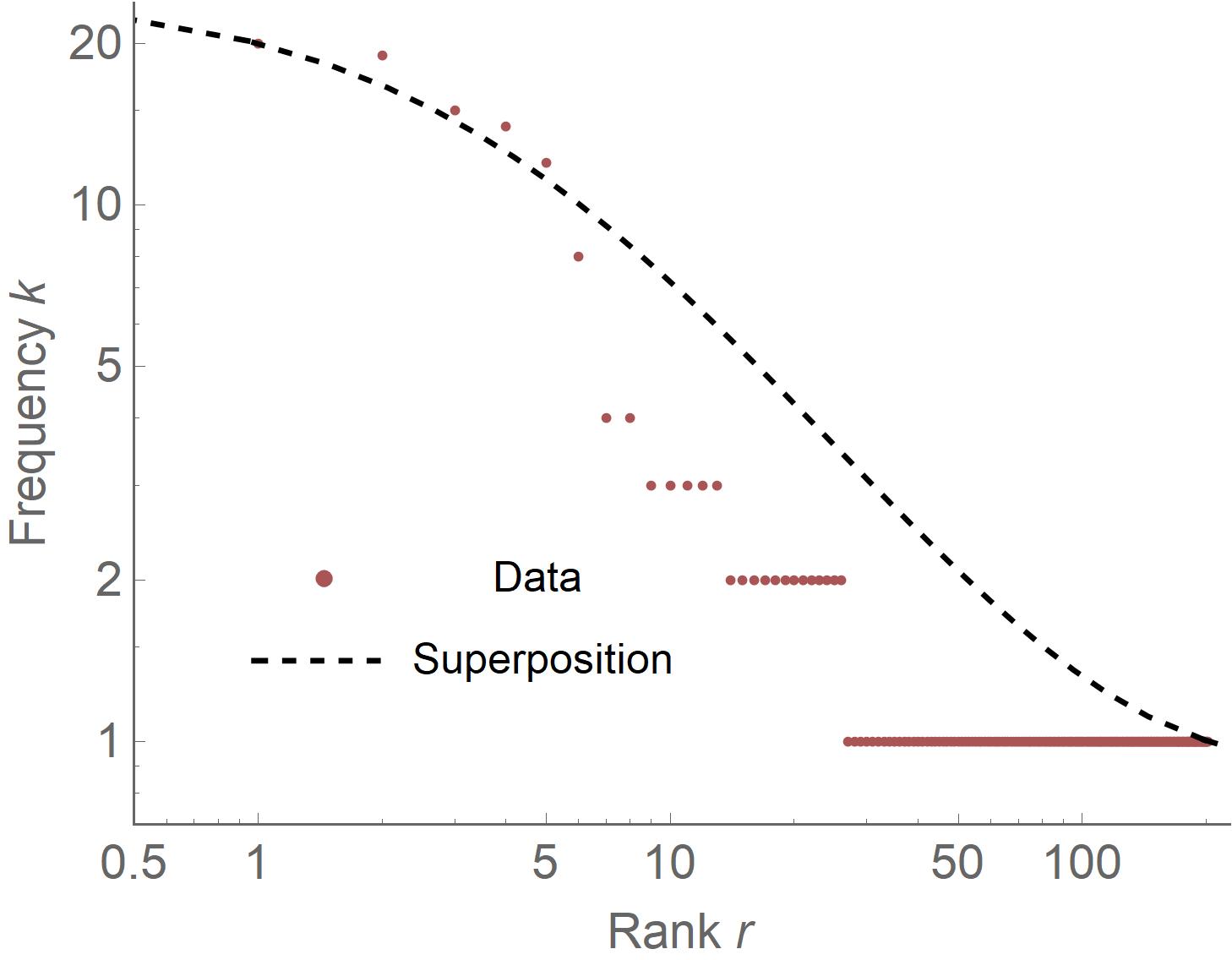}\label{fig:ZipfDiag_LC}}
\caption{\label{fig:listLesCorTitle} Interpretation of a list of less correlated titles. (a) Heaps' diagram (b) probability of word frequency (c) Zipf's diagram} 
\end{figure*}

We also analyze both lists of 20 titles of academic articles: one of a common specialized subject such as cosmology and another of randomly selected subjects. A correlation between the content of sentences affects the slope of Zipf’s law or Heaps’ exponent. If the titles are of the same field, which implies a greater correlation between them, then Heaps’ exponent and the slope in the histogram for the frequency, in particular for rare words, lower. If the titles are of less correlative themes, Heaps’ exponent tends to unity, i.e., there occur almost only new words in text, and the slope in the histogram increases as well. Figure~\ref{fig:listCorTitle} shows the result for the list of correlated titles and Figure~\ref{fig:listLesCorTitle} for that of randomly selected ones.  In Fig.~\ref{fig:listLesCorTitle} Heaps' exponent $\beta= 0.91$ and $A=3.23$ so that $A$ is beyond 2 which is at odds with the previous result between Heaps' and Zipf's laws: $A = 1 + \beta$ (see property~\ref{th:prop2}). However, the approximate power law is kept for rare words.
   
\begin{figure*}
\centering
\subfigure[]{\includegraphics[width=0.3\textwidth]{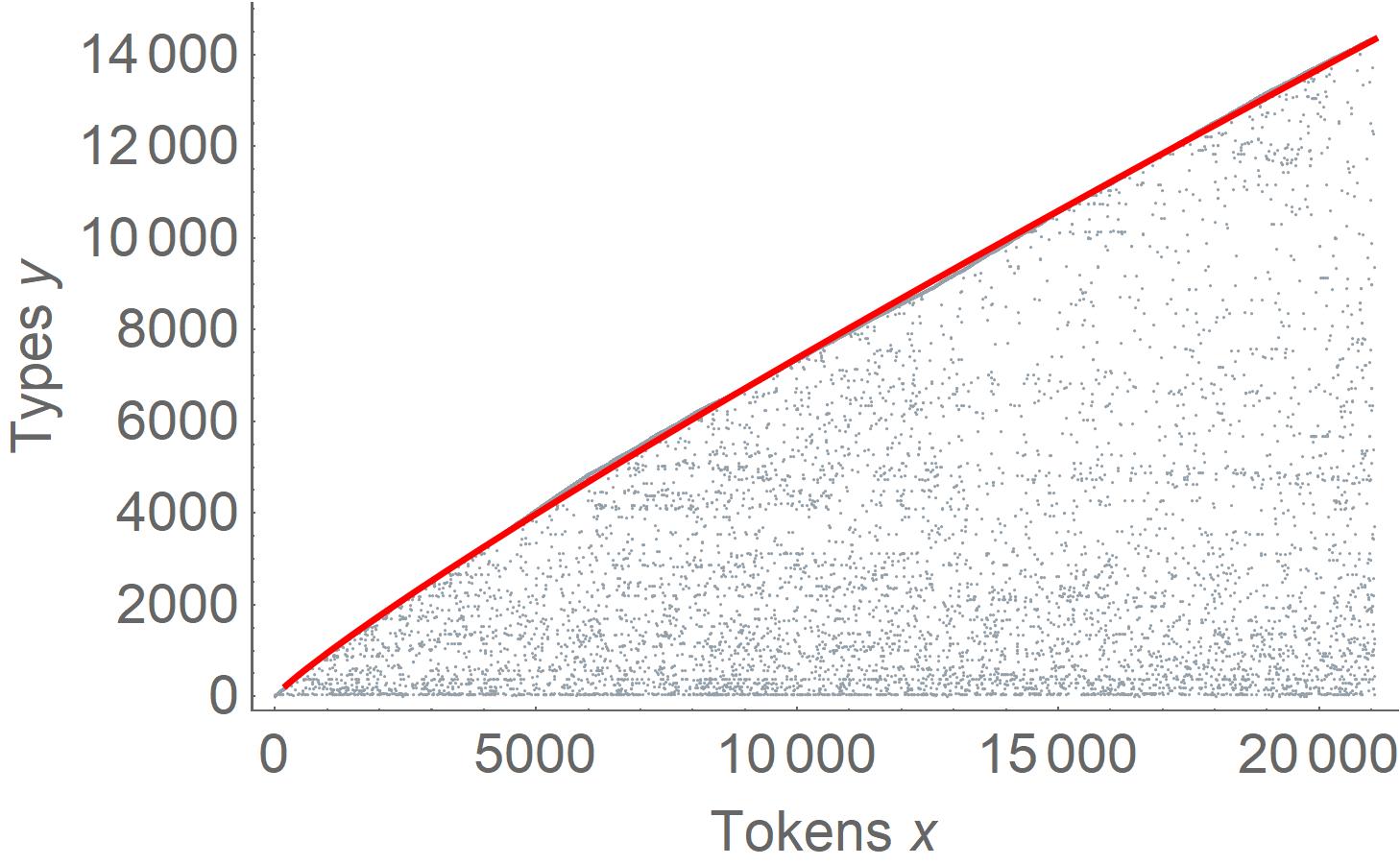}\label{fig:HeapsDiag_MD2}}  
\subfigure[]{\includegraphics[width=0.3\textwidth]{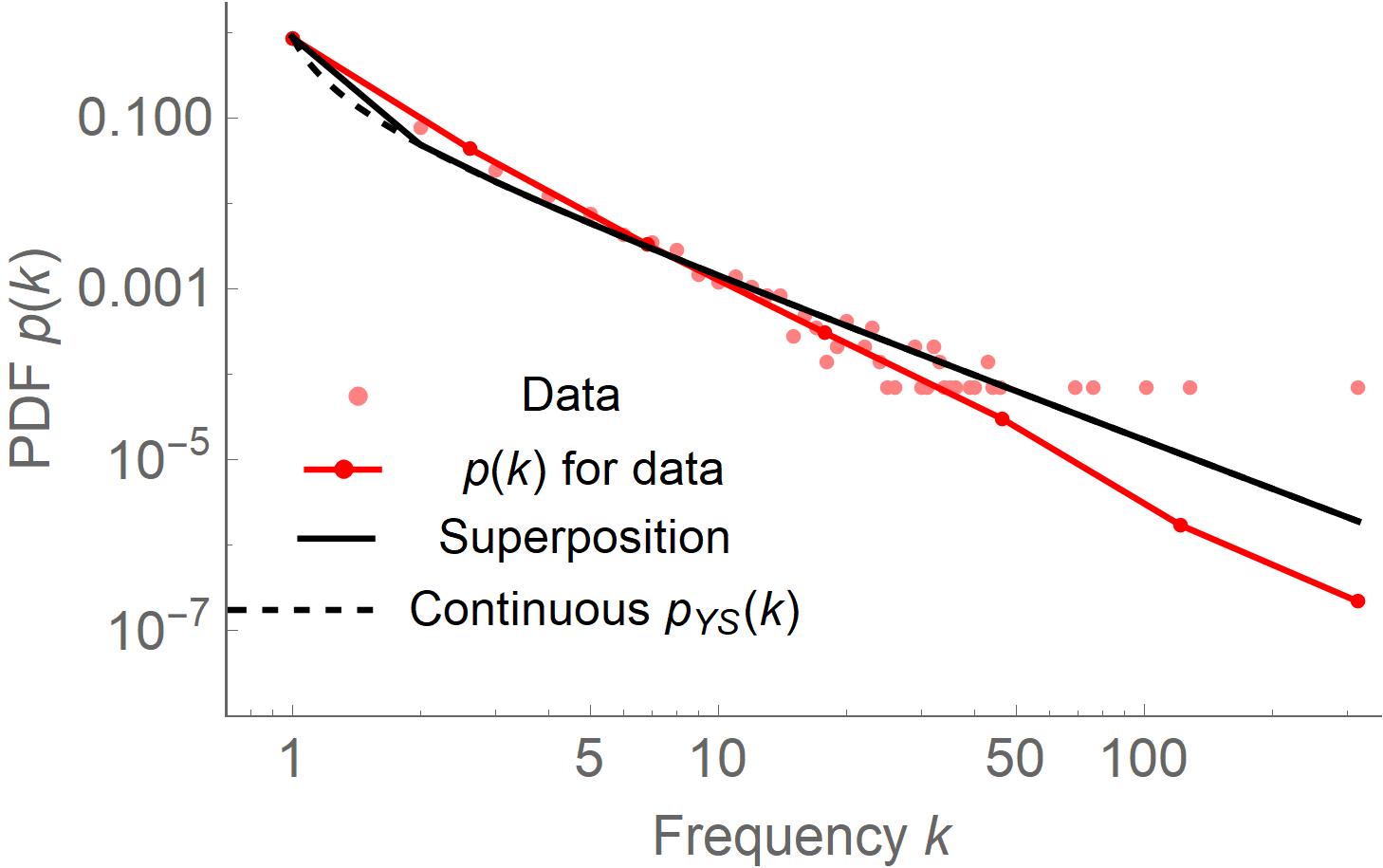}\label{fig:kfdiagram_MD2}} 
\caption{\label{fig:gram2} Interpretation of 2-grams of $Moby Dick$. (a) Heaps' diagram, (b) the probability of word frequency, $p(k)$} 
\end{figure*}

Property~\ref{th:prop2} of $A\rightarrow\infty$ in $\beta\rightarrow 1$ is observed in $n$-grams as well. When $n$ grows, the set of $n$-grams converges to a dictionary consisted of only ``new'' words, that is, $\beta\rightarrow 1$. Of course, in that case a steep exponent $A$ of $p(k)$ is observed for $k\approx1$, which means that $p(k=2) \ll p(k=1)$. For the excerpt of $Moby Dick$, while 1-gram gives $\beta = 0.74$, 2-grams gives even $\beta = 0.89$ and 3-grams $\beta = 0.98$ (even numerical inference of best-fit parameters is impossible because it seems that $\beta > 1$). Figure~\ref{fig:gram2} shows Heaps' and $p(k)$ diagrams for 2-grams. In spite of extreme condition, the $p(k)$ diagram shows a good coincidence with data and the superposition model.

\end{document}